\documentclass[a4paper,11pt]{article}
\usepackage[T1]{fontenc}
\usepackage{lmodern}
\usepackage{float}
\usepackage{array}
\usepackage{jheppub} 
\usepackage{lineno}
\usepackage{microtype}
\usepackage{verbatim}
\usepackage{amsmath}
\usepackage{mathtools}
\usepackage{amssymb}
\usepackage{mathrsfs}
\usepackage{graphicx}

\usepackage{needspace}
\hypersetup{hidelinks}

\usepackage{tikz}

\newcommand{\be}{\begin{equation}}
\newcommand{\ee}{\end{equation}}
\newcommand{\bea}{\begin{eqnarray}}
\newcommand{\eea}{\end{eqnarray}}

\makeatletter
\gdef\@fpheader{}
\makeatother
\begin{document}

\title{\boldmath 
Time-Dependent Classical and Quantum Integrability in Sigma Models: From the $\lambda$-Model to the Perturbed $SU(2)_1$ WZW Model}

\author[a,b]{Parameshwar R. Pasnoori}
\affiliation[a]{Department of Physics, University of Maryland, College Park, MD 20742, USA}
\affiliation[b]{Mani L. Bhaumik Institute for Theoretical Physics, Department of Physics and Astronomy, \\ University of California, Los Angeles, CA 90095, USA}

\emailAdd{parmesh12@g.ucla.edu}

\abstract{We use the Hamiltonian formulation of the $\lambda$-model with a time-dependent coupling to compare its classical Lax integrability with quantum integrability in the isotropic current-current deformation of the $SU(2)_1$ Wess-Zumino-Witten (WZW) model. Starting from the classical WZW current algebra at fixed level, we derive the equations of motion for the coupled left- and right-moving currents and construct a generalized Lax connection whose flatness determines the integrable time-dependent coupling strength. The equal-time Poisson brackets of the Lax matrix take the generalized Maillet form, and the corresponding $r/s$ kernels satisfy the extended classical Yang-Baxter relation. A coupling redefinition and a reparametrization of time relate this current-algebra construction to the $\lambda$-model with local couplings. At level one, non-Abelian bosonization identifies the deformed WZW model with the interacting isospin sector of the chiral-invariant $SU(2)$ Gross-Neveu model. In its exact generalized Bethe ansatz (GBA) solution, the quantum Yang-Baxter equation constrains the time-dependent coupling strength. In the universal weak-coupling regime, it has the same leading inverse-affine dependence on time as that imposed by classical Lax flatness associated with the $\lambda$-model. The common weak-coupling equation coincides with the one-loop renormalization-group equation of the corresponding static model when time is identified with the logarithm of the ultraviolet cutoff.

In the GBA construction, periodic boundary conditions lead to quantum Knizhnik-Zamolodchikov equations, whose off-shell Bethe ansatz solution determines the exact many-body wavefunction. We express this wavefunction in terms of a Yang-Yang functional at any nonzero drive rate. In a controlled limit of slow variation, the leading saddle-point equations reproduce the Bethe equations of the Hamiltonian at fixed time. We identify the additional conditions under which a state prepared in an instantaneous eigenstate follows it adiabatically. The instantaneous spectrum exhibits dimensional transmutation as the dimensionless interaction generates a mass gap for a bulk isospin excitation. Keeping this gap finite in the continuum limit fixes a characteristic time from which the gap decays exponentially on the decreasing-coupling branch. For fast driving, the fermionic Hamiltonian is described by a weakly perturbed $SU(2)_1$ WZW model for the isospin sector along with a decoupled charge sector, with the mass gap in the isospin sector tending to zero at fixed ultraviolet cutoff. We further discuss the comparison between classical and quantum integrability of higher-level WZW models and sigma models based on noncompact Lie groups.}

\maketitle
\flushbottom

\section{Introduction}
\label{sec:intro}

Nonlinear sigma models provide a common framework for classical integrability, quantum field theory, and string theory. Their fields take values in a target manifold, whose geometry determines the interactions. In integrable examples, the classical equations of motion can be written as the flatness condition for a Lax connection, from which one can construct conserved quantities under suitable boundary conditions. In string theory, the fields on the worldsheet describe the embedding of a string into the target spacetime, and the target-space metric and other background fields determine the worldsheet interactions. Integrable worldsheet sigma models provide methods for studying both classical string dynamics and the quantum spectrum \cite{tseytlin2026,Hoare_2022}. A prominent example arises in the AdS/CFT correspondence \cite{Maldacena}, where integrability appears both in the worldsheet theory of strings on $\mathrm{AdS}_5\times S^5$ and in planar four-dimensional $\mathcal N=4$ supersymmetric Yang-Mills theory. On the string side, a Lax description allows the construction of classical conserved quantities. On the gauge-theory side, the one-loop planar dilatation operator is represented by an integrable quantum spin-chain Hamiltonian, allowing anomalous dimensions of local operators to be calculated using the Bethe ansatz. These classical and quantum integrable structures provide methods for comparing string energies with the scaling dimensions of gauge-theory operators
\cite{NBeisert04,Arutyunov_2009,Gromov_2009,Kazakov_2004}. Related methods also apply to lower-dimensional AdS backgrounds \cite{Sfondrini_2015}.

The principal chiral model and symmetric-space sigma models provide simpler settings to study the Lax representation of classical integrable models \cite{Zakharov1973,EICHENHERR,Forger}. Their integrable deformations show how the target-space geometry can change while preserving classical integrability \cite{klimcikdef1,Delduc2,HoareSSSM}. The conformal Wess-Zumino-Witten (WZW) model admits both a sigma-model formulation in terms of a group-valued field and a description in terms of left- and right-moving currents \cite{WESSWZW,WITTENWZW,WittenBosonization1984}. Classically, the Poisson brackets of these currents give two commuting copies of the affine current algebra, one for each chirality. These classical currents also enter the Hamiltonian formulation of the integrable $SU(2)$ $\lambda$-model. Its sigma-model action is obtained by gauging a common $SU(2)$ symmetry of a principal chiral model and a WZW model and then eliminating the auxiliary gauge fields \cite{Sfetsos2014}. In the Hamiltonian formulation, the left and right currents retain their WZW Poisson brackets, while the deformation parameter enters the coefficients of the quadratic current terms, including the interaction between the two chiral sectors \cite{HollowoodMiramontesSchmidtt2014}. Upon quantization at level $k$, the WZW Hilbert space decomposes into a direct sum of tensor products of representations of the left- and right-moving affine current algebras. The current-current deformation thus admits a Hamiltonian formulation in terms of the WZW currents at both the classical and quantum levels. We use this formulation to ask how classical and quantum integrability constrain the time dependence of the coupling between the left and right currents. Hoare, Levine, and Tseytlin showed that, for several classes of sigma models including the $\lambda$-model, the requirement of a generalized Lax connection selects particular time dependences of the coupling \cite{Benhoare}. In this work, we develop a current-algebra formulation of classical integrability for a time-dependent isotropic current-current deformation of the $SU(2)$ WZW model and compare it with the generalized Bethe ansatz (GBA) formulation of quantum integrability for the fermionic Hamiltonian of the perturbed level-one model \cite{PasnooriGrossNeveu}.

First we consider the classical model and introduce time dependence through the coupling between the left and right currents. Starting from their Poisson brackets and the deformed Hamiltonian, we derive the classical equations of motion and construct a generalized Lax connection directly in terms of these currents. Its flatness condition determines the allowed time-dependent coupling strength compatible with integrability. We derive the equal-time Poisson brackets of the Lax matrix in generalized Maillet form. The position dependence of the spectral variable produces an additional derivative term in these brackets, and we verify that the resulting $r/s$ kernels satisfy the extended classical Yang-Baxter relation. We then relate this construction to the local-coupling $\lambda$-model of Hoare, Levine, and Tseytlin through an explicit coupling redefinition and reparametrization of time. This establishes the connection between their sigma-model construction and the classical current formulation used here, in which the Hamiltonian and interaction coefficient can be compared directly with their quantum counterparts.

For the quantum construction, we use non-Abelian bosonization to identify the deformed $SU(2)_1$ WZW model with the interacting isospin sector of the chiral-invariant $SU(2)$ Gross-Neveu model, a fermionic quantum field theory in two spacetime dimensions. In the GBA solution of this model \cite{PasnooriGrossNeveu}, requiring the bare two-particle $S$-matrices to satisfy the quantum Yang-Baxter equation determines the allowed time dependence of the interaction. We compare this independently derived condition on the time-dependent coupling strength imposed by quantum integrability with the one obtained from classical Lax flatness in their common universal weak-coupling regime. The GBA also gives the explicit many-body wavefunction satisfying the time-dependent Schr\"odinger equation. We show that this wavefunction admits an exact representation in terms of a Yang-Yang functional at arbitrary drive rate. We then analyze the wavefunction to extract the physical consequences of the drive.

For a classical theory with time-independent couplings in $1+1$ dimensions, a Lax connection has the form
\begin{align}
\mathscr L(w)=\mathscr L_t(t,x;w)\,dt+\mathscr L_x(t,x;w)\,dx,\qquad\mathscr L_\mu(t,x;w)=\sum_a \mathscr L_\mu^a(t,x;w)\,T_a .
\end{align}
Here $\mu=t,x$ labels the spacetime direction and $a$ labels a basis $T_a$ of a Lie algebra $\mathfrak g$. In the present $SU(2)$ setting, $a=1,2,3$. The generators act on an auxiliary representation space and carry the internal Lie-algebra index. Thus, at each spacetime point, the connection assigns a Lie-algebra-valued quantity to each of the two coordinate directions. The parameter $w$ is the spectral parameter and labels a family of such connections. Their coefficients are constructed from the physical fields so that requiring the connection to be flat for every $w$ reproduces the classical equations of motion. The curvature of this connection is given by 
\begin{align}
\mathscr F_{tx}(t,x;w)=\partial_t\mathscr L_x(t,x;w)-\partial_x\mathscr L_t(t,x;w)+[\mathscr L_t(t,x;w),\mathscr L_x(t,x;w)].
\label{introFlatness}
\end{align}
The connection is flat when \(\mathscr F_{tx}=0\). Since the curvature takes values in the same Lie algebra as the connection, this condition requires every component in the expansion \(\mathscr F_{tx}=\sum_a\mathscr F_{tx}^{a}T_a\) to vanish. The Lax representation is constructed so that flatness for every value of the spectral parameter \(w\) encodes the classical field equations.

To understand the significance of this condition, introduce a vector \(\Phi(t,x;w)\) in the auxiliary representation space on which the generators \(T_a\) act. For a given classical field configuration, consider the linear equations
\begin{align}
\bigl(\partial_t+\mathscr L_t(t,x;w)\bigr)\Phi(t,x;w)&=0,
\nonumber\\
\bigl(\partial_x+\mathscr L_x(t,x;w)\bigr)\Phi(t,x;w)&=0.
\label{introAuxiliary}
\end{align}
These equations specify how \(\Phi\) changes along the time and space directions. Their compatibility follows from
\begin{align}
[\partial_t+\mathscr L_t,\partial_x+\mathscr L_x]\Phi=\mathscr F_{tx}\Phi.
\end{align}
For each fixed spectral parameter $w$, one may specify the auxiliary vector $\Phi$ at a single spacetime point and use the Lax connection to parallel transport it to any other point in the region. If the region is simply connected and the connection is nonsingular, flatness ensures that every path joining the two points gives the same final value of $\Phi$. Parallel transport therefore extends the prescribed initial value to a consistent solution of both auxiliary equations throughout the region. The nonlinear field equations are thus expressed as the compatibility condition for this family of linear transport problems.

Now consider transporting the auxiliary vector once around the system of length $L$ at a fixed time $t$. Since the matrices $\mathscr L_x(t,x;w)$ at different spatial points need not commute, the resulting transport is described by a path-ordered exponential. This defines the monodromy matrix,
\begin{align}
T(w;t)=\mathcal P\exp\!\left[-\int_0^L dx\,\mathscr L_x(t,x;w)\right],\qquad\Phi(t,L;w)=T(w;t)\Phi(t,0;w),\label{pathorderedtransport}
\end{align}
where $\mathcal P$ denotes ordering along the spatial path. To determine how $T(w;t)$ changes with time, we use the compatibility of spatial parallel transport with the temporal auxiliary equation (\ref{introAuxiliary}). Differentiating the relation (\ref{pathorderedtransport}) and using the temporal equation at both endpoints gives
\begin{align}
\partial_t T(w;t)=-\mathscr L_t(t,L;w)\,T(w;t)+T(w;t)\,\mathscr L_t(t,0;w).
\end{align}
The first term accounts for the evolution of the auxiliary vector at the final endpoint and the second for its evolution at the initial endpoint. If the temporal component of the Lax connection is periodic,
$\mathscr L_t(t,L;w)=\mathscr L_t(t,0;w)$,
the two endpoint contributions give
\begin{align}
\partial_t T(w;t)=[T(w;t),\mathscr L_t(t,0;w)].
\end{align}
For each fixed spectral parameter $w$, the monodromy matrices at different times are related by a similarity transformation. To see this, let $U(t,t_0;w)$ denote parallel transport from $t_0$ to $t$ at $x=0$. The temporal auxiliary equation gives
\begin{align}
U(t,t_0;w)=\mathcal P_t\exp\!\left[-\int_{t_0}^{t}ds\,\mathscr L_t(s,0;w)\right],
\end{align}
where $\mathcal P_t$ denotes time ordering. By periodicity of $\mathscr L_t$, the same matrix describes temporal transport at $x=L$. Now transport the auxiliary vector from $(t_0,0)$ to $(t,L)$ along the edges of the spacetime rectangle. Transporting first in time and then in space gives $T(w;t)U(t,t_0;w)$, while transporting first in space and then in time gives $U(t,t_0;w)T(w;t_0)$. Flatness equates the two products, giving rise to the finite form of the zero-curvature condition,
\begin{align}
T(w;t)U(t,t_0;w)=U(t,t_0;w)T(w;t_0).
\end{align}
Multiplying on the right by $U(t,t_0;w)^{-1}$ gives
\begin{align}
T(w;t)=U(t,t_0;w)\,T(w;t_0)\,U(t,t_0;w)^{-1}.
\end{align}
Thus, for each fixed $w$, the classical monodromy matrix evolves by a similarity transformation. Its eigenvalues are therefore conserved, so the evolution is isospectral. Consequently, the spectral invariants of $T(w;t)$ are conserved. In particular, the traces of its powers satisfy
\begin{align}
\frac{d}{dt}\operatorname{tr}\!\left[T(w;t)^n\right]&=n\,\operatorname{tr}\!\left[T(w;t)^{n-1}[T(w;t),\mathscr L_t(t,0;w)]\right]=0,\qquad n=1,2,\ldots ,
\end{align}
where the final equality follows from cyclicity of the trace. Thus, the spectral family of monodromy matrices supplies conserved classical quantities for every fixed $w$.

We now allow a coupling $h_{\rm d}$ in the sigma-model Lagrangian to vary with time according to a prescribed function $h_{\rm d}(t)$. The resulting field equations can contain terms proportional to $\dot h_{\rm d}(t)$. Substituting $h_{\rm d}(t)$ into the static Lax connection also generates additional terms in its curvature. To retain a Lax description, these contributions must be accounted for so that the flatness condition reproduces the time-dependent field equations. Hoare, Levine, and Tseytlin extended the classical Lax construction to several classes of time-dependent sigma models by allowing the spectral argument of the static connection to vary over spacetime \cite{Benhoare}. Making the coupling dependence explicit, the generalized connection takes the schematic form
\begin{align}
\widetilde{\mathscr L}_\mu(t,x;w)=\mathscr L_\mu\bigl(t,x;z(t,x;w);h_{\rm d}(t)\bigr),\qquad \mu=t,x.
\end{align}
Here $z(t,x;w)$ replaces the original constant spectral argument, while $w$ labels the family of connections and is held fixed when taking spacetime derivatives. Derivatives of $z$ contribute additional terms to the curvature. Requiring the resulting flatness condition to reproduce the time-dependent field equations determines equations for $\partial_t z$ and $\partial_x z$. Their mutual compatibility then constrains the allowed time dependence of $h_{\rm d}(t)$.

In the examples studied in Ref.~\cite{Benhoare}, this compatibility condition gives the one-loop renormalization-group equation of the corresponding static model, after an appropriate normalization and orientation of time. Thus, the allowed dependence of the coupling on physical time has the same functional form as its dependence on the logarithm of the renormalization scale, up to a rescaling and possible reversal of the time variable. These time-dependent models also admit a geometric realization in higher-dimensional conformal sigma models with a null Killing direction and a dilaton. A light-cone gauge choice identifies a null target-space coordinate with worldsheet time, producing the explicitly time-dependent couplings of the two-dimensional model \cite{Benhoare}.

The gauge-theory approach to integrability provides a broader setting for this classical construction. Following Atiyah's proposal to obtain the Yang-Baxter equation from a higher-dimensional theory \cite{Atiyah1987}, Costello introduced a four-dimensional gauge-theory construction that was subsequently developed with Witten and Yamazaki \cite{costelloa,costellob,Costello1,Costello2}. In its simplest setting, this Chern-Simons theory is defined on the product of a real two-dimensional topological plane and a complex spectral plane. Each Wilson line extends along a curve in the topological plane at a specified point in the spectral plane, whose coordinate supplies its spectral parameter. When two such curves cross in the topological plane, the interaction between the Wilson lines through the gauge field defines a quantum $R$-matrix acting on the tensor product of their representation spaces. Topological invariance along the first plane relates different arrangements of three pairwise crossings, giving the quantum Yang-Baxter equation. Ordered products of these $R$-matrices define monodromy matrices, whose traces over the auxiliary representation space give transfer matrices. This provides a gauge-theory realization of the Yang-Baxter equations underlying the Bethe ansatz \cite{Costello1,Costello2}.

This four-dimensional framework also produces two-dimensional integrable field theories and their classical Lax connections through suitable surface defects \cite{costello3}. More recently, Komatsu, Sakamoto, Wallberg, and Yamazaki extended this construction by allowing the one-form entering the Chern-Simons action to depend on the two-dimensional spacetime coordinates. Their construction gives time-dependent classical integrable theories, including non-Abelian Thirring and $\lambda$-models, and relates the allowed coupling trajectories to one-loop renormalization-group flows \cite{KomatsuSakamotoWallbergYamazaki2026}. This provides a gauge-theory realization of the time-dependent classical models and their generalized Lax connections. Deriving the corresponding quantum $R$-matrices from Wilson-line crossings in the time-dependent gauge theory requires a further quantum analysis.

For the time-dependent current-current interaction, the GBA provides quantities against which such a gauge-theory construction can be tested. The wavefunction is constructed by ordering the particles in the configuration space. The amplitudes corresponding to different orderings are related by the bare $S$-matrices which exchange the ordering of the particles. The time-dependent Hamiltonian fixes the bare two-particle $S$-matrix for fermions of opposite chirality and the consistency of the wavefunction fixes the bare $S$-matrix corresponding to the exchange of fermions with the same chirality and gives rise to the quantum Yang-Baxter equations, which impose conditions on the time-dependent strengths. Each bare $S$-matrix contains a scalar factor and a quantum $R$-matrix acting on the two isospins.  Applying periodic boundary conditions to the wavefunction gives rise to matrix difference equations whose solutions determine the exact form of the wavefunction. A gauge-theory realization of the time-dependent quantum model would therefore have to reproduce the $R$-matrices, the scalar factors, and the matrix difference equations described above. Conversely, the gauge-theory framework can suggest further models and Yang-Baxter algebras to which the GBA may be applied.

To formulate the quantum problem explicitly, we seek many-body states satisfying the time-dependent Schr\"odinger equation,
\begin{align}
i\partial_t|\Psi(t)\rangle=H(t)|\Psi(t)\rangle.
\label{introQuantumEvolution}
\end{align}
At each fixed time, diagonalizing $H(t)$ determines its instantaneous eigenvalues and eigenvectors. Expanding an evolving state in this basis introduces additional couplings between its components through the time dependence of the basis vectors. Solving the instantaneous eigenvalue problem therefore does not by itself determine the time-dependent wavefunction. For the current-current deformation considered here, we use its fermionic representation to construct exact solutions of Eq.~(\ref{introQuantumEvolution}) by the GBA.

To explain how the GBA formalism helps one construct time-dependent wavefunctions, we first recall the standard Bethe ansatz formulation for theories with constant couplings and its relation to factorized scattering. In relativistic integrable field theories, the higher conserved charges constrain multiparticle scattering to be elastic and to factorize into two-particle processes \cite{ZAMOLODCHIKOV}. For particles with rapidities $\theta_i$, consistency of this factorization requires the two-particle scattering matrices to satisfy the quantum Yang-Baxter equation \cite{BAXTER,Yang}
\begin{align}
S_{12}(\theta_1,\theta_2)S_{13}(\theta_1,\theta_3)S_{23}(\theta_2,\theta_3)
={}&S_{23}(\theta_2,\theta_3)S_{13}(\theta_1,\theta_3)S_{12}(\theta_1,\theta_2).
\label{QYB}
\end{align}
This equation ensures that different sequences of pairwise scatterings connecting the same initial and final configurations give the same result. Together with symmetry, unitarity, and crossing relations, it constrains the exact physical \(S\)-matrices and their bound-state content \cite{ZAMOLODCHIKOV}.

The same Yang-Baxter structure underlies the Bethe-ansatz construction of many-body wavefunctions \cite{Bethe1931,Hulthen}. In the microscopic coordinate formulation, particles are ordered in configuration space, with an amplitude assigned to each ordering. Amplitudes related by exchanging neighboring particles are connected by bare $S$-matrices determined from the interactions in the Hamiltonian. An ordered product of these matrices describes the transport of a particle around the system. Imposing periodic boundary conditions then gives an eigenvalue problem that can be solved using a commuting family of transfer matrices.\footnote{In a standard convention, the bare two-particle $S$-matrices can be written as $S^{\mathrm{bare}}_{ij}=s_{ij}R_{ij}(\xi_i,\xi_j)$, where $s_{ij}$ is a scalar factor and the relation of the spectral parameters $\xi_i$ to the particle quantum numbers and coupling is model dependent. The quantum monodromy and transfer matrices are
\[
\mathcal M_0(w)=R_{0N}(w,\xi_N)\cdots R_{01}(w,\xi_1),
\qquad
\tau(w)=\operatorname{tr}_0\mathcal M_0(w),
\]
where $0$ denotes the auxiliary space. The Yang-Baxter relation implies $[\tau(w),\tau(w')]=0$, providing a generating family of commuting operators. With the regular normalization $R_{ij}(\xi,\xi)=P_{ij}$, evaluating $\tau(w)$ at $w=\xi_i$ reproduces the corresponding particle transport operator, with the scalar factors restored.}
The algebraic Bethe ansatz constructs their eigenvectors, with the eigenstate conditions yielding the Bethe equations that constrain the allowed rapidities of the particles \cite{SklyaninQISM,Yang,Thacker,LiebFu}. Solving these equations identifies the ground state, which serves as the physical vacuum, and the excitations above it. Their scattering phases can then be extracted from the Bethe-ansatz solution \cite{trieste}, yielding the physical \(S\)-matrices that appear in Zamolodchikov's factorized-scattering description. The microscopic construction thus provides a route from the bare interactions in the Hamiltonian to the spectrum and scattering of physical excitations above the physical vacuum. Examples include the \(SU(2)\) Gross-Neveu model and its higher-rank generalizations \cite{AndreiLowensteinSU(N)}, the anisotropic $U(1)$ Thirring model \cite{Japaridze}, and the sine-Gordon model \cite{SklyaninQISM,Thacker}. Bethe-ansatz methods give access to their spectra, physical scattering matrices, and correlation functions \cite{Thacker,DestriLowenstein,Japaridze,trieste,korepin1993quantum,WiegmannGN}. The Gross-Neveu model illustrates how asymptotic freedom and dynamical mass generation enter the same exact solution \cite{GrossNeveu,AndreiLowenstein79}. A related example of dimensional transmutation is the emergence of the Kondo scale \cite{Anderson,Andrei80,AndreiLowensein81}.

The GBA extends the nested coordinate Bethe ansatz to time-dependent interactions \cite{PasnooriKondo,PasnooriGrossNeveu,pasnoori2026GCBA}. In both constructions the wavefunction is a sum over spatial orderings of the particles, with a spin amplitude for each ordering. For a static Bethe eigenstate, separating the plane waves leaves a constant spin amplitude corresponding to each ordering. In contrast, in the time-dependent model, the amplitude is a function of the spacetime coordinates of the particles. When a right mover and a left mover meet, the time-dependent Schr\"odinger equation relates the amplitudes on either side through a bare two-particle $S$-matrix evaluated at the physical collision time. Since the particles of the same chirality have equal velocities, the matrices relating their orderings are not constrained by the Hamiltonian. They are chosen so that all sequences of pairwise interchanges between two orderings yield the same amplitude. For three particles this requirement gives rise to the quantum Yang-Baxter equation, and the dependence of the $S$-matrix on the spacetime coordinates of the particles constrains the allowed time-dependent strengths.

For a system of length $L$, starting from a reference amplitude, a transport operator can be constructed by moving a particle past all the others, producing an ordered product of their two-particle $S$-matrices. Applying periodic boundary conditions relates two values of the reference amplitude for which that particle's null coordinate differs by $L$ and gives rise to a set of matrix difference equations. Their consistency requires that transporting two particles around the system in either order give the same final amplitude, which corresponds to the discrete zero-curvature condition for the transport operators. Each bare $S$-matrix can be expressed as a scalar factor multiplying a rational quantum $R$-matrix acting on the particle spins. Separating the corresponding scalar and spin parts of the amplitude reduces the matrix difference equations to a set of scalar analytic difference equations and the quantum Knizhnik-Zamolodchikov (qKZ) equations, respectively. The qKZ equations can be solved using the off-shell Bethe ansatz \cite{BabujianFlume1994, rishetikhin1,rishetikhin2} which provides the explicit form of the reference amplitude \cite{pasnoori2025Kondo2,PasnooriGrossNeveu}; applying the appropriate products of $S$-matrices then determines the amplitudes for all other particle orderings and constructs a many-body wavefunction satisfying Eq.~(\ref{introQuantumEvolution}).

We can now formulate the comparison between the driving conditions selected by the GBA and by a generalized classical Lax connection in terms of the same WZW currents. We keep the level $k$ explicit on the classical side and specialize to $k=1$ for the quantum construction. Consider the classical WZW action for a group-valued field $h(t,x)\in SU(2)$,
\begin{align}
S_{{\rm WZW},k}[h]={}&-\frac{k}{8\pi}\int_\Sigma dt\,dx\,\operatorname{tr}\!\left(h^{-1}\partial_\mu h\,h^{-1}\partial^\mu h\right)+\frac{k}{12\pi}\int_B\operatorname{tr}\!\left[(\widetilde h^{-1}d\widetilde h)^{\wedge3}\right],
\label{introWZWAction}
\end{align}
where $\partial B=\Sigma$ and $\widetilde h$ extends $h$ into $B$. We use Minkowski signature $(+,-)$ and the trace in the fundamental representation. The canonical formulation of this action gives the left and right currents and their Poisson brackets \cite{Gawedzki1991,PapadopoulosSpence1992}. We introduce a time-dependent isotropic interaction through the Hamiltonian
\begin{align}
H_{\rm cc}(t)=\frac{2\pi}{k}\int dx\,\bigl[\mathbf J_R^2+\mathbf J_L^2+2q(t)\,\mathbf J_R\cdot\mathbf J_L\bigr].
\label{introClassicalHamiltonian}
\end{align}
Here the current Poisson brackets retain their WZW form, while $q(t)$ controls the interaction between the two chiral sectors. This formulation allows us to derive the current equations directly from Hamilton's equations and subsequently express the same interaction in the fermionic quantum theory. Requiring the classical equations to admit a rational generalized Lax connection constrains the coupling to satisfy
\begin{align}
\dot q=Kq^2(1-q),
\label{introClassicalProtocol}
\end{align}
where $K$ is an integration constant setting the rate and direction of the drive. The same compatibility equations determine the spacetime dependence of the spectral variable.

The Hamiltonian in Eq.~(\ref{introClassicalHamiltonian}) is closely related to that of the integrable group $\lambda$-model \cite{Sfetsos2014,HollowoodMiramontesSchmidtt2014,ItsiosSfetsosSiampos2014}. Although the $\lambda$-model action depends nonlinearly on its deformation parameter, its Hamiltonian can be expressed in terms of the same WZW currents. We show that the coupling redefinition $q=2\ell/(1+\ell^2)$, where $\ell$ denotes the $\lambda$-model coupling, together with a coupling-dependent reparametrization of time, relates the two Hamiltonians. Applying these transformations to the integrability condition of the $\lambda$-model with time-dependent coupling reproduces Eq.~(\ref{introClassicalProtocol}). This establishes the connection between our current-algebra derivation and the generalized Lax construction of Hoare, Levine, and Tseytlin.

For the static $\lambda$-model, the Hamiltonian in terms of WZW currents and the Maillet algebra of the Lax matrix are both known \cite{HollowoodMiramontesSchmidtt2014,ItsiosSfetsosSiamposTorrielli2014}. Hoare, Levine, and Tseytlin constructed a generalized Lax connection for the $\lambda$-model with a time-dependent coupling and carried out their explicit Hamiltonian analysis for the principal chiral model \cite{Benhoare}. In Sec.~\ref{sec:classicalMaillet}, we use the WZW current Poisson brackets to derive the corresponding Lax algebra for the time-dependent current-current model. Since the spectral variable now depends on position, its spatial derivative contributes an additional term to the equal-time Poisson bracket of the Lax matrix. We obtain the resulting bracket in generalized Maillet form and verify that its $r/s$ kernels satisfy the extended classical Yang-Baxter relation.

The current-algebra formulation also provides a direct route to the quantum theory. Upon quantization, the classical current Poisson brackets become affine current commutators, and the products of currents in the Hamiltonian require a normal-ordering prescription. For $SU(2)$, normal ordering replaces the classical Sugawara coefficient proportional to $1/k$ with the quantum coefficient proportional to $1/(k+2)$. At level one, the currents admit an exact representation in terms of canonical fermions. In this representation, the current-current interaction becomes the spin interaction of the chiral-invariant $SU(2)$ Gross-Neveu model, with a decoupled $U(1)$ charge sector completing the fermionic theory. 

The exact time-dependent solution of the Gross-Neveu model was obtained in Ref.~\cite{PasnooriGrossNeveu}. We review this solution using the conventions adopted here for the comparison with the classical $\lambda$-model.  In the GBA formalism, as mentioned above, fermions with opposite chirality interact when their spacetime coordinates coincide, and the time-dependent Schr\"odinger equation relates the amplitudes before and after the collision through a bare two-particle $S$-matrix evaluated at the collision time. This matrix depends on a parameter $c$, whose relation to the microscopic coupling is fixed by the regularization of the contact interaction. In the conventions used here, its small-coupling form is $c=1/(2g)+\mathcal O(g)$. We use this relation to compare the quantum and classical time-dependent strengths in the universal regime near the undeformed WZW fixed point, where both interaction strengths are small. Requiring these $S$-matrices to satisfy the Yang-Baxter equation gives the affine time dependence
\begin{align}
c(2t)=2\alpha t+\beta.
\label{introQuantumProtocol}
\end{align}
Let $u(t)$ denote the coefficient multiplying $\int dx\,\mathbf J_R\cdot\mathbf J_L$ in the Hamiltonian. The fermionic interaction gives $u=8g$, so matching this coefficient to its classical value $4\pi q/k$ gives $q=2kg/\pi$. With $K=-2\pi\alpha/k$, the classical flatness condition yields
\begin{align}
\dot g=-4\alpha g^2+\mathcal O(g^3),\qquad g(t)\simeq\frac{1}{4\alpha t+B},
\label{introWeakProtocol}
\end{align}
where $B$ is an integration constant. The same leading equation follows independently from Eq.~(\ref{introQuantumProtocol}) and the small-coupling relation between $c$ and $g$. Classical Lax flatness and quantum Yang-Baxter consistency therefore select the same inverse-affine time-dependent strength in this universal regime. The full relation $c(g)$ is regularization dependent and cannot be used to extend this correspondence beyond weak coupling. Accordingly, differences between the finite-coupling expressions do not by themselves establish a physical discrepancy between the two constructions. Section~\ref{sec:protocolcomparison} develops the comparison with these restrictions and the coupling and time conventions stated explicitly.

With the integrability-preserving time-dependent strength fixed, the off-shell Bethe-ansatz solution of the qKZ equations \cite{Smirnov_1986,Frenkel,rishetikhin1,rishetikhin2,CherednikqKZ} expresses the reference spin amplitude as a Jackson-type integral where the auxiliary rapidities are summed with prescribed weights leading to a weighted discrete superposition of Bethe vectors \cite{Babujian_1997}.  Using this, we find the amplitudes corresponding to all the orderings of the particles and their continuation under repeated transport around the system and present the explicit form of the resulting many-body wavefunction.

We then apply Poisson resummation to the exact solution of the qKZ equations, converting the discrete weighted sum over rapidities into an integral. Writing the Gamma-function weights in exponential form gives an exact representation of the wavefunction in terms of the Yang-Yang action. 
Starting from this exact representation, we consider a controlled limit of slow variation of the coupling. In this limit, the leading saddle-point equations of the Yang-Yang action give rise to the Bethe equations, which, in a specific long-time limit, reduce to the instantaneous Bethe equations. After normalization, the corresponding saddle-point wavefunction agrees, to leading order, with an instantaneous Bethe eigenstate multiplied by its dynamical and geometric phase factors. We further show that, for a state initially prepared in one of these eigenstates, transitions to other instantaneous eigenstates are suppressed in this regime corresponding to the slow variation of the coupling. We therefore identify this regime as adiabatic.

For the decreasing time-dependent coupling, we characterize this regime in terms of the positive drive rate $\alpha_0$ and a characteristic time $t_0\gg L$. The latter is determined by requiring a finite physical mass gap in the continuum scaling limit. Taking $t_0$ as the time at which the instantaneous eigenstate is prepared, we study the subsequent evolution at $t=t_0+\Delta t$ within the interval over which the adiabatic conditions remain satisfied. Within this regime, we consider a state prepared in the instantaneous ground state and use the instantaneous Bethe equations to determine the rest energy of an elementary isospin excitation (a spinon) \cite{GrossNeveu,AndreiLowenstein79,DestriLowenstein}. In the continuum scaling limit, this elementary mass gap is given by
\begin{align}
m(t)=2\Lambda\exp\!\left[-\frac{\pi}{g_{\rm B}(t)}\right],
\label{staticmassbos}
\end{align}
where $\Lambda$ is the ultraviolet cutoff and $g_{\rm B}(t)$ is the Bethe-ansatz coupling defined in Eq.~(\ref{BetheCoupling}). Its weak-coupling relation to the microscopic interaction is $g_{\rm B}=4g+\mathcal O(g^3)$. For the time-dependent coupling $g_{\rm B}^{-1}(t)=\alpha_0t+\beta/2$, fix a reference gap $m_0=m(t_0)$. In the continuum scaling regime the corresponding characteristic time is
\begin{align}
t_0=\frac{1}{\pi\alpha_0}\ln\!\left(\frac{2\Lambda}{m_0}\right)-\frac{\beta}{2\alpha_0}.
\end{align}
Thus, $t_0$ grows logarithmically with the cutoff. At fixed elapsed time $\Delta t$, the continuum limit retains the finite mass
\begin{align}
m(t_0+\Delta t)=m_0e^{-\pi\alpha_0\Delta t}.
\end{align}
The drive rate and evolution interval must also satisfy the adiabatic conditions identified above. Within this regime, the instantaneous spectrum describes the adiabatically evolving state, and the dimensionless interaction generates a finite, time-dependent mass scale through dimensional transmutation.

We also consider the fast-drive limit, $\alpha L\to+\infty$ with $L$ and $\beta$ fixed. On any fixed time interval bounded away from $t=0$ on the positive-time branch, the coupling becomes uniformly small. The charge sector remains decoupled, while the isospin sector approaches the $SU(2)_1$ WZW fixed point with a weak current-current interaction at large but finite drive rate. As this interaction vanishes, the instantaneous mass gap for a bulk spinon tends to zero at fixed ultraviolet cutoff. Comparing the gap with its value at the characteristic time $t_0$ shows how increasing the drive rate at fixed positive time brings the system into the same weak-coupling regime reached at late times along the decreasing-coupling trajectory. At small but finite coupling, the rational Gaudin structure already appears in the quasi-classical expansion of the $R$-matrices entering the qKZ transport. For $\eta=1/(\alpha L)\ll 1$ and spectral differences $v$ bounded away from zero, the leading isospin-dependent term in each $R$-matrix is proportional to $\eta\,\Omega_{ij}/v$. The kernel $\Omega_{ij}/v$ satisfies the classical Yang-Baxter equation and gives the commuting Gaudin Hamiltonians 
\begin{align} H_i^{\rm G}=\sum_{j\ne i}\frac{\Omega_{ij}}{(z_i-z_j)}.\end{align}
At the WZW fixed point, these Hamiltonians govern the coordinate dependence of conformal blocks through the differential Knizhnik-Zamolodchikov equations \cite{KnizhnikZamolodchikov,FeiginFrenkelReshetikhin1994,BabujianFlume1994,rishetikhin2}. In the GBA, each $R$-matrix is evaluated at the spectral difference prescribed by periodic transport, and the coordinate shift remains finite throughout the quasi-classical expansion. The resulting discrete KZ equations thus contain the same rational kernel. Relating their solutions to WZW conformal blocks requires an asymptotic analysis of the full GBA wavefunction for a specified initial state.

The current formulation undertaken in this work also suggests extending the construction to the $SU(2)_k$ WZW model. Deriving the corresponding GBA wavefunctions would allow us to investigate their time-dependent couplings, instantaneous Bethe equations, and time-dependent mass scales. These quantum results could then be compared with the sigma-model description, with the large-$k$ limit providing a controlled regime for studying the geometry. Extensions to sigma models based on noncompact Lie groups require further analysis of the current representations, excitation spectra, and admissible states. We discuss these questions together with the physical observables accessible through the present level-one construction.

The paper is organized as follows. In Section~\ref{sec:classical}, we construct a generalized Lax connection for the interacting WZW currents, derive its Maillet brackets, and determine the time dependence of the coupling required by Lax flatness. A coupling redefinition and a reparametrization of time relate this construction to the $\lambda$-model with local couplings. In Section~\ref{sec:quantumbridge}, we quantize the current Hamiltonian and use non-Abelian bosonization to identify the perturbed $SU(2)_1$ WZW model with the interacting isospin sector of the chiral-invariant $SU(2)$ Gross-Neveu model. In Section~\ref{sec:Wavefunction}, we use the GBA to determine how quantum Yang-Baxter consistency constrains the time dependence of the interaction. We compare this result with the condition imposed by classical Lax flatness in the universal weak-coupling regime and then impose periodic boundary conditions to construct the exact many-body wavefunction.  In Section~\ref{sec:Analysis}, we derive an exact Yang-Yang representation of the wavefunction and identify the limit in which its saddle-point equations reproduce the instantaneous Bethe equations. In Section~\ref{sec:massgap}, we use the instantaneous Bethe spectrum to derive the time-dependent mass gap and show how requiring a finite gap in the continuum limit fixes a characteristic time. Finally, Section~\ref{sec:fast} examines the fast-drive limit, in which the current-current interaction becomes weak and the isospin sector approaches the $SU(2)_1$ WZW fixed point.

\section{Classical current dynamics and a time-dependent Lax connection}
\label{sec:classical}
In this section we study the time-dependent isotropic interaction between the left- and right-moving currents of the classical $SU(2)$ WZW model. Keeping the level $k$ and the current Poisson brackets fixed, we derive the equations of motion from the time-dependent Hamiltonian. Following the local-coupling approach of Hoare, Levine, and Tseytlin \cite{Benhoare}, we construct a generalized Lax connection whose flatness constrains the allowed time dependence of the interaction. We also calculate the Poisson brackets of the generalized Lax matrix and derive its Maillet $r/s$ algebra and extended classical Yang-Baxter relation. We then relate this construction to the $\lambda$-model through an explicit coupling redefinition and reparametrization of time. Section~\ref{sec:protocolcomparison} compares the resulting time-dependent coupling with the one obtained from Yang-Baxter consistency in the quantum GBA, using their common normalization near the WZW fixed point.

\subsection{The Wess-Zumino phase space and the time-dependent Hamiltonian}
\label{sec:classicalcurrents}

We use the $SU(2)$ WZW action in Eq.~(\ref{introWZWAction}), with spacetime metric $\eta_{\mu\nu}=\operatorname{diag}(1,-1)$. To fix the normalization of the currents, we choose the Lie-algebra generators $t_a=-i\sigma^a/2$, which satisfy
\begin{align}
[t_a,t_b]=\epsilon^{abc}t_c,\qquad\operatorname{tr}(t_at_b)=-\frac{1}{2}\delta^{ab},
\end{align}
where the trace is taken in the fundamental representation. The WZW field $h:\Sigma\to SU(2)$ maps the two-dimensional worldsheet $\Sigma$ into the group manifold $SU(2)$. Its kinetic term is determined by the group metric, which is invariant under both left and right multiplication. To define the Wess-Zumino term, we choose an oriented three-dimensional manifold $B$ whose boundary is the worldsheet, $\partial B=\Sigma$, and extend $h$ to a field $\widetilde h$ on $B$ that agrees with $h$ on $\Sigma$. The Wess-Zumino term integrates the closed three-form $H_3$ on $SU(2)$ over $B$ using this extension. Locally, $H_3=dB_2$, where $B_2$ is the two-form potential on the target space. Our choice of orientation and sign for the Wess-Zumino term fixes the chirality convention for the currents below. Changing the extension shifts the action by an integer multiple of $2\pi k$, so the quantum phase $e^{iS}$ is independent of this choice for integer $k$ \cite{WESSWZW,WITTENWZW,WittenBosonization1984}. 

The canonical formulation of the WZW action gives currents associated with left and right multiplication of the group field $h$. Their definitions include the Wess-Zumino contribution to the canonical momentum. With $\partial_\pm=(\partial_t\pm\partial_x)/2$, the currents of the undeformed model can be written as
\begin{align}
J_L^at_a=-\frac{k}{4\pi}h^{-1}\partial_+h,\qquad J_R^at_a=\frac{k}{4\pi}\partial_-h\,h^{-1}.
\label{classicalWZWGroupCurrents}
\end{align}
When introducing the current-current interaction, we retain the definitions of these currents in terms of $h$ and its canonical momentum. The modified Hamiltonian changes the relation between the momentum and $\partial_t h$, so the expressions in Eq.~(\ref{classicalWZWGroupCurrents}) apply specifically to the undeformed model. The equal-time brackets thus retain the WZW form \cite{Gawedzki1991,PapadopoulosSpence1992},
\begin{align}
\{J_R^a(x),J_R^b(y)\}
 &=\epsilon^{abc}J_R^c(x)\delta(x-y)
   -\frac{k}{4\pi}\delta^{ab}\partial_x\delta(x-y),\nonumber\\
\{J_L^a(x),J_L^b(y)\}
 &=\epsilon^{abc}J_L^c(x)\delta(x-y)
   +\frac{k}{4\pi}\delta^{ab}\partial_x\delta(x-y),\nonumber\\
\{J_R^a(x),J_L^b(y)\}&=0.
\label{classicalKM}
\end{align}
These brackets define two commuting classical affine current algebras. The terms proportional to $\partial_x\delta(x-y)$ are their central terms, with coefficients fixed by the WZW level $k$ and signs corresponding to the two chiralities.

We introduce a time-dependent isotropic interaction between the two currents through the Hamiltonian
\begin{align}
H_{\rm cc}(t)&=\frac{2\pi}{k}\int dx\, \left[\mathbf J_R^2+\mathbf J_L^2+2q(t)\,\mathbf J_R\cdot\mathbf J_L\right],\nonumber\\u(t)&=\frac{4\pi}{k}q(t),\qquad H_{\rm int}(t)=u(t)\int dx\,\mathbf J_R\cdot\mathbf J_L.
\label{classicalH}
\end{align}
Here $q(t)$ is the dimensionless classical coupling, and $u(t)$ denotes the coefficient multiplying the current product. For positive level $k$, the quadratic form in the Hamiltonian is positive definite when $|q|<1$. We will relate this Hamiltonian to the $\lambda$-model below through an explicit coupling redefinition and reparametrization of time. To express the current equations compactly, define
\begin{align}
R=J_R^at_a,\qquad L=J_L^at_a.
\end{align}
Using the Poisson brackets in Eq.~(\ref{classicalKM}), Hamilton's equation $\partial_t J=\{J,H_{\rm cc}(t)\}$ gives
\begin{align}
\partial_t R&=-\partial_xR-u[R,L]-q\partial_xL,\nonumber\\
\partial_t L&=\phantom{-}\partial_xL+u[R,L]+q\partial_xR.
\label{classicalEOM}
\end{align}
The commutator terms arise from the Lie-algebra part of the current brackets, while the spatial derivative terms follow from their central terms. At $q=0$, the left- and right-moving currents propagate independently. A nonzero interaction couples them through both the commutator and the spatial derivatives of the opposite current.

All explicit time dependence in Eq.~(\ref{classicalEOM}) enters through $q(t)$ and $u(t)$ because the currents are fixed functions of the canonical variables. The derivative $\dot q$ enters the flatness condition when we differentiate the coefficients of the Lax connection.

\subsection{Generalized Lax connection and the time-dependent coupling}
\label{sec:classicalLax}

We seek a generalized Lax connection $\mathscr L_\mu(t,x;w)$ for the current equations~(\ref{classicalEOM}), where $w$ is a constant spectral parameter. The corresponding auxiliary linear system is
\begin{align}
(\partial_x+\mathscr L_x)\Phi=0,\qquad(\partial_t+\mathscr L_t)\Phi=0,
\label{classicalAuxiliary}
\end{align}
and its compatibility requires
\begin{align}
\mathscr F_{tx}=\partial_t\mathscr L_x-\partial_x\mathscr L_t+[\mathscr L_t,\mathscr L_x]=0.
\label{classicalCurvature}
\end{align}
We take the connection to be linear in the currents,
\begin{align}
\mathscr L_x=a(t,x;w)R+b(t,x;w)L,\qquad \mathscr L_t=c(t,x;w)R+d(t,x;w)L.
\label{classicalLaxAnsatz}
\end{align}
Substituting the current equations into Eq.~(\ref{classicalCurvature}) gives
\begin{align}
\mathscr F_{tx}={}&(-a+qb-c)\partial_xR+(b-qa-d)\partial_xL\nonumber\\
&+(\partial_t a-\partial_x c)R+(\partial_t b-\partial_x d)L\nonumber\\
&+\big[u(b-a)+cb-da\big][R,L].
\label{classicalCurvatureExpansion}
\end{align}
For the curvature to vanish for arbitrary current profiles satisfying Eq.~(\ref{classicalEOM}), the coefficients must obey
\begin{align}
c=-a+qb,\qquad & d=b-qa,\nonumber\\
\partial_t a=\partial_x c,\qquad &\partial_t b=\partial_x d,\nonumber\\
u(b-a)+q(a^2+b^2&)-2ab=0.
\label{classicalLaxConditions}
\end{align}
The first line determines the temporal coefficients $c,d$ in terms of $a,b$. The remaining relations constrain $a,b$ algebraically and determine their spacetime dependence. We solve the algebraic condition by introducing a spectral variable $y=y(t,x;w)$ and writing
\begin{align}
D(t,x;w)=(1-q)y^2-(1+q),\nonumber\\
a=\frac{u(y-1)}{D},\qquad b=\frac{u(y+1)}{D}.
\label{classicalRationalParam}
\end{align}
Substituting these expressions into the first line of Eq.~(\ref{classicalLaxConditions}) gives the connection
\begin{align}
\mathscr L_x&=\frac{u}{D}\big[(y-1)R+(y+1)L\big],\nonumber\\
\mathscr L_t&=\frac{u}{D} \big[\big(1+q-(1-q)y\big)R +\big(1+q+(1-q)y\big)L\big].
\label{classicalExactLax}
\end{align}
For a constant coupling, $y$ can be independent of $t$ and $x$. When the coupling varies, derivatives of the coefficients generate additional curvature terms proportional to $\dot q$. Allowing $y$ to depend on spacetime provides the terms needed to cancel them, while $w$ remains the constant parameter labeling the family of connections. Substituting Eq.~(\ref{classicalRationalParam}) into the differential conditions in Eq.~(\ref{classicalLaxConditions}), with $u=4\pi q/k$, yields
\begin{align}
\partial_xy&=\frac{\dot q\,(y^2-1)}{q(1-q)D},\qquad\partial_ty=(1-q)y\,\partial_xy.
\label{classicalSpectralPDE}
\end{align}
On a branch where $q\neq0,1$ and $D\neq0$, these equations determine the two derivatives of $y$. Their compatibility requires the mixed derivatives to agree. Evaluating their difference using Eq.~(\ref{classicalSpectralPDE}) gives
\begin{align}
\partial_t(\partial_xy)-\partial_x(\partial_ty)=\frac{y^2-1}{q^2(1-q)^2D} \left[q(1-q)\ddot q-(2-3q)\dot q^{\,2}\right].
\label{classicalSpectralMismatch}
\end{align}
For a nondegenerate spectral family, with $y^2-1$ not identically zero, compatibility therefore requires
\begin{align}
q(1-q)\ddot q-(2-3q)\dot q^{\,2}=0.
\label{classicalCompatibility}
\end{align}
This equation has the first integral
\begin{align}
\frac{d}{dt}\left[\frac{\dot q}{q^2(1-q)}\right]=0,\qquad\dot q=Kq^2(1-q),
\label{classicalProtocol}
\end{align}
where $K$ is an integration constant setting the rate and direction of the drive. Equation~(\ref{classicalProtocol}) determines the time-dependent couplings for which the current equations admit the generalized Lax connection constructed above. Integrating Eq.~(\ref{classicalProtocol}) gives
\begin{align}
\ln\left|\frac{q(t)}{1-q(t)}\right|-\frac1{q(t)}=Kt+B,
\label{classicalIntegratedProtocol}
\end{align}
where $B$ is fixed by the initial value of the coupling. On the branch $0<q<1$, choosing $K<0$ gives a coupling that decreases toward the undeformed WZW fixed point at $q=0$. The associated spectral variable $y(t,x;w)$ is determined implicitly by
\begin{align}
\frac{1-q(t)}{q(t)}y(t,x;w)-\ln\frac{y(t,x;w)-1}{y(t,x;w)+1}=Kx+w.
\label{classicalImplicitSpectral}
\end{align}
On a fixed branch of the logarithm, differentiating this relation with respect to $x$ and $t$, and using Eq.~(\ref{classicalProtocol}), yields
\begin{align}
\partial_xy=\frac{Kq(y^2-1)}{D},\qquad\partial_ty=(1-q)y\,\partial_xy.
\label{classicalSolvedSpectralPDE}
\end{align}
These equations reproduce Eq.~(\ref{classicalSpectralPDE}) for the coupling trajectory obtained above. Together, Eqs.~(\ref{classicalExactLax}), (\ref{classicalIntegratedProtocol}), and (\ref{classicalImplicitSpectral}) give a locally flat family with the full classical coupling dependence. Conversely, before imposing the equations of motion, the curvature is $a$ times the residual of the $R$ equation plus $b$ times the residual of the $L$ equation, with all terms in each equation taken to the left. Since $a/b=(y-1)/(y+1)$ varies with $w$, flatness for the whole family implies both current equations. For $K=0$ on this branch, the coupling is constant and $y$ can depend on $w$ alone, recovering the static family. With periodic boundary conditions on the currents, the shift $x\to x+L$ in Eq.~(\ref{classicalImplicitSpectral}) gives
\begin{align}
y(t,x+L;w)=y(t,x;w+KL),
\label{classicalSpectralGluing}
\end{align}
provided the logarithmic branches are continued consistently. A translation around the system therefore shifts the spectral label by $KL$. Even when the currents are periodic, the generalized Lax connection need not be periodic at fixed $w$. To construct conserved quantities from the monodromy matrix, one must therefore match the Lax connection at $x=L$ and spectral label $w$ to the connection at $x=0$ and shifted label $w+KL$ \cite{Benhoare}.

\subsection{Maillet brackets and the classical Yang-Baxter algebra}
\label{sec:classicalMaillet}

We now use the WZW current algebra to determine the equal-time Poisson bracket of the Lax matrix $\mathscr L_x$. The central terms in Eq.~(\ref{classicalKM}) produce a derivative of the delta function. This makes the bracket non-ultralocal and leads to the two kernels $r$ and $s$ in Maillet's formulation \cite{Maillet1986}. In the time-dependent model the coefficients $a$ and $b$ also depend on position through $y(t,x;w)$, so the treatment of this derivative matters.

We use tensor notation to distinguish the two auxiliary spaces. For any Lie-algebra-valued matrix $X$, write $X_1=X\otimes I$ and $X_2=I\otimes X$, and define
\begin{align}
C_{12}=\sum_{a=1}^3t_a\otimes t_a,\qquad [C_{12},X_1+X_2]=0.
\label{classicalMailletCasimir}
\end{align}
The normalization of $C_{12}$ is fixed by the generators $t_a=-i\sigma^a/2$ used above. We work at a common time $t$, suppressing it in the intermediate expressions, and set $a_i(x)=a(t,x;w_i)$ and $b_i(x)=b(t,x;w_i)$. The spectral labels $w_i$ are held fixed. Since $q(t)$ and $y(t,x;w_i)$ are prescribed scalar functions, their Poisson brackets with the currents vanish. Substituting $\mathscr L_x=aR+bL$ into Eq.~(\ref{classicalKM}) gives
\begin{align}
\{\mathscr L_{x1}(x;w_1),\mathscr L_{x2}(x';w_2)\}
={}&[C_{12},a_1(x)a_2(x)R_1(x)
             +b_1(x)b_2(x)L_1(x)]\delta(x-x')\nonumber\\
&+\frac{k}{4\pi}
\big[b_1(x)b_2(x')-a_1(x)a_2(x')\big]
C_{12}\partial_x\delta(x-x').
\label{classicalLaxPoissonDirect}
\end{align}
The first line follows from the Lie-algebra part of the current brackets. The second line contains their central terms. To evaluate its coefficient at a single position, we use the distributional identity
\begin{align}
f(x')\partial_x\delta(x-x')=f(x)\partial_x\delta(x-x')+\partial_xf(x)\delta(x-x'),
\label{classicalMailletDistribution}
\end{align}
so that moving the coefficient at $x'$ to $x$ produces the additional term
\begin{align}
\frac{k}{4\pi}\big[b_1\partial_xb_2-a_1\partial_xa_2\big]C_{12}\delta(x-x').
\label{classicalMailletDerivativeTerm}
\end{align}
To express the result in the usual transport convention, define
\begin{align}
U(t,x;w)=-\mathscr L_x(t,x;w),\qquad \partial_x\Phi=U\Phi.
\label{classicalTransportGenerator}
\end{align}
Thus $U$ is the generator appearing in the path-ordered exponential in Eq.~(\ref{pathorderedtransport}); the sign follows from the auxiliary equations~(\ref{classicalAuxiliary}). Introduce the scalar function
\begin{align}
F(t;y)=\frac{u(t)(y^2-1)}{(1-q(t))y^2-(1+q(t))},\qquad y_i=y(t,x;w_i),\qquad F_i=F(t;y_i).
\label{classicalMailletScalar}
\end{align}
For distinct $y_1,y_2$ away from the poles of the Lax parametrization, define
\begin{align}
r^+_{12}(t,x;w_1,w_2)&=\frac{F_1}{y_1-y_2}C_{12},
&r^-_{12}(t,x;w_1,w_2)&=\frac{F_2}{y_1-y_2}C_{12},\nonumber\\
r_{12}&=\frac12(r^+_{12}+r^-_{12}),
&s_{12}&=\frac12(r^+_{12}-r^-_{12}).
\label{classicalMailletKernels}
\end{align}
Under interchange of the auxiliary spaces and spectral labels, $r_{21}(w_2,w_1)=-r_{12}(w_1,w_2)$ and $s_{21}(w_2,w_1)=s_{12}(w_1,w_2)$. The kernels depend explicitly on spacetime, but not on the canonical fields. The identity $a_1a_2=(F_1a_2-F_2a_1)/(y_1-y_2)$, together with its counterpart for $b$, reproduces the coefficients of $R_1$ and $L_1$ in the current bracket. For the central terms one obtains
\begin{align}
-2s_{12}=\frac{k}{4\pi}(b_1b_2-a_1a_2)C_{12},\;\;\;
\partial_x r^-_{12}=\frac{k}{4\pi}\big(b_1\partial_xb_2-a_1\partial_xa_2\big)C_{12}.
\label{classicalMailletIdentities}
\end{align}
The second identity uses the spectral equation~(\ref{classicalSolvedSpectralPDE}), which takes the form $\partial_xy_i=kK F_i/(4\pi)$. Combining these identities with the current terms gives the generalized Maillet bracket
\begin{align}
\{U_1(x;w_1),U_2(x';w_2)\}
={}&\Big([r^-_{12},U_1(x;w_1)]
       +[r^+_{12},U_2(x;w_2)]
       +\partial_xr^-_{12}\Big)\delta(x-x')\nonumber\\
&-2s_{12}\partial_x\delta(x-x').
\label{classicalDrivenMaillet}
\end{align}
Every kernel on the right is evaluated at $(t,x;w_1,w_2)$, and $\partial_x$ acts at fixed $t,w_1,w_2$. Thus the position dependence of $y$ fixes the coefficient $\partial_xr^-_{12}$ as well as $s_{12}$. To recover the bracket for $\mathscr L_x$ itself, substitute $U=-\mathscr L_x$ on the right. This reverses the signs of the two commutator terms and leaves the central terms as written.

We next examine the classical Yang-Baxter relation associated with these kernels. At a common spacetime point, with the spectral arguments on each pair understood, they satisfy
\begin{align}
[r^+_{13},r^-_{12}]
+[r^+_{23},r^+_{12}]
+[r^+_{23},r^+_{13}]=0.
\label{classicalExtendedCYBE}
\end{align}
This is the extended classical Yang-Baxter relation appropriate to a non-ultralocal Maillet algebra \cite{Maillet1986,ItsiosSfetsosSiamposTorrielli2014}. It involves both $r^+$ and $r^-$ because the coefficient $s$ of the derivative term is generally nonzero. To verify it, use the invariance identity in Eq.~(\ref{classicalMailletCasimir}) to reduce the three tensor commutators to one common invariant tensor. The remaining scalar coefficient is proportional to
\begin{align}
F_1F_2\biggl[-\frac{1}{(y_1-y_3)(y_1-y_2)}+\frac{1}{(y_2-y_3)(y_1-y_2)}-\frac{1}{(y_2-y_3)(y_1-y_3)}\biggr]=0.
\label{classicalCYBEVerification}
\end{align}
The cancellation holds at each $(t,x)$, so the explicit spacetime dependence of the spectral variables preserves this identity. The full Poisson bracket, including its distributional derivative terms, satisfies the Jacobi identity because Eq.~(\ref{classicalDrivenMaillet}) was obtained exactly from the WZW current brackets. The pointwise identity~(\ref{classicalExtendedCYBE}) exhibits its classical Yang-Baxter structure.

For $K=0$, the coupling and spectral variables are independent of spacetime, and $\partial_xr^-_{12}=0$. Equation~(\ref{classicalDrivenMaillet}) then reduces to the static Maillet bracket of the current model, whose relation to the established $\lambda$-model algebra \cite{ItsiosSfetsosSiamposTorrielli2014} follows from the coupling and Hamiltonian identifications in the next subsection. Hoare, Levine, and Tseytlin obtained coordinate-dependent Lax Poisson brackets for the PCM with local couplings \cite{Benhoare}; the calculation above gives the corresponding explicit algebra for the time-dependent WZW-current Hamiltonian considered here.

The result is an equal-time, local Poisson algebra. Constructing brackets of monodromy matrices also requires a prescription for coincident endpoints because of the derivative of the delta function \cite{Maillet1986}. For a system of length $L$ with periodic currents, the spectral shift in Eq.~(\ref{classicalSpectralGluing}) must additionally be included in the boundary analysis before asserting conservation and mutual Poisson commutativity of global charges. We can therefore compare the Yang-Baxter identity for the local Maillet kernels in Eq.~(\ref{classicalMailletKernels}) with the quantum Yang-Baxter equation for the $R$-matrices of the GBA. Deriving those quantum matrices by quantizing the classical Poisson algebra remains a separate question.

\subsection{Relation to the established \texorpdfstring{$\lambda$}{lambda}-model construction}
\label{sec:classicalLambdaBridge}

We now relate the current Hamiltonian~(\ref{classicalH}) to the integrable group $\lambda$-model. It is constructed by gauging a common symmetry of a principal chiral model and a WZW model, and then fixing the principal chiral field to the identity \cite{Sfetsos2014,HollowoodMiramontesSchmidtt2014}. We denote the deformation parameter by $\ell$, reserving $q$ for the coupling in Eq.~(\ref{classicalH}). In terms of the principal chiral model coupling $\kappa^2$, these parameters satisfy $\ell=k/(k+\kappa^2)$. The auxiliary fields remaining after gauge fixing allow us to derive the sigma-model action and express its Hamiltonian in terms of the WZW currents. We denote the sigma-model time by $\tau$ and set $\partial_\pm=(\partial_\tau\pm\partial_x)/2$. Let $A_\pm\in\mathfrak{su}(2)$ denote the auxiliary fields, distinct from the Lax connection $\mathscr L_\mu$. With the trace convention $\operatorname{tr}(t_at_b)=-\delta^{ab}/2$ introduced above, the action is
\begin{align}
S_\ell[h,A]=S_{{\rm WZW},k}[h]-\frac{k}{\pi}\int d\tau\,dx\,\operatorname{tr}\Big[&A_+\partial_-h\,h^{-1}-A_-h^{-1}\partial_+h\nonumber\\&-h^{-1}A_+hA_-+\ell^{-1}A_+A_-\Big].
\label{classicalLambdaParent}
\end{align}
Since the action contains no derivatives of $A_\pm$, their equations of motion are algebraic. Defining
\begin{align}
B_+=h^{-1}\partial_+h,\qquad B_-=\partial_-h\,h^{-1},\qquad \operatorname{Ad}_hX=hXh^{-1},
\end{align}
these equations take the form
\begin{align}
\big(\ell^{-1}-\operatorname{Ad}_{h^{-1}}\big)A_+&=B_+,
&
\big(\ell^{-1}-\operatorname{Ad}_h\big)A_-&=-B_-.
\end{align}
Eliminating the auxiliary fields gives
\begin{align}
S_\ell[h]=S_{{\rm WZW},k}[h]-\frac{k}{\pi}\int d\tau\,dx\,\operatorname{tr}\!\left[B_+\big(\ell^{-1}-\operatorname{Ad}_h\big)^{-1}B_-\right].
\label{classicalLambdaSigmaAction}
\end{align}
The dependence of the inverse operator on $h$ modifies both the target-space metric and the antisymmetric tensor. The leading deformation at small $\ell$ has Lagrangian density $-k\ell\,\operatorname{tr}(B_+B_-)/\pi$, corresponding to the WZW current-current perturbation. The full inverse operator retains all higher powers of $\ell$, giving the nonlinear coupling dependence of the finite $\lambda$-deformation \cite{Sfetsos2014,ItsiosSfetsosSiampos2014}. To compare this action with Eq.~(\ref{classicalH}), we express its Hamiltonian in terms of the canonical currents. As in the preceding subsections, the comparison uses their definitions in terms of the group field and canonical momentum. Their expressions in terms of group-field derivatives and auxiliary fields are
\begin{align}
\mathcal I_+&=-\frac{k}{2\pi}
\big(B_++h^{-1}A_+h-A_-\big),\nonumber\\
\mathcal I_-&=\frac{k}{2\pi}
\big(B_- -hA_-h^{-1}+A_+\big).
\label{classicalLambdaCanonicalCurrents}
\end{align}
Using the auxiliary-field equations, these become
\begin{align}
\mathcal I_\pm=-\frac{k}{2\pi}\big(\ell^{-1}A_\pm-A_\mp\big),\qquad A_\pm=-\frac{2\pi\ell}{k(1-\ell^2)}\big(\mathcal I_\pm+\ell\mathcal I_\mp\big).
\label{classicalLambdaConstraints}
\end{align}
After imposing the constraints and eliminating the auxiliary fields and their conjugate momenta, the Dirac brackets of $\mathcal I_\pm$ coincide with their original WZW Poisson brackets \cite{HollowoodMiramontesSchmidtt2014}. The relation to our current normalization is
\begin{align}
\mathcal I_+=2L=2J_L^at_a,\qquad \mathcal I_-=2R=2J_R^at_a.
\end{align}
To compare components with Ref.~\cite{HollowoodMiramontesSchmidtt2014}, introduce its normalized basis $T_a=\sqrt2\,t_a$, for which $\operatorname{tr}(T_aT_b)=-\delta^{ab}$ and $[T_a,T_b]=\sqrt2\,\epsilon^{abc}T_c$. Writing $\mathcal I_\pm=\mathcal I_\pm^aT_a$ then gives
\begin{align}
\mathcal I_+^a=\sqrt2\,J_L^a,\qquad \mathcal I_-^a=\sqrt2\,J_R^a.
\end{align}
This rescaling converts the central coefficient $k/(2\pi)$ in the component brackets of $\mathcal I_\pm$ to $k/(4\pi)$ in Eq.~(\ref{classicalKM}), while leaving the level $k$ unchanged. Substituting Eq.~(\ref{classicalLambdaConstraints}) into the reduced Hamiltonian and using these normalizations gives
\begin{align}
H_{\ell}=\frac{2\pi}{k(1-\ell^2)}\int dx\,\left[(1+\ell^2)(\mathbf J_R^2+\mathbf J_L^2)+4\ell\,\mathbf J_R\cdot\mathbf J_L\right].
\label{classicalLambdaH}
\end{align}
Thus $H_\ell$ and $H_{\rm cc}$ are expressed in terms of currents satisfying the same affine Poisson brackets. The reduced symplectic structure is independent of $\ell$, so prescribing $\ell=\ell(\tau)$ changes the Hamiltonian without changing those brackets. Comparing Eqs.~(\ref{classicalH}) and~(\ref{classicalLambdaH}), we obtain
\begin{align}
q=\frac{2\ell}{1+\ell^2},\qquad\rho=\frac{1-\ell^2}{1+\ell^2},\qquad H_{\rm cc}=\rho H_\ell.
\label{classicalLambdaMap}
\end{align}
On the branch $0<\ell<1$, we have $\rho=\sqrt{1-q^2}>0$. The factor $\rho$ is accounted for by the time reparametrization
\begin{align}
\frac{d\tau}{dt}=\rho(t),
\end{align}
with $\ell$ evaluated at $\tau(t)$. Under this change of time coordinate, the equations generated by $H_\ell$ in time $\tau$ become those generated by $H_{\rm cc}$ in time $t$. To express the same transformation for the Lax connection, denote the $\lambda$-model connection by $\mathscr L^{(\lambda)}_\mu$. Its one-form transforms according to
\begin{align}
\mathscr L_t(t,x;w)
&=\rho(t)\,\mathscr L^{(\lambda)}_\tau(\tau(t),x;w),\\
\mathscr L_x(t,x;w)
&=\mathscr L^{(\lambda)}_x(\tau(t),x;w).
\end{align}
To identify the spectral coordinates in these expressions, write $\mathscr L^{(\lambda)}_\pm=(\mathscr L^{(\lambda)}_\tau\pm\mathscr L^{(\lambda)}_x)/2$. Substituting Eqs.~(\ref{classicalExactLax}), (\ref{classicalLambdaConstraints}), and (\ref{classicalLambdaMap}) then gives
\begin{align}
\mathscr L^{(\lambda)}_\pm
 &=\frac{1+z^{\pm1}}{1+\ell}\,A_\pm,
&
z&=\frac{1+\ell+(1-\ell)y}{1+\ell-(1-\ell)y}.
\end{align}
This is the group $\lambda$-model Lax connection of Ref.~\cite{Benhoare} in the light-cone convention used here. In particular, the coupling and time maps identify $\mathscr L_x$ at corresponding spectral arguments. Since the equal-time current brackets agree, the Maillet algebra in Sec.~\ref{sec:classicalMaillet} also gives the bracket of the time-dependent $\lambda$-model Lax matrix in this normalization. For the group $\lambda$-model, the local-coupling condition obtained in Ref.~\cite{Benhoare} can be written as
\begin{align}
\frac{d\ell}{d\tau}=b\,\frac{\ell^2}{(1+\ell)^2},
\label{classicalHLTProtocol}
\end{align}
where the constant $b$ incorporates the normalization and orientation of time. Applying the coupling and time transformations gives
\begin{align}
\frac{dq}{dt}=\frac{dq}{d\ell}\frac{d\ell}{d\tau}\frac{d\tau}{dt}=\frac b2q^2(1-q).
\label{classicalHLTMatch}
\end{align}
This reproduces Eq.~(\ref{classicalProtocol}) with $K=b/2$. Our Lax construction therefore reformulates the local-coupling $\lambda$-model result in the canonical current variables used for the quantum comparison. The correspondence specifies the classical action, current brackets, coupling normalization and the time coordinate. Establishing an all-orders quantum equivalence with the level-one fermionic theory would require a separate quantum analysis, which we do not undertake in the current work.

A related classical fermionic construction was obtained in Ref.~\cite{KomatsuSakamotoWallbergYamazaki2026}, where the massless non-Abelian Thirring model admits a Lax connection for an inverse-affine time-dependent interaction. Its currents are local bilinears of canonical fermion fields. Since the fundamental fermion brackets are proportional to $\delta(x-y)$ and these bilinears contain no spatial derivatives, their current algebra contains no derivatives of the delta function and is therefore ultralocal. The WZW formulation used here is bosonic: its currents contain the group-field momentum and spatial derivatives of the group field, and their Poisson brackets include the central derivative terms in Eq.~(\ref{classicalKM}). This difference is consistent with non-Abelian bosonization, which relates the quantum fermionic and bosonic theories \cite{WITTENWZW,WittenBosonization1984}. In the fermionic description, the regulated quantum current commutators contain a central term proportional to $\partial_x\delta(x-y)$, known as the Schwinger term. In the bosonized WZW description, the same central extension is already encoded in the classical current algebra. Bosonization thus identifies the two quantum current algebras and accounts for the different origins of their central terms in the two formulations.

The time-dependent couplings can be compared independently of this difference in their classical realization. After matching the coupling and time conventions, the inverse-affine time dependence of the fermionic coupling agrees with the leading weak-coupling form of Eq.~(\ref{classicalProtocol}). The explicit finite-coupling classical relation established above is with the local-coupling $\lambda$-model, which also appears in the gauge-theory construction of Ref.~\cite{KomatsuSakamotoWallbergYamazaki2026}. The coupling redefinition and reparametrization of time derived above connect our current-current deformation of the WZW Hamiltonian to that model.

\section{Quantization and the fermionic model}
\label{sec:quantumbridge}

We now formulate the fermionic quantum field theory corresponding to the classical Hamiltonian studied in the preceding section. This requires specifying the affine current algebra and the normal ordering of the quadratic current operators. At level one, a fermionic representation realizes this algebra and identifies the current-current perturbation with the interaction of the chiral-invariant $SU(2)$ Gross-Neveu model. This fermionic quantum field theory is the starting point for the GBA construction in Sec.~\ref{sec:Wavefunction}, where we derive the time-dependent coupling strength that preserves quantum integrability and solve the time-dependent Schr\"odinger equation. Its current Hamiltonian will also be used in Sec.~\ref{sec:fast} to discuss the approach to the WZW model in the fast-drive limit.

\subsection{From classical currents to quantum operators}

Quantization replaces the classical Poisson brackets in Eq.~(\ref{classicalKM}) by affine current commutators at the same level $k$. The central derivative term is inherited from the classical WZW current algebra, and its coefficient remains fixed as the interaction varies. Defining the quadratic current operators additionally requires normal ordering. Requiring the undeformed Hamiltonian to generate chiral propagation with unit velocity gives the quantum Sugawara coefficient $2\pi/(k+2)$. The spin Hamiltonian, including the time-dependent interaction, is therefore
\begin{align}
H_{s,k}(t)=\frac{2\pi}{k+2}\int_0^Ldx\,\bigl(:\mathbf J_R^2:+:\mathbf J_L^2:\bigr)+u(t)\int_0^Ldx\,\mathbf J_R\cdot\mathbf J_L .
\label{quantumCurrentHamiltonian}
\end{align}
The colons denote normal ordering in the WZW model. The shift $k\to k+2$ follows from the affine algebra and normal ordering, with $2$ the dual Coxeter number of $SU(2)$. It ensures that the quantum Hamiltonian has the same unit propagation velocity used in the classical construction.



At level one, the spin currents admit a representation in terms of right- and left-moving two-component fermions, with velocities $v_R=1$ and $v_L=-1$. Here spin denotes the internal $SU(2)$ isospin of the fermion doublet. We set $\hbar=v_F=1$ and impose periodic boundary conditions on a system of length $L$. Repeated spin indices $a,b=\uparrow,\downarrow$ and adjoint indices $A,B,C=1,2,3$ are summed. Using the same spatial Fourier convention for both chiralities, we write
\begin{align}
\psi_\chi(x)=\frac1{\sqrt L}\sum_{r\in\mathbb Z}e^{2\pi irx/L}c_{\chi,r},\qquad\{c_{\chi,ra},c_{\chi',sb}^\dagger\}=\delta_{\chi\chi'}\delta_{rs}\delta_{ab},\qquad \chi=R,L.
\label{bridgeFermionModes}
\end{align}
The one-particle energies are $v_\chi 2\pi r/L$. Filling the negative-energy modes therefore gives the massless reference occupations $f_R(r)=\theta(-r)$ and $f_L(r)=\theta(r)$. Normal ordering of fermion bilinears with respect to this reference, denoted by $:\,\cdot\,:_0$, is defined by
\begin{align}
:c_{\chi,ra}^\dagger c_{\chi,sb}:_0=c_{\chi,ra}^\dagger c_{\chi,sb}-\delta_{rs}\delta_{ab}f_\chi(r)I.
\label{bridgeFixedOrdering}
\end{align}
We choose the zero-energy modes to be empty. The central term derived below is independent of this choice. The filled sea specifies the subtraction used to define the continuum operators. The coordinate Bethe construction instead uses an empty Fock reference to assemble finite-particle wavefunctions, while the algebraic Bethe ansatz uses a polarized reference vector for their spin degrees of freedom. With point splitting or an equivalent mode regulator understood, the charge and spin currents are
\begin{align}
\rho_\chi(x)&=:\psi^\dagger_{\chi a}(x)\psi_{\chi a}(x):_0,\nonumber\\
J_\chi^A(x)&=:\psi^\dagger_{\chi a}(x)s^A_{ab}\psi_{\chi b}(x):_0,\qquad s^A=\frac{\sigma^A}{2}.
\label{fastcurrents}
\end{align}
To derive the current algebra, introduce the Fourier modes
\begin{align}
J_{\chi,m}^A=\sum_r:c_{\chi,r}^\dagger s^A c_{\chi,r+m}:_0,\qquad J_\chi^A(x)=\frac1L\sum_m e^{2\pi imx/L}J_{\chi,m}^A.
\label{bridgeCurrentModes}
\end{align}
Evaluating their commutator with the same regulator gives
\begin{align}
[J_{\chi,m}^A,J_{\chi,n}^B]
&=i\epsilon^{ABC}J_{\chi,m+n}^C
+\frac{\delta^{AB}}2\delta_{m+n,0}
\sum_r[f_\chi(r)-f_\chi(r+m)]
\nonumber\\
&=i\epsilon^{ABC}J_{\chi,m+n}^C
+\frac{\kappa_\chi m}{2}\delta^{AB}\delta_{m+n,0},
\qquad \kappa_R=1,\quad\kappa_L=-1.
\label{bridgeSchwingerModes}
\end{align}
For $m>0$, the right-moving occupation difference counts $m$ modes across the Fermi surface. With the zero mode empty, these are $r=-m,\ldots,-1$, and the sum equals $m$. The left-moving filling gives the opposite sign. The factor $1/2$ follows from $\operatorname{tr}(s^As^B)=\delta^{AB}/2$. Thus the central term arises from the regulated sum over occupied fermion modes, a contribution that is absent from the commutator of finite-dimensional spin matrices alone. Using
\begin{align}
\frac{1}{2L^2}\sum_m m\,e^{2\pi im(x-y)/L}=-\frac{i}{4\pi}\partial_x\delta_L(x-y),
\end{align}
where $\delta_L$ is the periodic delta function, we obtain
\begin{align}
[J_R^A(x),J_R^B(y)]
&=i\epsilon^{ABC}J_R^C(x)\delta(x-y)-\frac{i}{4\pi}\delta^{AB}\partial_x\delta(x-y),
\nonumber\\
[J_L^A(x),J_L^B(y)]
&=i\epsilon^{ABC}J_L^C(x)\delta(x-y)+\frac{i}{4\pi}\delta^{AB}\partial_x\delta(x-y),
\nonumber\\
[J_R^A(x),J_L^B(y)]&=0.
\label{fastKM}
\end{align}
Here and below, the subscript on the periodic delta function is suppressed. These are the level-one affine $SU(2)$ commutation relations, with the same chirality conventions as the classical brackets. Under interacting unitary evolution, the currents continue to satisfy this algebra at equal time because conjugation evolves the current operator and leaves its central term unchanged. The level is fixed by the fermionic representation and remains one as the interaction approaches zero.

\subsection{Sugawara decomposition and the Gross-Neveu Hamiltonian}
\label{sec:fastWZW}

Consider the free fermion Hamiltonian, with unit propagation velocity and the normal-ordering prescription specified above,
\begin{align}
H_0=-i\int_0^Ldx\,
\bigl(:\psi^\dagger_{Ra}\partial_x\psi_{Ra}:_0-:\psi^\dagger_{La}\partial_x\psi_{La}:_0\bigr).
\label{bridgeFreeFermionHamiltonian}
\end{align}
We now use the currents defined in Eq.~(\ref{fastcurrents}) to express this Hamiltonian as a sum of charge and spin contributions. The spin currents satisfy the level-one algebra derived in the preceding subsection. Using the same regulator, the charge currents satisfy
\begin{align}
[\rho_\chi(x),\rho_{\chi'}(y)]
&=-\frac{i\kappa_\chi}{\pi}\,\delta_{\chi\chi'}\partial_x\delta(x-y),\nonumber\\
[\rho_\chi(x),J_{\chi'}^A(y)]&=0,
\label{bridgeChargeAlgebra}
\end{align}
with $\kappa_R=1$ and $\kappa_L=-1$, as in Eq.~(\ref{bridgeSchwingerModes}). The fermion theory therefore contains commuting $U(1)$ charge and $SU(2)_1$ spin current algebras. Non-Abelian bosonization expresses this separation directly at the level of the Hamiltonian \cite{WITTENWZW,WittenBosonization1984}. Within a fixed sector of chiral charges and spatial boundary conditions,
\begin{align}
H_0&=H_c+H_{\mathrm{WZW}},\nonumber\\
H_c&=\frac{\pi}{2}\int_0^Ldx\,
\bigl(:\rho_R^2:_0+:\rho_L^2:_0\bigr),\nonumber\\
H_{\mathrm{WZW}}&=\frac{2\pi}{3}\int_0^Ldx\,\bigl(:\mathbf J_R^2:_0+:\mathbf J_L^2:_0\bigr),
\label{fastSugawara}
\end{align}
up to an additive constant in each sector. The first contribution is the Hamiltonian of a free compact $U(1)$ boson. The second is the Sugawara Hamiltonian of the $SU(2)_1$ WZW model, with the coefficient obtained by setting $k=1$ in Eq.~(\ref{quantumCurrentHamiltonian}). The squares of the currents in Eq.~(\ref{fastSugawara}) use current-mode normal ordering in the massless reference representation defined above. Equivalently, they are obtained by separating the current insertions and subtracting the singular terms before taking the coincident-point limit. With these definitions, Eq.~(\ref{fastSugawara}) is an identity of regulated operators \cite{WITTENWZW,WangFurusakiStarykh2020,WittenBosonization1984}.

The central charges provide a further check of the decomposition. Each massless Dirac fermion contributes one unit, so the original fermion doublet has central charge two. The charge and spin contributions are
\begin{align}
c_{\mathrm{charge}}=1,\qquad c_{\mathrm{spin}}=\left.\frac{3k}{k+2}\right|_{k=1}=1.
\label{fastCentralCharges}
\end{align}
Their sum reproduces the central charge of the free fermion theory. Thus the decomposition identifies both the current algebras and the corresponding energy operators of the two sectors.

We now introduce the isotropic interaction between the right- and left-moving spin currents. Since the charge currents commute with the spin currents, this interaction leaves the free charge Hamiltonian unchanged. Moving the left-moving creation operator past the right-moving annihilation operator in Eq.~(\ref{fastcurrents}) gives
\begin{align}
\mathbf J_R\cdot\mathbf J_L
&=-\frac14\sigma^A_{ab}\sigma^A_{cd}:\psi^\dagger_{Ra}\psi^\dagger_{Lc} \psi_{Rb}\psi_{Ld}:_0.
\label{bridgeSpinInteraction}
\end{align}
In the convention used below, the coefficient of this quartic operator is $-2g(t)$, where $g(t)$ is spatially uniform. The full fermionic Hamiltonian is therefore
\begin{align}
H_{\rm GN}(t)=\int_0^Ldx\,\Big[&-i:\psi^\dagger_{Ra}\partial_x\psi_{Ra}:_0+i:\psi^\dagger_{La}\partial_x\psi_{La}:_0
\nonumber\\
&-2g(t)\,\sigma^A_{ab}\sigma^A_{cd}:\psi^\dagger_{Ra}\psi^\dagger_{Lc} \psi_{Rb}\psi_{Ld}:_0\Big].
\label{Hamiltonian}
\end{align}
This is the chiral-invariant $SU(2)$ Gross-Neveu Hamiltonian. Using Eq.~(\ref{bridgeSpinInteraction}), it can equivalently be written as
\begin{align}
H_{\rm GN}(t)&=H_0+u(t)V=H_c+H_{\mathrm{WZW}}+u(t)V,
\nonumber\\
V&=\int_0^Ldx\,\mathbf J_R\cdot\mathbf J_L,\qquad u(t)=8g(t).
\label{bridgeCurrentHamiltonian}
\end{align}
This derives the interaction normalization introduced earlier. The interacting spin Hamiltonian is precisely Eq.~(\ref{quantumCurrentHamiltonian}) at $k=1$, while $H_c$ supplies the additional free charge sector of the fermionic realization. At the undeformed WZW fixed point, the right- and left-moving spin currents have conformal weights $(1,0)$ and $(0,1)$, respectively. Their product therefore has weights $(1,1)$, making the current-current interaction marginal by its fixed-point scaling dimension. 

In the corresponding static model, the perturbation is marginally relevant for $g>0$. The coupling grows toward the infrared and generates a mass gap in the spin sector. The operator correspondence above holds for a prescribed time-dependent coupling $g(t)$. In the following section we use the GBA to solve the time-dependent Schr\"odinger equation for the Hamiltonian in Eq.~(\ref{Hamiltonian}) and derive the quantum integrability condition on $g(t)$ from the consistency of the many-body wavefunction. We then construct the wavefunction for the resulting time-dependent coupling and analyze it to obtain physical quantities in Sec.~\ref{sec:Analysis}. The equivalent current form of the Hamiltonian, Eq.~(\ref{bridgeCurrentHamiltonian}), will be used in Sec.~\ref{sec:fast} to analyze the fast-drive limit.

\section{The quantum solution from the generalized Bethe ansatz}
\label{sec:Wavefunction}

We now use the fermionic Hamiltonian constructed in Sec.~\ref{sec:quantumbridge} to determine the time-dependent coupling compatible with quantum integrability using the generalized Bethe ansatz (GBA) and construct the corresponding many-body wavefunction. We follow the solution of Ref.~\cite{PasnooriGrossNeveu}. The wavefunction contains an amplitude for each ordering of the particles. Substituting it into the time-dependent Schr\"odinger equation relates amplitudes that differ by the ordering of a right-moving and a left-moving particle through a bare two-particle $S$-matrix. Requiring the amplitudes to be consistent when three or more particles are reordered gives the Yang-Baxter equations and constrains the time dependence of the interaction. Once these relations are established, all amplitudes can be expressed in terms of one reference amplitude. Periodic boundary conditions give matrix difference equations whose solution determines the reference amplitude and thus the complete wavefunction. 

The fermion fields obey periodic boundary conditions,
\begin{align}
\psi_{Ra}(0)=\psi_{Ra}(L),\qquad \psi_{La}(0)=\psi_{La}(L).
\label{pbc}
\end{align}
Each term in Eq.~(\ref{Hamiltonian}) creates and annihilates the same number of fermions of each chirality. The numbers of left- and right-moving fermions are thus separately conserved. They are given by
\begin{align}
N_L&=\sum_a\int_0^Ldx\,\psi_{La}^\dagger(x)\psi_{La}(x),\nonumber\\N_R&=\sum_a\int_0^Ldx\,\psi_{Ra}^\dagger(x)\psi_{Ra}(x).
\label{number}
\end{align}
We therefore construct the state in a sector containing $N_L$ left movers and $N_R$ right movers, with $N=N_L+N_R$, and require
\begin{align}
i\partial_t|N_L,N_R;t\rangle =H_{\rm GN}(t)|N_L,N_R;t\rangle.
\label{SE}
\end{align}
Label the left-moving particles by $1,\ldots,N_L$ and the right-moving particles by $N_L+1,\ldots,N$. The label of a particle continues to specify the same particle as its position in the ordering changes. In coordinate space, write
\begin{align}
|N_L,N_R;t\rangle={}&\sum_{\boldsymbol\sigma}\int_{[0,L]^N}d^Nx\,\bigl[\mathcal A F^{1\ldots N,\{\chi_i\}}(\boldsymbol x,t)\bigr]_{\boldsymbol\sigma}\prod_{i=1}^N\psi_{\chi_i\sigma_i}^\dagger(x_i)|0\rangle.
\label{npwf1}
\end{align}
Here $\boldsymbol\sigma=(\sigma_1,\ldots,\sigma_N)$ denotes the spin indices, $\chi_i=R,L$ specifies the chirality, and $|0\rangle$ is the empty Fock state. The operator $\mathcal A$ antisymmetrizes the particle coordinates and their spin and chirality labels. An overall normalization is understood. Substitution into Eq.~(\ref{SE}) gives the $N$-particle Schr\"odinger equation
\begin{align}
i\partial_tF=\left[-i\sum_{i=1}^Nv_{\chi_i}\partial_{x_i}+2g(t)\sum_{\substack{i<j\\\chi_i\ne\chi_j}}\delta(x_i-x_j)\boldsymbol\sigma_i\cdot\boldsymbol\sigma_j\right]F,\qquad v_R=1,\quad v_L=-1.
\label{senp}
\end{align}
The interaction coefficient $2g(t)$ follows from the quartic term in Eq.~(\ref{Hamiltonian}). The delta functions show that the interaction acts when a right mover and a left mover occupy the same position. When their positions are distinct, the equation reduces to that of free fermions.

\subsection{Ansatz wavefunction and the Yang-Baxter equations}

To construct the interacting wavefunction, we first solve the free fermion problem. This determines the coordinate dependence that can be used wherever the particles are separated, since the delta-function interaction in Eq.~(\ref{senp}) vanishes in those regions. With velocities $v_L=-1$ and $v_R=1$, introduce
\begin{align}
\bar z_l&=x_l+t,\qquad l=1,\ldots,N_L,\nonumber\\z_r&=x_r-t,\qquad r=N_L+1,\ldots,N,\nonumber\\\boldsymbol z&=(\bar z_1,\ldots,\bar z_{N_L},z_{N_L+1},\ldots,z_N).
\label{complete_null_coordinates}
\end{align}
For expressions applying to either chirality, we write $z_i^{\rm null}=x_i-v_{\chi_i}t$ for the corresponding component of $\boldsymbol z$. A spin-valued function $f(\boldsymbol z)$ satisfies
\begin{align}
\left(\partial_t+\sum_{i=1}^N v_{\chi_i}\partial_{x_i}\right)f(\boldsymbol z)=0,
\end{align}
and therefore solves the free Schr\"odinger equation. To include the time-dependent interaction, we assign a separate spin amplitude $f^Q(\boldsymbol z)$ to each spatial ordering $Q$ and combine these contributions into the many-body wavefunction. Let $Q\in S_N$ denote a permutation, with the associated ordering $x_{Q(1)}<\cdots<x_{Q(N)}$. The product of Heaviside functions
\begin{align}
\Theta_Q(\boldsymbol x)=\prod_{a=1}^{N-1}\theta\bigl(x_{Q(a+1)}-x_{Q(a)}\bigr)
\label{complete_ordering_projector}
\end{align}
selects that ordering. We take the ansatz
\begin{align}
F^{1\ldots N,\{\chi_i\}}_{\boldsymbol\sigma}(\boldsymbol x,t)=\sum_{Q\in S_N}\Theta_Q(\boldsymbol x)f^{Q,\{\chi_i\}}_{\boldsymbol\sigma}(\boldsymbol z).
\label{npwf2}
\end{align}
Each $f^Q$ is a vector in $(\mathbb C^2)^{\otimes N}$, with one spin space for each labeled particle. In the ordinary coordinate Bethe ansatz, plane-wave factors can be separated from spin amplitudes that are constant within an ordering. Here the spin amplitudes themselves may depend on $\boldsymbol z$. This dependence allows the wavefunction to satisfy the Schr\"odinger equation with a time-dependent interaction. We suppress the spin indices and the fixed chirality assignment when writing relations between these vectors.

Consider adjacent particles $j$ and $k$, with $j$ right-moving and $k$ left-moving. When derivatives in Eq.~(\ref{senp}) act on the Heaviside functions, they produce delta functions at $x_j=x_k$. Equating the coefficient of $\delta(x_j-x_k)$ to zero relates the amplitudes on the two sides of the collision,
\begin{align}
f^{\ldots kj\ldots}(\boldsymbol z)=S^{jk}(z_j,\bar z_k)f^{\ldots jk\ldots}(\boldsymbol z).
\label{rel1N}
\end{align}
All other particles have the same ordering on both sides. In a forward collision, the right mover initially lies to the left of the left mover, so Eq.~(\ref{rel1N}) takes the ordering $(j,k)$ before the collision to $(k,j)$ afterward. When the right mover $j$ and the left mover $k$ meet, $x_j=x_k$ and their null coordinates satisfy $\bar z_k-z_j=2t_c$, where $t_c$ is the time of the collision. The interaction strength at that instant is therefore $g(t_c)=g\bigl((\bar z_k-z_j)/2\bigr)$. Applying the time-dependent Schr\"odinger equation at $x_j=x_k$ gives Eq.~(\ref{rel1N}), which relates the isospin amplitudes on either side of the collision through the bare two-particle $S$-matrix, which at this coupling takes the form
\begin{align}
S^{jk}(z_j,\bar z_k)=e^{i\phi(d)}\frac{ic(d)I^{jk}+P^{jk}}{ic(d)+1}.
\label{smatlr}
\end{align}
Here $P^{jk}$ interchanges the two spin states and $I^{jk}$ is the identity on their tensor-product space. We use the notation  
\begin{align}
d=\bar z_k-z_j=2t_c,\qquad \gamma(d)=g(d/2).
\end{align}
As in the standard coordinate Bethe ansatz, this is the bare $S$-matrix of the constituent fermions. To determine $c(d)$ and $\phi(d)$ in terms of $\gamma(d)$, one must specify how the delta-function interaction is regularized. The wavefunction can be discontinuous at $x_j=x_k$, so its value in the product $\delta(x_j-x_k)F$ requires a prescription. We write
\begin{align}
c(d)&=\mathcal C_{\mathfrak r}\bigl(\gamma(d)\bigr),\qquad e^{i\phi(d)}=e^{i\Phi_{\mathfrak r}(\gamma(d))},
\label{contactSchemeDefinition}
\end{align}
where $\mathfrak r$ labels this choice. The same regularization dependence occurs in the time-dependent anisotropic Kondo model and its isotropic limit \cite{PasnooriAKM}. For the symmetric prescription $\theta(0)=1/2$, the wavefunction at the interaction point is the average of its limiting values. Let $F_-$ and $F_+$ denote these values for $x_j-x_k<0$ and $x_j-x_k>0$, respectively. Integrating Eq.~(\ref{senp}) across
$x_j=x_k$, with relative velocity $v_R-v_L=2$, gives
\begin{align}
-2i(F_+-F_-)+\gamma\,\boldsymbol\sigma_j\cdot\boldsymbol\sigma_k(F_++F_-)=0.
\label{symmetricMatchingEquation}
\end{align}
Solving for $F_+$ yields
\begin{align}
S_{\rm sym}^{jk}
=\left(I^{jk}+\frac{i\gamma}{2}\boldsymbol\sigma_j\cdot\boldsymbol\sigma_k\right)^{-1}\left(I^{jk}-\frac{i\gamma}{2}\boldsymbol\sigma_j\cdot\boldsymbol\sigma_k\right).
\label{symmetricContactJump}
\end{align}
The operator $\boldsymbol\sigma_j\cdot\boldsymbol\sigma_k$ has eigenvalues $1$ and $-3$ in the triplet and singlet sectors. Comparing the two eigenvalues of Eq.~(\ref{symmetricContactJump}) with Eq.~(\ref{smatlr}) therefore gives
\begin{align}
\mathcal C_{\rm sym}(\gamma)
&=\frac{1-3\gamma^2/4}{2\gamma},\qquad e^{i\Phi_{\rm sym}(\gamma)}=\frac{1-i\gamma/2}{1+i\gamma/2}.
\label{symmetricContactMap}
\end{align}
A different prescription replaces the delta function by a narrow smooth interaction and solves the first-order equation through that region before taking its width to zero. This gives
\begin{align}
S_{\rm exp}^{jk}=\exp\!\left[-i\gamma\,\boldsymbol\sigma_j\cdot\boldsymbol\sigma_k\right].
\label{exponentialContactJump}
\end{align}
The triplet and singlet phases are now $e^{-i\gamma}$ and $e^{3i\gamma}$, respectively. Expressing them in the same parametrization gives
\begin{align}
\mathcal C_{\rm exp}(\gamma)=\cot(2\gamma),\qquad e^{i\Phi_{\rm exp}(\gamma)}=e^{-i\gamma}.
\label{cotangentContactMap}
\end{align}
The factor of two in the cotangent argument follows from the interaction coefficient and the relative velocity in Eq.~(\ref{senp}) and changes with the normalization of the microscopic coupling. The rational and cotangent formulas therefore describe two specified regularizations of the same delta-function problem. The choice of regularization fixes both the relation between $c$ and $g$ and the accompanying scalar phase. Both prescriptions give
\begin{align}
\mathcal C_{\mathfrak r}(\gamma)=\frac{1}{2\gamma}+\mathcal O(\gamma)
\label{universalContactMap}
\end{align}
at weak coupling. Their higher-order terms and scalar phases depend on the prescription. We use this leading relation when comparing microscopic couplings near the WZW fixed point. The exact wavefunction will instead be expressed in terms of $c$ and the scalar phase appropriate to the chosen prescription. For example, the square root obtained by inverting Eq.~(\ref{symmetricContactMap}) at finite coupling is specific to the symmetric prescription.

These physically determined two-particle relations must fit into a consistent many-body wavefunction. To connect all its ordering sectors, we also need relations between amplitudes that differ by an interchange of particles with the same chirality. Such particles have equal velocities and do not collide, and the Hamiltonian contains no interaction between them. The Schr\"odinger equation therefore does not directly fix these additional $S$-matrices. Within the GBA, they are chosen so that different sequences of pairwise interchanges give the same final amplitude, consistently with the opposite-chirality $S$-matrices already determined above. Write
\begin{align}
f^{\ldots ji\ldots}(\boldsymbol z)=S'^{ij}(z_i^{\rm null},z_j^{\rm null})f^{\ldots ij\ldots}(\boldsymbol z),\qquad \chi_i=\chi_j.
\label{sameChiralityAmplitude}
\end{align}
For two right movers $i,j$ and a left mover $k$, consistency of the two possible sequences of three pairwise interchanges requires
\begin{align}
S'^{ij}S^{ik}S^{jk}=S^{jk}S^{ik}S'^{ij},
\label{mixedChiralityYB}
\end{align}
with the arguments of each matrix supplied by its particle labels. The opposite-chirality factors contain the interaction at their respective collision times. Their Yang-Baxter consistency consequently constrains the time dependence of the coupling as well as the $S$-matrix relating the two right-moving particles. Within this pairwise rational $S$-matrix construction, the parameter of
$S'^{ij}$ is the difference $c(\bar z_k-z_i)-c(\bar z_k-z_j)$. This difference must depend only on the coordinates of $i$ and $j$. Differentiating with respect to $\bar z_k$ therefore requires $c'(\bar z_k-z_i)=c'(\bar z_k-z_j)$ for arbitrary coordinates.
For a smooth time-dependent coupling, $c'$ is constant, giving
\begin{align}
c(d)=\alpha d+\beta.
\label{intconst}
\end{align}
The same argument for two left movers gives the corresponding matrix for that chirality. The Yang-Baxter equation fixes the ratio of the coefficients of $I^{ij}$ and $P^{ij}$. Choosing unit amplitude in the triplet channel gives
\begin{align}
S'^{ij}(z_i^{\rm null},z_j^{\rm null})=\frac{i\alpha(z_j^{\rm null}-z_i^{\rm null})I^{ij}+P^{ij}}{i\alpha(z_j^{\rm null}-z_i^{\rm null})+1},\qquad \chi_i=\chi_j.
\label{sameChiralityExchange}
\end{align}
These matrices also satisfy the Yang-Baxter equations when all three particles have the same chirality. The constant $\alpha$ has dimensions of inverse time, or inverse length in our units, and $\beta$ is dimensionless. At the scattering time, $d=2t$. Combining Eqs.~(\ref{intconst}) and (\ref{universalContactMap}) gives
\begin{align}
g(t)\simeq\frac{1}{4\alpha t+2\beta}=\frac{1}{4(\alpha t+\beta/2)}.
\label{intstrength}
\end{align}
On the positive weak-coupling branch, $\alpha>0$ describes a decreasing interaction and $\alpha<0$ an increasing interaction. Equation~(\ref{intconst}) gives the exact constraint on the parameter entering the bare $S$-matrix, and Eq.~(\ref{intstrength}) expresses it in terms of the microscopic coupling near the WZW fixed point.

The role of the Yang-Baxter equations is now explicit. Two particle orderings can be connected by different sequences of adjacent interchanges. Each sequence gives a product of bare $S$-matrices acting on the initial amplitude, and these products must produce the same final amplitude. The Yang-Baxter equations equate the two sequences involving three particles. Matrices acting on disjoint pairs commute, and reversing an interchange uses the inverse matrix. Together, these relations make the construction consistent for arbitrary $N$. We choose the ordering $Q_0=(N,\ldots,1)$ as the reference and set
\begin{align}
f_0(\boldsymbol z)=f^{Q_0,\{\chi_i\}}(\boldsymbol z).
\label{referenceAmplitudeDefinition}
\end{align}
All other amplitudes can now be obtained from $f_0$ by products of the bare $S$-matrices. The reference amplitude $f_0$ can be determined by applying the periodic boundary conditions. Before venturing into this, let us compare the conditions on the time-dependent couplings imposed by the quantum Yang-Baxter equations with those imposed by the flatness of the generalized Lax connections of the classical $\lambda$-model obtained above.

\subsection{Comparing classical and quantum integrability-preserving time-dependent couplings}
\label{sec:protocolcomparison}

The classical and quantum constructions have independently determined conditions on the time-dependent coupling strength. To compare them, we use the same normalization of the currents and the same physical time, and distinguish the microscopic coupling from the parameter entering the bare $S$-matrix. In the quantum construction, Eqs.~(\ref{contactSchemeDefinition}) and
(\ref{intconst}) give
\begin{align}
C(t)\equiv c(2t)=2\alpha t+\beta,\qquad C(t)=\mathcal C_{\mathfrak r}\bigl(g(t)\bigr).
\label{comparisonContactProtocol}
\end{align}
The Yang-Baxter equations fix the first relation. The second converts it to the microscopic coupling using the chosen regularization of the delta-function interaction. Wherever
$\mathcal C_{\mathfrak r}$ is differentiable, differentiation gives
\begin{align}
\mathcal C_{\mathfrak r}'(g)\dot g=2\alpha.
\label{comparisonQuantumDrive}
\end{align}
At finite bare coupling this conversion depends on the regularization, as illustrated by Eqs.~(\ref{symmetricContactMap}) and (\ref{cotangentContactMap}). Near the WZW fixed point, the positive-coupling branch has $g\to0$ and $C\to\infty$. The common leading relation is
\begin{align}
g(t)=\frac{1}{2C(t)}+\mathcal O\!\left(C(t)^{-3}\right),
\label{comparisonQuantumBranch}
\end{align}
which reproduces Eq.~(\ref{intstrength}). The coefficient of the subleading term is prescription dependent. This distinction between a bare coupling and the parameter appearing in a bare $S$-matrix also arises in the anisotropic Kondo problem \cite{PasnooriAKM}. The distinction also matters in the static model where the weak-coupling limit fixes how the bare parameters in different schemes can be compared, while the exact Bethe-ansatz solution determines nonperturbative observables, including the excitation spectrum and its dynamically generated mass gap.

The fermionic Hamiltonian gives $u=8g$, while the classical Hamiltonian in Eq.~(\ref{classicalH}) has $u=4\pi q/k$.
Consequently, $q=2kg/\pi$. The transformation from the $\lambda$-model time to the physical time $t$ of the current Hamiltonian was established in Sec.~\ref{sec:classicalLambdaBridge}. Expressing the classical
condition~(\ref{classicalProtocol}) in terms of $g$ and setting
$K=-2\pi\alpha/k$ then gives
\begin{align}
\dot g_{\rm Lax}=-4\alpha g^2\left(1-\frac{2k}{\pi}g\right).
\label{comparisonClassicalDrive}
\end{align}
This retains the full coupling dependence of the classical Hamiltonian. As mentioned above, we have kept the level $k$ explicit in the classical construction, while for the quantum construction we specialize to $k=1$. In the universal weak-coupling regime, the two conditions become
\begin{align}
\dot g_{\rm Lax}&=-4\alpha g^2+\mathcal O(g^3),\nonumber\\
\dot g_{\rm GBA}&=-4\alpha g^2+o(g^2).
\label{comparisonWeakDrive}
\end{align}
For both explicit prescriptions in Eqs.~(\ref{symmetricContactMap}) and~(\ref{cotangentContactMap}), the first correction to $\dot g_{\rm GBA}$ is of order $g^4$. Matching the initial coupling gives the same leading inverse-affine dependence on time in the two constructions. Beyond this leading behavior, the classical correction comes from the factor $1-q$ in the Lax condition, whereas the quantum conversion to $g$ depends on $\mathcal C_{\mathfrak r}$. To interpret any difference at this order as a quantum correction to the classical Lax connection, one would first need to define its quantum counterpart and match the coupling prescriptions, including the Sugawara normalization. We now impose periodic boundary conditions to determine the reference amplitude $f_0$ and construct the corresponding many-body wavefunction.

\Needspace{7\baselineskip}
\subsection{Periodic boundary conditions and the quantum Knizhnik-Zamolodchikov equations}
\label{sec:periodicqKZ}

Periodic boundary conditions relate the reference amplitude $f_0=f^{N\ldots1}$ at different values of its arguments. To derive these relations, we take a right mover through the system to the right and a left mover through it to the left. With either choice, interchanging an oppositely moving pair takes its ordering from $(r,l)$ to $(l,r)$, the order before and after a physical collision. The corresponding bare matrix is therefore $S^{rl}$ in Eq.~(\ref{rel1N}). Interchanges of particles with the same chirality use the matrices $S'$ determined by the Yang-Baxter equations.

When the transported particle reaches the boundary, we identify $x_r=L$ with $x_r=0$ for a right mover and $x_l=0$ with $x_l=L$ for a left mover. These identifications are imposed at fixed physical time and with the other particle coordinates fixed. Their action on the arguments of the amplitudes is
\begin{align}
z_r\longrightarrow z_r-L,\qquad \bar z_l\longrightarrow\bar z_l+L.
\label{periodicForwardShifts}
\end{align}
For a pair consisting of the transported particle and a particle of the opposite chirality, either identification increases $d=\bar z_l-z_r$ by $L$. Applying the shifts to all right movers and all left movers gives precisely the change of null coordinates induced by $t\to t+L$ at fixed particle positions. This relates the periodic continuation to the time dependence of the wavefunction. To construct each continuation, we must keep track of the interchanges before and after the boundary identification, since only the latter use the shifted coordinate of the transported particle.

Let $Q$ specify an ordering of all particles except $i$. The two orientations of the boundary identification are
\begin{align}
f^{iQ}(\boldsymbol z-L\boldsymbol e_i)
&=f^{Qi}(\boldsymbol z),\nonumber\\
f^{Qi}(\boldsymbol z+L\boldsymbol e_i)
&=f^{iQ}(\boldsymbol z).
\label{periodicOrderingIdentification}
\end{align}
Here $\boldsymbol e_i$ selects the argument of particle $i$. The first line identifies $x_i=L$ with $x_i=0$ and is used for forward transport of a right mover; the second identifies $x_i=0$ with $x_i=L$ and is used for a left mover. The arguments and spin spaces remain attached to the particle labels, while the superscripts specify their spatial ordering. Define $T_i\boldsymbol z=\boldsymbol z-L\boldsymbol e_i$, so these two identifications use $T_i$ and $T_i^{-1}$, respectively.

For one right mover and one left mover the reference ordering is $(r,l)$. To apply periodicity to the right mover, take it past the left mover and identify $x_r=L$ with $x_r=0$. To apply periodicity to the left mover, take it past the right mover and identify $x_l=0$ with $x_l=L$. Both constructions use the same interchange $S^{rl}(z_r,\bar z_l)$ and restore the reference ordering at the boundary. Hence
\begin{align}
f_0(\bar z_l,z_r-L)
&=S^{rl}(z_r,\bar z_l)f_0(\bar z_l,z_r),\nonumber\\
f_0(\bar z_l+L,z_r)
&=S^{rl}(z_r,\bar z_l)f_0(\bar z_l,z_r).
\label{periodicTwoParticleForward}
\end{align}
The bare matrix in both equations is evaluated at the same collision time $t_c=(\bar z_l-z_r)/2$. Reversing one of these continuations requires its inverse matrix and the corresponding shifted arguments.

Now consider the case of three particles. Take particle $1$ to be left-moving and particles $2,3$ to be right-moving, so that $\boldsymbol z=(\bar z_1,z_2,z_3)$ and $f_0=f^{321}$.
To transport particle $2$, first interchange it with particle $1$. The ordering becomes $(3,1,2)$ and the corresponding amplitude is
\begin{align}
f^{312}(\boldsymbol z)=S^{21}(z_2,\bar z_1)f^{321}(\boldsymbol z).
\label{periodicThreeParticleBefore}
\end{align}
Particle $2$ is now the rightmost particle. Applying Eq.~(\ref{periodicOrderingIdentification}) places it at the left end of the system and changes its argument to $z_2-L$:
\begin{align}
f^{231}(T_2\boldsymbol z)=f^{312}(\boldsymbol z).
\label{periodicThreeParticleBoundary}
\end{align}
Finally, interchange particles $2$ and $3$ to restore the reference ordering. This interchange acts on an amplitude whose argument for particle $2$ is already $z_2-L$, and hence
\begin{align}
f^{321}(T_2\boldsymbol z)
&=S'^{23}(z_2-L,z_3)f^{231}(T_2\boldsymbol z)\nonumber\\
&=S'^{23}(z_2-L,z_3)S^{21}(z_2,\bar z_1)
f^{321}(\boldsymbol z).
\label{periodicThreeParticleTransport}
\end{align}
The first interchange uses the original arguments, whereas the interchange after the boundary identification uses the shifted argument. Since $S'^{23}$ depends on the coordinate difference,
\begin{align}
S'^{23}(z_2-L,z_3)=S'^{23}(z_2,z_3+L).
\label{periodicDifferenceShift}
\end{align}
Only the coordinate of particle $2$ has changed. The $+L$ in the second expression is an equivalent way of writing the difference $z_3-(z_2-L)$. This explains why a shift appears in one factor of Eq.~(\ref{periodicThreeParticleTransport}) and not in the other.

For $N$ particles, denote the right-mover transport operator by $Z_j$, so that $f_0(T_j\boldsymbol z)=Z_j(\boldsymbol z)f_0(\boldsymbol z)$. It is obtained by returning to the reference ordering after the boundary identification. To construct it for $j>N_L$, start from $(N,\ldots,1)$ and interchange $j$ successively with $j-1,\ldots,N_L+1$, followed by $N_L,\ldots,1$. The resulting ordering is $(N,\ldots,j+1,j-1,\ldots,1,j)$, and all these interchanges use the original arguments. Periodicity then gives the ordering
$(j,N,\ldots,j+1,j-1,\ldots,1)$ with $z_j$ replaced by $z_j-L$. Restoring $(N,\ldots,1)$ requires interchanging $j$ with $N,\ldots,j+1$, now at the shifted argument. Therefore
\begin{align}
Z_j(\boldsymbol z)={}&S'^{j,j+1}(z_j-L,z_{j+1})\cdots S'^{jN}(z_j-L,z_N)\nonumber\\
&\times S^{j1}(z_j,\bar z_1)\cdots
S^{jN_L}(z_j,\bar z_{N_L})\nonumber\\
&\times S'^{j,N_L+1}(z_j,z_{N_L+1})\cdots
S'^{j,j-1}(z_j,z_{j-1}).
\label{periodicFullTransport}
\end{align}
The rightmost factor acts first, and an empty product is the identity. Thus the factors with labels smaller than $j$ act before the boundary identification, while those with labels larger than $j$ act afterward. The latter may equivalently be written as $S'^{jk}(z_j,z_k+L)$, as in Eq.~(\ref{periodicDifferenceShift}). For this right-moving particle, every interchange with a left mover occurs before the boundary identification. Separating the scalar phases in Eq.~(\ref{smatlr}) therefore gives
\begin{align}
Z_j(\boldsymbol z)=\left[\prod_{l=1}^{N_L}e^{i\phi(\bar z_l-z_j)}\right]Z'_j(\boldsymbol z),\qquad j=N_L+1,\ldots,N,
\label{diffeq2}
\end{align}
where $Z'_j$ contains the spin matrices. For a left mover $l\leq N_L$, we instead move toward the left end of the ordering. Starting from $(N,\ldots,1)$, interchange $l$ successively with $l+1,\ldots,N_L$ and then with $N_L+1,\ldots,N$. Particle $l$ is now leftmost. Periodicity moves it to the right end and replaces $\bar z_l$ by $\bar z_l+L$. Finally, interchange it with $1,\ldots,l-1$ to restore the reference ordering. Writing $Z_l^{(+)}$ for this forward transport gives
\begin{align}
f_0(T_l^{-1}\boldsymbol z)
&=Z_l^{(+)}(\boldsymbol z)f_0(\boldsymbol z),\nonumber\\
Z_l^{(+)}(\boldsymbol z)
={}&S'^{l-1,l}(\bar z_{l-1},\bar z_l+L)\cdots
S'^{1l}(\bar z_1,\bar z_l+L)\nonumber\\
&\times S^{Nl}(z_N,\bar z_l)\cdots
S^{N_L+1,l}(z_{N_L+1},\bar z_l)\nonumber\\
&\times S'^{N_L,l}(\bar z_{N_L},\bar z_l)\cdots
S'^{l+1,l}(\bar z_{l+1},\bar z_l).
\label{periodicLeftForwardTransport}
\end{align}
Here too the rightmost factor acts first. The first line of the product is performed after the boundary identification; the remaining factors use the original arguments. Every opposite-chirality interchange is a forward collision before the boundary identification and contributes the $S$-matrix $S^{rl}(z_r,\bar z_l)$. Thus
\begin{align}
Z_l^{(+)}(\boldsymbol z)=\left[\prod_{r=N_L+1}^{N}e^{i\phi(\bar z_l-z_r)}\right]Z_l^{\prime(+)}(\boldsymbol z),
\label{periodicLeftForwardPhase}
\end{align}
where $Z_l^{\prime(+)}$ contains the spin matrices. The spin parameter and scalar phase in every such factor are evaluated at $2t_c=\bar z_l-z_r$, with the regularization fixed above.

For the off-shell solution below, it is convenient to express both chiralities with the same negative coordinate shift. For a left mover, evaluate Eq.~(\ref{periodicLeftForwardTransport}) at $T_l\boldsymbol z$ and invert the complete product:
\begin{align}
Z_l(\boldsymbol z)=\bigl[Z_l^{(+)}(T_l\boldsymbol z)\bigr]^{-1}.
\label{periodicLeftOrientationConversion}
\end{align}
The inverse reverses the order of the factors as well as each individual interchange. Both chiralities then obey
\begin{align}
f_0(T_i\boldsymbol z)=Z_i(\boldsymbol z)f_0(\boldsymbol z),\qquad T_i\boldsymbol z=\boldsymbol z-L\boldsymbol e_i.
\label{diffeq1}
\end{align}
For a left mover, this is the reverse of the forward continuation just derived. In this common convention the interchanges with every right mover take place after the boundary identification, when the argument of particle $l$ is $\bar z_l-L$. Restoring the reference ordering now changes $(l,r)$ to $(r,l)$ and therefore uses $[S^{rl}(z_r,\bar z_l-L)]^{-1}$. Consequently,

\begin{align}
Z_l(\boldsymbol z)=\left[\prod_{r=N_L+1}^{N}e^{-i\phi(\bar z_l-z_r-L)}\right]Z'_l(\boldsymbol z),\qquad l=1,\ldots,N_L.
\label{periodicLeftScalarPhase}
\end{align}
The local two-particle relation assigns the earlier collision time $t_c'=(\bar z_l-L-z_r)/2$ to these shifted arguments. Both the spin parameter and the scalar phase must be evaluated at $2t_c'$, using the same regularization. The inverse matrices and earlier arguments are therefore consequences of reversing the left-mover continuation to obtain Eq.~(\ref{diffeq1}); the original forward transport in Eq.~(\ref{periodicLeftForwardTransport}) uses the direct collision matrices.

The equations for different particles must be mutually consistent. Applying the equation for $k$ and then the equation for $j$ gives $Z_j(T_k\boldsymbol z)Z_k(\boldsymbol z)f_0$: the first operation has changed the arguments at which the second transport operator is evaluated. Reversing the order must produce the same amplitude, and hence
\begin{align}
Z_j(T_k\boldsymbol z)Z_k(\boldsymbol z)=Z_k(T_j\boldsymbol z)Z_j(\boldsymbol z).
\label{transportop}
\end{align}
This identity holds as an operator relation. The Yang-Baxter equations interchange the two possible orders for three particles, factors on disjoint pairs commute, and a matrix followed by its inverse cancels. Equation~(\ref{transportop}) is the zero-curvature condition for the matrix difference equations. The positive-shift version of Eq.~(\ref{diffeq1}) follows by evaluating it at $T_i^{-1}\boldsymbol z$ and using $Z_i(T_i^{-1}\boldsymbol z)^{-1}$.

We next separate the scalar and spin parts of these equations. Write
\begin{align}
f_0(\boldsymbol z)=A^{Q_0}(\boldsymbol z)\prod_{r=N_L+1}^{N}\prod_{l=1}^{N_L}h(\bar z_l-z_r).
\label{separation}
\end{align}
For a right mover, $z_r\to z_r-L$ changes every difference $d=\bar z_l-z_r$ to $d+L$. The scalar factor in Eq.~(\ref{diffeq2}) is therefore reproduced if
\begin{align}
h(d+L)=e^{i\phi(d)}h(d).
\label{scalarDifferenceEquation}
\end{align}
Forward transport of a left mover gives $\bar z_l\to\bar z_l+L$ and therefore also $d\to d+L$, reproducing the direct phases in Eq.~(\ref{periodicLeftForwardPhase}). In the common negative-shift convention, $\bar z_l\to\bar z_l-L$ gives $d\to d-L$, and the same scalar equation yields $h(d-L)/h(d)=e^{-i\phi(d-L)}$, exactly the factor in Eq.~(\ref{periodicLeftScalarPhase}). Thus one scalar function implements periodicity for both chiralities and in either orientation. For the contact prescription used in Ref.~\cite{PasnooriGrossNeveu}, we take its scalar solution and analytic prescription. If the regularization is changed, the phase $e^{i\Phi_{\mathfrak r}}$ in Eq.~(\ref{contactSchemeDefinition}) specifies the new right-hand side of Eq.~(\ref{scalarDifferenceEquation}) and hence the equation for $h$. In either case, multiplication by an admissible $L$-periodic function of $d$ preserves the scalar equation. The general theory of such difference equations is discussed in Ref.~\cite{ruijisenars}.

To express the remaining spin equations in the standard rational form, introduce the XXX $R$-matrix for $\alpha\ne0$,
\begin{align}
R^{ij}(\lambda)=\frac{i\lambda I^{ij}+\alpha^{-1}P^{ij}}{i\lambda+\alpha^{-1}}.
\label{Rmatrix}
\end{align}
The spectral argument $\lambda$ has dimensions of length and is fixed here by the particle coordinates. Using $c(d)=\alpha d+\beta$, the bare $S$-matrices become
\begin{align}
S^{ij}(z_i,\bar z_j)
&=e^{i\phi(\bar z_j-z_i)}
R^{ij}(\bar z_j-z_i+\beta/\alpha),\nonumber\\
S'^{ij}(z_i^{\rm null},z_j^{\rm null})
&=R^{ij}(z_j^{\rm null}-z_i^{\rm null}),
\qquad \chi_i=\chi_j.
\label{relsmatrmat}
\end{align}
In the first line $i$ is right-moving and $j$ is left-moving. Reversing an interchange uses the inverse matrix, with $R^{ij}(\lambda)^{-1}=R^{ji}(-\lambda)$. The static case $\alpha=0$ is described by the unscaled $S$-matrices before introducing $\alpha^{-1}$. Substituting Eq.~(\ref{separation}) into the periodicity equations and using Eq.~(\ref{scalarDifferenceEquation}) leaves
\begin{align}
A^{Q_0}(T_i\boldsymbol z)=Z'_i(\boldsymbol z)A^{Q_0}(\boldsymbol z).
\label{qkz}
\end{align}
For a right mover $j$, Eq.~(\ref{periodicFullTransport}) gives
\begin{align}
Z'_j(\boldsymbol z)
={}&R^{j,j+1}(z_{j+1}+L-z_j)\cdots
R^{jN}(z_N+L-z_j)\nonumber\\
&\times R^{j1}(\bar z_1-z_j+\beta/\alpha)\cdots
R^{jN_L}(\bar z_{N_L}-z_j+\beta/\alpha)\nonumber\\
&\times R^{j,N_L+1}(z_{N_L+1}-z_j)\cdots
R^{j,j-1}(z_{j-1}-z_j).
\label{qkzop}
\end{align}
For a left mover in the common negative-shift convention, the opposite-chirality factors take the form
\begin{align}
[S^{rl}(z_r,\bar z_l-L)]^{-1}={}e^{-i\phi(\bar z_l-z_r-L)}\: R^{lr}(z_r-\bar z_l+L-\beta/\alpha).
\label{periodicLeftInverseMatrix}
\end{align}
This fixes the signs of both the shift and the constant offset in the reoriented left-mover spin transport operator. Both chiralities can now be written in a common dimensionless form. Define
\begin{align}
w_r&=\frac{z_r}{L}-\frac{\beta}{2\alpha L},\qquad r=N_L+1,\ldots,N,\nonumber\\\bar w_l&=\frac{\bar z_l}{L}+\frac{\beta}{2\alpha L},\qquad l=1,\ldots,N_L,
\label{relGNqKZ}
\end{align}
and
\begin{align}
\boldsymbol\zeta&=(\bar w_1,\ldots,\bar w_{N_L},w_{N_L+1},\ldots,w_N),\qquad \eta=\frac{1}{\alpha L},\nonumber\\\widehat R^{ij}(v)&=R^{ij}(Lv)=\frac{ivI^{ij}+\eta P^{ij}}{iv+\eta}.
\label{dimensionlessQKZData}
\end{align}
The constant offsets absorb the $\beta/\alpha$ terms in the opposite-chirality matrices. In the forward-propagation convention the coordinate steps are $w_r\to w_r-1$ for right movers and $\bar w_l\to\bar w_l+1$ for left movers. After the reorientation in Eq.~(\ref{periodicLeftOrientationConversion}), $T_i$ acts as $\zeta_i\to\zeta_i-1$ for either chirality. Denoting the reference spin amplitude in these variables by $A_0(\boldsymbol\zeta)$, we obtain
\begin{align}
A_0(\boldsymbol\zeta-\boldsymbol e_i)=Z'_i(\boldsymbol\zeta)&A_0(\boldsymbol\zeta),
\label{qkzoverview}
\\Z'_i(\boldsymbol\zeta)=\widehat R^{i,i+1}(\zeta_{i+1}-\zeta_i+1)\cdots &\widehat R^{iN}(\zeta_N-\zeta_i+1)\nonumber\\\qquad\times\widehat R^{i1}(\zeta_1-\zeta_i)&\cdots\widehat R^{i,i-1}(\zeta_{i-1}-\zeta_i).
\label{uniformQKZTransport}
\end{align}
In this common orientation, the factors with $k<i$ act before the boundary identification and depend on $\zeta_k-\zeta_i$. Those with $k>i$ act afterward and depend on $\zeta_k-(\zeta_i-1)$. Thus the $+1$ inside these matrix arguments comes from subtracting the shifted coordinate; it is distinct from the positive coordinate step in the original left-mover equation. Equation~(\ref{uniformQKZTransport}) retains the ordering and argument shifts derived above, with the left-mover continuation rewritten in the common negative orientation. Note that we are using $Z'_i(\boldsymbol z)$ and $Z'_i(\boldsymbol\zeta)$ for the same operator in the two coordinate systems. Equations~(\ref{qkzoverview}) and~(\ref{uniformQKZTransport}) are nothing but the quantum Knizhnik-Zamolodchikov (qKZ) system for the rational $R$-matrix with difference step $-1$. These equations first appeared in the study of sine-Gordon form factors \cite{Smirnov_1986} and were subsequently formulated through quantum affine algebras \cite{Frenkel}. Their solutions have been studied in Refs.~\cite{Smirnov_1986,Frenkel,rishetikhin1,Babujian_1997,JimboMiwa1995,Cherednik2006,Tarasov:1994bb}. In the following subsection we use the off-shell Bethe-ansatz construction \cite{Babujian_1997,rishetikhin1} to solve these equations.


\subsection{Explicit form of the exact wavefunction}
\label{sec:explicitqKZsolution}

The $S$-matrices which relate different amplitudes in the wavefunction together with the periodic boundary conditions gave rise to matrix difference equations, whose solutions determine the reference amplitude. In this section we provide their solution and the explicit form of the reference amplitude $A^{Q_0}(\boldsymbol\zeta)=A_0(\boldsymbol\zeta)$, which is obtained using the off-shell Bethe ansatz \cite{Babujian_1997,rishetikhin1,PasnooriGrossNeveu}. The construction has two ingredients. First, products of spin-lowering operators generate vectors in the $N$-particle spin space. Second, a weighted sum of these vectors is chosen to satisfy the coordinate shifts required by the qKZ equations. We follow the derivation in Appendix~A of Ref.~\cite{PasnooriGrossNeveu}, retaining the steps that explain how the transport operator determines the weights.

Start from the fully polarized vector $|\Omega\rangle=|\uparrow\rangle_1\otimes\cdots\otimes |\uparrow\rangle_N$. Introduce a two-dimensional auxiliary spin space $a$ and the ordinary monodromy matrix
\begin{align}
T_a^{(0)}(u;\boldsymbol\zeta)&=\widehat R^{1a}(\zeta_1-u)\cdots \widehat R^{Na}(\zeta_N-u)\nonumber\\&=\begin{pmatrix}A^{\rm mon}_0(u)&B_0(u)\\ C_0(u)&D_0(u)
\end{pmatrix}_{\!a}.
\label{offShellMonodromy}
\end{align}
The rows and columns refer to the auxiliary spin, and each entry acts on the $N$ particle spins. We write $A^{\rm mon}_0(u)$ for the diagonal entry to distinguish it from the spin amplitude $A_0(\boldsymbol\zeta)$. The factors are ordered as written, with the rightmost factor acting first. The entry $B_0(u)$ lowers the total spin projection by one, and its commutation relation $[B_0(u),B_0(v)]=0$ follows from the Yang-Baxter equation. Thus
\begin{align}
|\boldsymbol u;\boldsymbol\zeta\rangle=\prod_{m=1}^{M}B_0(u_m)|\Omega\rangle,\qquad S^z=\frac N2-M,
\label{offShellBetheVector}
\end{align}
is a Bethe vector in a sector of fixed spin projection. The parameters $u_m$ are auxiliary rapidities; its dependence on the physical coordinates enters through $\boldsymbol\zeta$ in Eq.~(\ref{offShellMonodromy}).

The qKZ equation compares this construction at $\boldsymbol\zeta$ and $\boldsymbol\zeta-\boldsymbol e_i$. We therefore also need the shifted monodromy matrix
\begin{align}
T_a^{(i)}(u;\boldsymbol\zeta)=T_a^{(0)}(u;\boldsymbol\zeta-\boldsymbol e_i).
\label{offShellShiftedMonodromy}
\end{align}
Only the factor belonging to particle $i$ changes: its argument is $\zeta_i-1-u$. Denote its entries by $A^{\rm mon}_i(u),B_i(u),C_i(u),D_i(u)$ in the same order as above. Consequently, $\prod_m B_i(u_m)|\Omega\rangle$ is precisely the Bethe vector $|\boldsymbol u;\boldsymbol\zeta-\boldsymbol e_i\rangle$ required at the shifted coordinates.

To determine how the transport operator acts on the unshifted Bethe vector, introduce the $Q$-monodromy matrix
\begin{align}
T_a^{Q_i}(v;\boldsymbol\zeta)={}&\widehat R^{1a}(\zeta_1-v)\cdots\widehat R^{ia}(\zeta_i-v)\nonumber\\&\times\widehat R^{i+1,a}(\zeta_{i+1}-v+1)\cdots\widehat R^{Na}(\zeta_N-v+1).
\label{offShellQMonodromy}
\end{align}
Its factors with labels greater than $i$ contain the shifts derived in the preceding subsection. At $v=\zeta_i$, the $i$th factor is $\widehat R^{ia}(0)=P^{ia}$. Taking the auxiliary trace then gives exactly the ordered transport operator in Eq.~(\ref{uniformQKZTransport}). Write
\begin{align}
T_a^{Q_i}(\zeta_i;\boldsymbol\zeta)&=\begin{pmatrix}A_{Q_i}&B_{Q_i}\\ C_{Q_i}&D_{Q_i}\end{pmatrix}_{\!a},\nonumber\\Z'_i(\boldsymbol\zeta)&=\operatorname{tr}_aT_a^{Q_i}(\zeta_i;\boldsymbol\zeta)=A_{Q_i}+D_{Q_i}.
\label{offShellQTrace}
\end{align}

To evaluate the action of $Z'_i=A_{Q_i}+D_{Q_i}$ on a Bethe vector, we must commute $A_{Q_i}$ and $D_{Q_i}$ through the operators $B_0(u)$ that construct it. Write the rational $R$-matrix as
$\widehat R(v)=b(v)I+c_R(v)P$, where
\begin{align}
b(v)=\frac{iv}{iv+\eta},\qquad c_R(v)=\frac{\eta}{iv+\eta}.
\label{offShellRcoefficients}
\end{align}
Commuting the $Q$-monodromy in space $a$ past the ordinary monodromy in a second auxiliary space $b$ shifts the latter from $\boldsymbol\zeta$ to $\boldsymbol\zeta-\boldsymbol e_i$. This is the mixed Yang-Baxter relation of Ref.~\cite{Babujian_1997}, given in Eq.~(A23) of Ref.~\cite{PasnooriGrossNeveu}. Suppressing the common argument $\boldsymbol\zeta$, we write this relation as
\begin{align}
T_a^{Q_i}(\zeta_i)\,T_b^{(0)}(u)\,\widehat R^{ab}(\zeta_i-1-u)&=\widehat R^{ab}(\zeta_i-u)\,T_b^{(i)}(u)\,T_a^{Q_i}(\zeta_i).\tag{\ref*{offShellRcoefficients}a}\label{offShellMixedYangBaxter}
\end{align}
The two arguments of $\widehat R^{ab}$ follow from the ordering of factors in Eq.~(\ref{offShellQMonodromy}). Moving $\widehat R^{ab}$ from right to left through the product, we first encounter the particles with labels $k>i$, for which the auxiliary parameter in the $Q$-monodromy is $\zeta_i-1$. At particle $i$, the factor $\widehat R^{ia}(0)=P^{ia}$ interchanges spaces $i$ and $a$. It transfers the shift to the $i$th factor of $T_b^{(0)}(u)$, producing $T_b^{(i)}(u)$. For the remaining particles $k<i$, the auxiliary parameter is $\zeta_i$. At every site $k\ne i$, the rearrangement follows from the ordinary Yang-Baxter equation.

To extract the relation for $A_{Q_i}B_0(u)$, take the matrix element $\langle\uparrow\uparrow|\cdot|\uparrow\downarrow\rangle$ of Eq.~(\ref{offShellMixedYangBaxter}) in the ordered auxiliary spaces $(a,b)$. For $D_{Q_i}B_0(u)$, first multiply the same equation on the left by $\widehat R^{ab}(u-\zeta_i)$ and use $\widehat R^{ab}(-v)\widehat R^{ab}(v)=I$. Taking the $\langle\downarrow\uparrow|\cdot|\downarrow\downarrow\rangle$ matrix element then gives the second relation. Solving for the two products yields

\begin{align}
A_{Q_i}B_0(u)&=\frac{B_i(u)A_{Q_i}}{b(\zeta_i-u-1)}-\frac{c_R(\zeta_i-u-1)}{b(\zeta_i-u-1)}B_{Q_i}A^{\rm mon}_0(u),\nonumber\\D_{Q_i}B_0(u)&=\frac{B_i(u)D_{Q_i}}{b(u-\zeta_i)}-\frac{c_R(u-\zeta_i)}{b(u-\zeta_i)}B_{Q_i}D_0(u).
\label{offShellQCommutations}
\end{align}
In the first term of each line, $B_i$ produces the required vector at the shifted coordinates. The other term contains $B_{Q_i}$. For arbitrary rapidities such terms survive in the action on a single Bethe vector, so we must arrange for them to cancel in the sum that defines the qKZ amplitude. Following the off-shell construction, we sum over auxiliary rapidities with a scalar weight $W$:
\begin{align}
A^{Q_0}(\boldsymbol\zeta)=\sum_{\boldsymbol\ell\in\mathbb Z^M}W(\boldsymbol\zeta,\boldsymbol u)|\boldsymbol u;\boldsymbol\zeta\rangle,\qquad u_m=\widetilde u_m-\ell_m.
\label{summation}
\end{align}
The constants $\widetilde u_m$ specify the auxiliary summation lattices. The sum is understood with a prescription permitting the unit-shift reindexing used below. We determine $W$ by requiring the terms containing
only shifted $B_i$ operators to reproduce $A^{Q_0}(\boldsymbol\zeta-\boldsymbol e_i)$ and the remaining terms to cancel in the sum.

Commuting $A_{Q_i}+D_{Q_i}$ through the product in Eq.~(\ref{offShellBetheVector}) using Eq.~(\ref{offShellQCommutations}) gives these two types of contributions. On the polarized vector, $A_{Q_i}|\Omega\rangle=|\Omega\rangle$ and $D_{Q_i}|\Omega\rangle=0$. The second equality follows because $b(0)=0$ in the $i$th factor. The contribution with the required shifted Bethe vector therefore has coefficient $\prod_m b(\zeta_i-u_m-1)^{-1}$. Its scalar weight must satisfy
\begin{align}
W(\boldsymbol\zeta-\boldsymbol e_i,\boldsymbol u)=W(\boldsymbol\zeta,\boldsymbol u)\prod_{m=1}^{M}\frac{1}{b(\zeta_i-u_m-1)}.
\label{offShellExternalWeightShift}
\end{align}
Both this requirement and cancellation of the additional terms are achieved with the factorized weight
\begin{align}
W(\boldsymbol\zeta,\boldsymbol u)=\prod_{j=1}^{N}\prod_{m=1}^{M}p(\zeta_j-u_m)\prod_{m<n}\tau(u_m-u_n),
\label{offShellFactorizedWeight}
\end{align}
provided the scalar functions obey
\begin{align}
p(v+1)&=b(v)p(v),\nonumber\\
\frac{\tau(v)}{b(v)}&=\frac{\tau(v+1)}{b(-v-1)}.
\label{offShellWeightShifts}
\end{align}
The first relation immediately gives Eq.~(\ref{offShellExternalWeightShift}). The second controls the factors involving pairs of auxiliary rapidities when the additional terms are combined. More explicitly, let $U^A_{i,m}(\boldsymbol\zeta,\boldsymbol u)$ and $U^D_{i,m}(\boldsymbol\zeta,\boldsymbol u)$ denote the weighted additional contributions associated with rapidity $u_m$ from $A_{Q_i}$ and $D_{Q_i}$, respectively. The commutation relations and Eq.~(\ref{offShellWeightShifts}) give
\begin{align}
U^A_{i,m}(\boldsymbol\zeta,\boldsymbol u-\boldsymbol e_m) =-U^D_{i,m}(\boldsymbol\zeta,\boldsymbol u),
\label{offShellPairCancellation}
\end{align}
where $\boldsymbol e_m$ here selects an auxiliary rapidity. Since $u_m\to u_m-1$ is a reindexing $\ell_m\to\ell_m+1$ of the same lattice, these contributions cancel in Eq.~(\ref{summation}). The surviving terms are exactly the sum at $\boldsymbol\zeta-\boldsymbol e_i$. This proves the qKZ relation for each $i$; the detailed commutations yielding Eq.~(\ref{offShellPairCancellation}) are given in Refs.~\cite{Babujian_1997,PasnooriGrossNeveu}.

An explicit meromorphic solution of the scalar recurrences is
\begin{align}
p(v)=\frac{\Gamma(v)}{\Gamma(v-i\eta)},\qquad \tau(v)=\frac{v\Gamma(v-i\eta)}{\Gamma(v+i\eta+1)}.
\label{offShellScalarWeights}
\end{align}
The identity $\Gamma(v+1)=v\Gamma(v)$ verifies Eq.~(\ref{offShellWeightShifts}). Substituting these functions into the weighted sum gives the reference spin amplitude, which takes the following form
\begin{align}
A^{Q_0}(\boldsymbol\zeta) ={}&\sum_{\boldsymbol\ell\in\mathbb Z^M} \left[\prod_{m=1}^{M}B_0(u_m)\right]|\Omega\rangle \prod_{j=1}^{N}\prod_{m=1}^{M} \frac{\Gamma(\zeta_j-u_m)}{\Gamma(\zeta_j-u_m-i\eta)}
\nonumber\\&\times\prod_{m<n}\frac{(u_m-u_n)\Gamma(u_m-u_n-i\eta)}{\Gamma(u_m-u_n+i\eta+1)},\qquad u_m=\widetilde u_m-\ell_m.
\label{ansatzstatefinalform}
\end{align}
Here $\zeta_j=\bar w_j$ for a left mover and $\zeta_j=w_j$ for a right mover, with the coordinates and $\eta$ defined in Eqs.~(\ref{relGNqKZ}) and (\ref{dimensionlessQKZData}). All rapidities and arguments in this expression are dimensionless. In contrast to the static model, here the auxiliary rapidities are thus summed rather than fixed by the static Bethe equations. This sum in Equation~(\ref{ansatzstatefinalform}) is a Jackson-type sum in which each auxiliary rapidity takes the values $u_m=\widetilde u_m-\ell_m$, with $\ell_m\in\mathbb Z$ and the offsets $\widetilde u_m$ chosen so that the summand is finite at each lattice point. The Gamma-function weights $p(v)$ and $\tau(v)$ used here give one solution of Eq.~(\ref{offShellWeightShifts}). Since these equations relate values one unit apart, multiplying either weight by a function of period one preserves its recurrence. For a given choice of weights, one may also multiply every amplitude in the wavefunction by the same scalar function of the particle coordinates $\boldsymbol z$. This common factor preserves the two-particle $S$-matrix relations and, if it is $L$-periodic in each argument, the periodic boundary conditions. One example is $\exp(i\sum_j k_j z_j^{\rm null})$ with $k_jL\in2\pi\mathbb Z$. Such a factor can be included when specifying the state.

Note that the above construction applies at any nonzero drive rate $\alpha$ and uses no expansion in $\alpha L$. Multiplying Eq.~(\ref{ansatzstatefinalform}) by the scalar factors in Eq.~(\ref{separation}) gives $f_0$, and the bare $S$-matrices then determine every other amplitude corresponding to different orderings. The next subsection displays this reconstruction explicitly.

\subsection{Structure of the exact wavefunction}
\label{sec:wavefunctionstructure}

The preceding subsection gives the spin amplitude for the reference ordering $Q_0=(N,\ldots,1)$. We now use it to answer a physical question. When the particles return to the same positions after going around the system, how does the wavefunction record the interactions they have undergone? The positions can always be specified in the interval $0\leq x_i<L$, but a given pair can meet many times. The two particles generally encounter a different interaction strength each time they meet. The arguments of their bare $S$-matrix must therefore identify the time of each encounter. This dependence can be read directly from the shifts in the qKZ equations and the accompanying scalar difference equation.

\paragraph{Two particles on successive trips around the system.}
Consider a right mover $r$ and a left mover $l$ in the reference ordering $x_r<x_l$ at time $t$. Put $a=z_r=x_r-t$ and $b=\bar z_l=x_l+t$. At time $s$ their positions, followed through the ends of the system, are $a+s$ and $b-s$. They meet whenever these positions agree modulo $L$. Writing $d=b-a$, we obtain the successive collision times
\begin{align}
t_{c,j}=\frac{d+jL}{2}=t+\frac{x_l-x_r+jL}{2},\qquad j=0,1,2,\ldots .
\label{structure_successive_collisions}
\end{align}
The first encounter after the observation time $t$ occurs at $t_{c,0}$. The time to the next encounter is $L/2$, since the relative speed is two. Write $S(d)$ for the full bare matrix $S^{rl}(a,b)$ of Eq.~(\ref{smatlr}), including its scalar phase, with the coupling evaluated at $t=d/2$. At the next encounter the coupling is $g((d+L)/2)$ and the matrix is $S(d+L)$.

These same arguments occur in the two periodicity relations of Eq.~(\ref{periodicTwoParticleForward}). Taking the right mover through the right end sends $a\to a-L$, while taking the left mover through the left end sends $b\to b+L$. Either step increases $b-a$ by $L$. Apply first the right-mover relation to the full reference amplitude $f_0$ and then the left-mover relation at the shifted argument. This gives
\begin{align}
 f_0(b+L,a-L)= S(d+L)\,f_0(b,a-L)= S(d+L)S(d)\,f_0(b,a).
 \label{structure_pair_one_trip}
\end{align}
The rightmost matrix acts at $t_{c,0}$ and the one to its left acts at $t_{c,1}$. During the time interval $L$, the right mover has travelled a distance $L$ to the right and the left mover a distance $L$ to the left. Both have returned to their original positions in the interval. At these positions the null coordinates at time $t+L$ are precisely $(b+L,a-L)$. Equation~(\ref{structure_pair_one_trip}) therefore gives the amplitude one trip later. After $m$ trips it becomes
\begin{align}
f_0(b+mL,a-mL)
 &= S\bigl(d+(2m-1)L\bigr)\cdots S(d+L)S(d)\,f_0(b,a),
 \qquad m\geq 1 .
\label{structure_pair_many_trips}
\end{align}
A shift of one particle's argument advances the collision-time argument for this pair by $L/2$. Shifting both arguments in their forward directions advances the common observation time by $L$. For a constant interaction the same bare matrix occurs at every encounter. For the time-dependent interaction the ordered product samples the coupling at each of the times in Eq.~(\ref{structure_successive_collisions}).

\paragraph{Periodic transport for $N$ particles.}
For several particles, the off-shell sum in Eq.~(\ref{ansatzstatefinalform}) gives the reference spin amplitude $A_0$. Multiplying by the solution $h$ of the scalar difference equation~(\ref{scalarDifferenceEquation}) restores the scalar phases of the opposite-chirality matrices. With the dimensionless coordinates $\boldsymbol\zeta(\boldsymbol z)$ of Eqs.~(\ref{relGNqKZ}) and~(\ref{dimensionlessQKZData}), the full reference amplitude is
\begin{align}
f_0(\boldsymbol z)
&=\mathcal H_0(\boldsymbol z)
A_0\bigl(\boldsymbol\zeta(\boldsymbol z)\bigr),\nonumber\\
\mathcal H_0(\boldsymbol z)
&=\prod_{r=N_L+1}^N\prod_{l=1}^{N_L}h(\bar z_l-z_r).
\label{complete_reference_block}
\end{align}
Let $T_i\boldsymbol z=\boldsymbol z-L\boldsymbol e_i$. At fixed time, identifying $x_r=L$ with $x_r=0$ gives the shift $T_r$, while identifying $x_l=0$ with $x_l=L$ gives $T_l^{-1}$. These are the steps for forward transport in Sec.~\ref{sec:periodicqKZ},
\begin{align}
f_0(T_r\boldsymbol z)
&=Z_r(\boldsymbol z)f_0(\boldsymbol z),\nonumber\\
f_0(T_l^{-1}\boldsymbol z)
&=Z_l^{(+)}(\boldsymbol z)f_0(\boldsymbol z),
\qquad
Z_l^{(+)}(\boldsymbol z)
=\bigl[Z_l(T_l^{-1}\boldsymbol z)\bigr]^{-1}.
\label{complete_forward_transport}
\end{align}
Here $Z_l^{(+)}$ uses the direct bare matrices for encounters with right movers. The last equality follows by reversing the transport of the left mover and evaluating its inverse at the shifted arguments. It allows the qKZ solution to use a common negative shift for both chiralities. Including the scalar factors, the resulting equation is
\begin{align}
T_i\boldsymbol z=\boldsymbol z-L\boldsymbol e_i,
\qquad
f_0(T_i\boldsymbol z)=Z_i(\boldsymbol z)f_0(\boldsymbol z).
\label{complete_one_step}
\end{align}
The scalar multiplier in this equation follows from the ratio of the reference scalar factors before and after the shift,
\begin{align}
Z_i(\boldsymbol z)
&=\Xi_i(\boldsymbol z)Z'_i(\boldsymbol z),\nonumber\\
\Xi_i(\boldsymbol z)
&=\frac{\mathcal H_0(T_i\boldsymbol z)}
{\mathcal H_0(\boldsymbol z)}.
\label{complete_scalar_transport}
\end{align}
Here $Z'_i$ acts on the reference spin amplitude. Writing out the scalar multipliers gives
\begin{align}
\Xi_r(\boldsymbol z)
&=\prod_{l=1}^{N_L}e^{i\phi(\bar z_l-z_r)},
\qquad r=N_L+1,\ldots,N,\nonumber\\
\Xi_l(\boldsymbol z)
&=\prod_{r=N_L+1}^{N}e^{-i\phi(\bar z_l-z_r-L)},
\qquad l=1,\ldots,N_L.
\label{complete_chiral_scalar_transport}
\end{align}
The first line is the phase for forward right-moving transport. The second line uses the reversed left-moving transport, for which $\bar z_l-z_r$ decreases by $L$. In the forward direction $\bar z_l$ increases by $L$, and the corresponding left-mover phase is $\prod_{r=N_L+1}^N e^{i\phi(\bar z_l-z_r)}$, as in Eq.~(\ref{periodicLeftForwardPhase}). To follow the same particle through successive trips, we apply its transport matrix at the arguments reached on the preceding step. For example, two negative shifts give
\begin{align}
f_0(T_i^2\boldsymbol z)
=Z_i(T_i\boldsymbol z)Z_i(\boldsymbol z)f_0(\boldsymbol z).
\label{complete_two_steps}
\end{align}
For any positive integer $p$, the product for $p$ negative shifts and the inverse relation for a single positive shift read
\begin{align}
 f_0(T_i^p\boldsymbol z)
 &= Z_i(T_i^{p-1}\boldsymbol z)\cdots Z_i(T_i\boldsymbol z)
 Z_i(\boldsymbol z)f_0(\boldsymbol z),\nonumber\\
 f_0(T_i^{-1}\boldsymbol z)
 &= Z_i(T_i^{-1}\boldsymbol z)^{-1}f_0(\boldsymbol z).
 \label{complete_iterated_transport}
\end{align}
Iteration of the second line gives positive shifts of any order, including repeated forward transport of a left mover. Each factor in these products carries the arguments appropriate to that step, just as the two successive factors in Eq.~(\ref{structure_pair_one_trip}) do.

Let $\boldsymbol n\in\mathbb Z^N$ specify the shifts of all the particle coordinates. Define the transport of the reference amplitude by
\begin{align}
f_0(\boldsymbol z-L\boldsymbol n)
=\mathcal U_{\boldsymbol n}(\boldsymbol z)f_0(\boldsymbol z),
\qquad \mathcal U_{\boldsymbol 0}=I.
\label{complete_winding_transport}
\end{align}
A positive $n_r$ counts forward trips of a right mover, and a negative $n_l$ counts forward trips of a left mover. The product defining $\mathcal U_{\boldsymbol n}$ uses the matrices $Z_i$ and their shifted inverses. The consistency relation~(\ref{transportop}),
\begin{align}
Z_i(T_j\boldsymbol z)Z_j(\boldsymbol z)
=Z_j(T_i\boldsymbol z)Z_i(\boldsymbol z),
\label{complete_flatness}
\end{align}
ensures that the product gives the same amplitude for every order of performing the prescribed shifts. If the shifts $\boldsymbol n$ are followed by $\boldsymbol m$, the second transport starts at $\boldsymbol z-L\boldsymbol n$. Thus
\begin{align}
\mathcal U_{\boldsymbol n+\boldsymbol m}(\boldsymbol z)
=\mathcal U_{\boldsymbol m}(\boldsymbol z-L\boldsymbol n)
\mathcal U_{\boldsymbol n}(\boldsymbol z).
\label{complete_transport_cocycle}
\end{align}
These formulas hold wherever the required matrices are defined and invertible, with continuation as specified by the analytic prescription for the reference amplitude.

A trip in which every particle moves in its own direction has $\boldsymbol n=\boldsymbol v$, where $v_i=v_{\chi_i}$ is $+1$ for a right mover and $-1$ for a left mover. On following the motion during a time $L$, the displacement of particle $i$ is $Lv_i$. When we express its position in $[0,L)$, its null coordinate at the later time is $x_i-v_i(t+L)$. Hence, at fixed positions,
\begin{align}
\boldsymbol z(\boldsymbol x,t+L)
=\boldsymbol z(\boldsymbol x,t)-L\boldsymbol v.
\label{complete_collective_time_shift}
\end{align}
Equation~(\ref{complete_winding_transport}) with $\boldsymbol n=\boldsymbol v$ now relates $f_0$ at $t+L$ to $f_0$ at $t$. On the second trip the transport is $\mathcal U_{\boldsymbol v}(\boldsymbol z-L\boldsymbol v)$, so the amplitude after two trips contains the product $\mathcal U_{\boldsymbol v}(\boldsymbol z-L\boldsymbol v)\mathcal U_{\boldsymbol v}(\boldsymbol z)$. This gives the transformation of the reference amplitude under a finite shift of time. Although the positions repeat, the drive generally has $g(t+L)\ne g(t)$ and successive trips involve different scattering matrices. Each elementary identification relates amplitudes at a common time. Their product for all particles has a direct interpretation in the evolution governed by Eq.~(\ref{SE}): during the interval from $t$ to $t+L$, each oppositely moving pair meets twice, and equal-chirality particles retain their relative positions. Matrices $S'$ that occur when an individual shift is written in the reference ordering implement the Yang-Baxter rearrangements of these amplitudes. The bare matrices $S^{rl}$ at the successive collision times give the physical scattering in that interval.

\paragraph{The wavefunction for all orderings and shifts.}
To obtain the amplitude for a spatial ordering $Q$ at the observation time, choose a sequence of adjacent interchanges $$Q_0\to Q_1\to\cdots\to Q_m=Q.$$ Let $S_a$ be the bare two-particle $S$-matrix for the $a$th interchange in this sequence. For an opposite-chirality pair, $(r,l)\to(l,r)$ uses $S^{rl}$ evaluated with the coupling at their collision time. Particles of equal chirality travel at the same velocity. Their amplitudes are related by $S'$ in Eq.~(\ref{sameChiralityExchange}), chosen to satisfy the Yang-Baxter equations with $S^{rl}$. Reversing an interchange uses its inverse. The full ordering product and its action are
\begin{align}
\mathcal S_Q(\boldsymbol z)
&=S_m(\boldsymbol z)\cdots
S_1(\boldsymbol z),
\qquad \mathcal S_{Q_0}=I,\nonumber\\
f^Q(\boldsymbol z)
&=\mathcal S_Q(\boldsymbol z)f_0(\boldsymbol z).
\label{complete_ordering_amplitudes}
\end{align}
Each factor acts on the spin spaces of its two particles and is evaluated using their null coordinates. The Yang-Baxter equations, commutativity for disjoint pairs, and the inverse relations ensure independence of the chosen sequence. Writing $\mathcal S_Q=\omega_Q\mathcal R_Q$ separates the product of scalar phases $\omega_Q$ from the ordered matrices $\mathcal R_Q$ acting in spin spaces. The corresponding factorization is
\begin{align}
f^Q=\mathcal H_Q A_Q,\qquad
\mathcal H_Q=\omega_Q\mathcal H_0,\qquad
A_Q=\mathcal R_Q A_0.
\label{complete_sector_scalars}
\end{align}
The scalar factor relative to $\mathcal H_0$ is fixed by the phases of the bare $S$-matrices, while the spin amplitude is obtained by acting on $A_0$ with the ordered product of $R$-matrices.

After $\boldsymbol n$ shifts, the reference amplitude must first be transported to the required arguments. The matrices relating it to ordering $Q$ are then evaluated at those same arguments. Combining Eqs.~(\ref{complete_ordering_amplitudes}) and~(\ref{complete_winding_transport}) gives
\begin{align}
f^Q(\boldsymbol z-L\boldsymbol n)
={}&\mathcal S_Q(\boldsymbol z-L\boldsymbol n)
\mathcal U_{\boldsymbol n}(\boldsymbol z)
\mathcal H_0(\boldsymbol z)
A_0\bigl(\boldsymbol\zeta(\boldsymbol z)\bigr).
\label{complete_all_blocks}
\end{align}
For example, the elementary transport for $f^Q$ is
\begin{align}
Z_i^{(Q)}(\boldsymbol z)
&=\mathcal S_Q(T_i\boldsymbol z)
Z_i(\boldsymbol z)\mathcal S_Q(\boldsymbol z)^{-1},
\nonumber\\
f^Q(T_i\boldsymbol z)
&=Z_i^{(Q)}(\boldsymbol z)f^Q(\boldsymbol z).
\label{complete_sector_transport}
\end{align}
The inverse factor converts $Q$ to the reference ordering before the shift, and the first factor restores $Q$ after it. For one forward trip of every particle, the same reasoning gives the transformation of the amplitude in its observed ordering,
\begin{align}
f^Q\bigl(\boldsymbol z(\boldsymbol x,t+L)\bigr)
={}&\mathcal S_Q(\boldsymbol z-L\boldsymbol v)\,
\mathcal U_{\boldsymbol v}(\boldsymbol z)\,
\mathcal S_Q(\boldsymbol z)^{-1}\,
 f^Q(\boldsymbol z),
 \qquad \boldsymbol z=\boldsymbol z(\boldsymbol x,t).
\label{structure_time_shift_ordering}
\end{align}

The amplitudes $f^Q$ describe all orderings of a single state at the common time $t$. In the physical configuration domain $\mathcal C=[0,L)^N$, the functions $\Theta_Q(\boldsymbol x)$ defined in Eq.~(\ref{complete_ordering_projector}) select the appropriate ordering. The coordinate wavefunction is therefore
\begin{align}
F_{\mathcal C}(\boldsymbol x,t)
=\sum_{Q\in S_N}\Theta_Q(\boldsymbol x)
\mathcal S_Q(\boldsymbol z)\mathcal H_0(\boldsymbol z)
A_0\bigl(\boldsymbol\zeta(\boldsymbol z)\bigr).
\label{complete_cut_wavefunction}
\end{align}
In this form we evaluate the off-shell spin amplitude and its scalar factor directly at the arguments reached at time $t$. Equivalently, the periodic transport equations determine the reference amplitude at all shifts of those arguments for which the chosen solution is defined. With periodic boundary conditions, we represent the position of each particle by $0\leq x_i<L$, so that $\boldsymbol x\in\mathcal C=[0,L)^N$. At fixed particle positions, advancing the time from $t$ to $t+L$ changes each right-moving argument $z_r=x_r-t$ to $z_r-L$ and each left-moving argument $\bar z_l=x_l+t$ to $\bar z_l+L$. To keep track of these shifts, we choose $D=[0,L)^N$ as a reference domain for the null coordinates. The periodic transport equations then determine the reference amplitude $f_0$ at the shifted arguments from its values in $D$, with the ordered products of bare $S$-matrices accounting for the interactions on successive trips. Denote the restriction of $f_0=\mathcal H_0 A_0$ to $D$ by $f_D$. For each $\boldsymbol n\in\mathbb Z^N$ put $\boldsymbol y_{\boldsymbol n}=\boldsymbol z+L\boldsymbol n$, and let $\mathbf 1_D$ be one on $D$ and zero outside it. There is a unique $\boldsymbol n$ for which $\boldsymbol y_{\boldsymbol n}\in D$. Starting from that point, $\mathcal U_{\boldsymbol n}(\boldsymbol y_{\boldsymbol n})$ transports the reference amplitude to $\boldsymbol y_{\boldsymbol n}-L\boldsymbol n=\boldsymbol z$. Equation~(\ref{complete_cut_wavefunction}) can thus be written with the shifts displayed explicitly,
\begin{align}
F_{\mathcal C}(\boldsymbol x,t)
={}&\sum_{Q\in S_N}\sum_{\boldsymbol n\in\mathbb Z^N}
 \Theta_Q(\boldsymbol x)\,\mathbf 1_D(\boldsymbol y_{\boldsymbol n})\,
 \mathcal S_Q(\boldsymbol z)\,
 \mathcal U_{\boldsymbol n}(\boldsymbol y_{\boldsymbol n})\,
 f_D(\boldsymbol y_{\boldsymbol n}).
\label{structure_wavefunction_signed_shifts}
\end{align}
Only the selected term in the sum over $\boldsymbol n$ is evaluated. When $t$ advances by $L$ at fixed $\boldsymbol x$, its index changes from $\boldsymbol n$ to $\boldsymbol n+\boldsymbol v$, and the point $\boldsymbol y_{\boldsymbol n}$ in the reference intervals is recovered. Equation~(\ref{complete_transport_cocycle}) then supplies the scatterings of the next trip from the new starting arguments. The sum displays where each continued amplitude enters the wavefunction. The ordered product in that amplitude carries the interactions from the successive trips.

When a right mover and a left mover coincide, the limiting amplitudes satisfy the matching condition used to derive the opposite-chirality bare matrix. At the ends of the interval, let $\widehat Q$ denote the ordering of the particles other than $i$. Placing $i$ at $x_i=0$ gives $i\widehat Q$, and placing it at $x_i=L$ gives $\widehat Q i$. The physical boundary condition is
\begin{align}
f^{i\widehat Q}(\ldots,z_i^{\rm null},\ldots)
=f^{\widehat Q i}(\ldots,z_i^{\rm null}+L,\ldots)
\qquad z_i^{\rm null}=-v_{\chi_i}t.
\label{complete_seam_sewing}
\end{align}
The left-hand side is evaluated at $x_i=0$ and the right-hand side at $x_i=L$, with $t$ and all other positions fixed. The displacement by $L$ shifts $z_i^{\rm null}$ by $L$ and $\zeta_i$ by one. Matching these two ordering amplitudes gives the qKZ and scalar difference equations used above. Inserting the coordinate wavefunction of Eq.~(\ref{structure_wavefunction_signed_shifts}) into the many-body state gives
\begin{align}
|N_L,N_R;t\rangle
={}&\mathcal N\sum_{\boldsymbol\sigma}
\int_{[0,L)^N}d^Nx\,
\bigl[\mathcal A F_{\mathcal C}(\boldsymbol x,t)
\bigr]_{\boldsymbol\sigma}
\prod_{i=1}^N\psi^\dagger_{\chi_i\sigma_i}(x_i)|0\rangle.
\label{complete_fock_state}
\end{align}
Here $\mathcal A$ implements the fermionic antisymmetrization of Eq.~(\ref{npwf1}), the fields obey Eq.~(\ref{pbc}), and $\mathcal N$ normalizes the chosen nonzero normalizable solution. Each physical coordinate is integrated over one interval. All continued values are determined by that solution, so the integers $\boldsymbol n$ specify its arguments and carry no independent state multiplicity.

The off-shell solution also contains an integer sum. For a fixed external shift $\boldsymbol n$, its indices $\boldsymbol\ell\in\mathbb Z^M$ run over auxiliary rapidities $\widetilde u_a-\ell_a$. If $\mathcal V$ denotes the summand of Eq.~(\ref{ansatzstatefinalform}), including the Bethe vector and its Gamma-function weights, the shifted reference amplitude reads
\begin{align}
f_0(\boldsymbol z-L\boldsymbol n)
={}&\mathcal H_0(\boldsymbol z-L\boldsymbol n)
\sum_{\boldsymbol\ell\in\mathbb Z^M}
\mathcal V\!\left(
\boldsymbol\zeta(\boldsymbol z)-\boldsymbol n,\,
\widetilde{\boldsymbol u}-\boldsymbol\ell\right).
\label{complete_two_lattices}
\end{align}
The shift $\boldsymbol n$ changes the external arguments of the complete sum. Reindexing the auxiliary rapidities proves the difference equations through cancellations between summands. In Sec.~\ref{sec:Analysis} we apply Poisson resummation to this auxiliary sum to obtain the Yang-Yang representation. The scalar phases, the amplitudes corresponding to different orderings, and the transport operators derived here enter that representation for every nonzero drive rate for which the chosen sum is defined.

\section{The Yang-Yang representation and the instantaneous spectrum}
\label{sec:Analysis}

In the preceding sections, we constructed the exact time-dependent wavefunction using the generalized Bethe ansatz. We now use this wavefunction to determine when a state prepared in an instantaneous eigenstate follows that eigenstate as the coupling varies. Our main focus is the adiabatic regime. As we shall show, the dimensionless current-current coupling gives rise, through dynamical dimensional transmutation, to a time-dependent mass gap in the instantaneous isospin spectrum. We also discuss the fast-drive regime, in which the interaction becomes weak. At fixed ultraviolet cutoff, the mass gap tends to zero as one approaches the gapless $SU(2)_1$ WZW model.

Our starting point is the reference amplitude in Eq.~(\ref{ansatzstatefinalform}), which solves the qKZ equations and is expressed as a Jackson-type sum over the auxiliary rapidities $u_1,\ldots,u_M$. To make its connection with the standard Bethe ansatz explicit, we apply Poisson resummation to the discrete sum. This gives an exact integral representation in terms of a Yang-Yang functional at arbitrary drive rate. We then identify a regime of slow variation near a characteristic time $t_0$ in which the saddle-point equations reduce to the instantaneous Bethe equations. We derive these equations independently from the Hamiltonian at fixed time and examine when transitions between its instantaneous eigenstates are suppressed.

\subsection{The exact Yang-Yang functional of the qKZ state}

We start from the reference amplitude $A_0(\boldsymbol\zeta)$, using the dimensionless coordinates $\boldsymbol\zeta=(\bar w_1,\ldots,\bar w_{N_L},w_{N_L+1},\ldots,w_N)$ defined in Eqs.~(\ref{relGNqKZ}) and~(\ref{dimensionlessQKZData}). In the off-shell solution, the scalar weight associated with $M$ auxiliary rapidities $\boldsymbol u$ is
\begin{align}
\mathcal W_\eta(\boldsymbol u;\boldsymbol\zeta)
={}&\prod_{a=1}^M\prod_{j=1}^N
\frac{\Gamma(\zeta_j-u_a)}{\Gamma(\zeta_j-u_a-i\eta)}
\prod_{a<b}
\frac{(u_a-u_b)\Gamma(u_a-u_b-i\eta)}
{\Gamma(u_a-u_b+i\eta+1)}.
\label{YYexactweight}
\end{align}
For each auxiliary rapidity $u_a$, the product over $j$ includes each physical fermion once, with $\zeta_j$ taking the value appropriate to its chirality. The second product runs over pairs of auxiliary rapidities. The weight multiplies the spin vector
\begin{align}
\mathcal B_\eta(\boldsymbol u;\boldsymbol\zeta)
=\left[\prod_{a=1}^M B_0(u_a)\right]|\Omega\rangle.
\label{Betheprefactor}
\end{align}
The dependence of this vector on $\eta$ and on the particle coordinates enters through the operators $B_0(u_a)$ of the auxiliary monodromy matrix in Eq.~(\ref{offShellMonodromy}). At this stage the rapidities are off shell, ranging over the lattice specified by the analytic prescription for the sum. The transfer-matrix eigenvalue equations enter later, when we identify the rapidities at a contributing saddle. In the full coordinate wavefunction, the scalar factor $\mathcal H_0$, the ordered products of bare $S$-matrices, and the transport operators for successive trips around the system all act on the spin amplitude after the auxiliary-rapidity sum has been performed. We may therefore work directly with $A_0$ and restore these factors as in Sec.~\ref{sec:wavefunctionstructure}.

To pass from the discrete sum to integrals, first consider a single rapidity on a lattice with offset $\widetilde u$. For a function $f$ satisfying the convergence conditions for Poisson summation, we have
\begin{align}
\sum_{l\in\mathbb Z}f(\widetilde u-l)
=\sum_{m\in\mathbb Z}e^{2\pi im\widetilde u}
\int_{\mathcal C}du\,f(u)e^{-2\pi imu}.
\label{Poisson}
\end{align}
The contour $\mathcal C=\widetilde u+\mathbb R$ runs in the direction of increasing real part. For example, with $M=1$ and real $\eta\ne0$, the site factors in Eq.~(\ref{YYexactweight}) give a leading power $|u|^{iN\eta}$ as $u\to-\infty$, while each nonzero one-spin-flip component of $B_0(u)|\Omega\rangle$ is of order $1/|u|$. The resulting absolute sum has a harmonic tail. We therefore apply Eq.~(\ref{Poisson}) in a domain of convergence, or with a regulator, and continue both sides with the same prescription that permits reindexing the qKZ sum. When we deform a contour we track the poles of the product of the Gamma-function weight and the Bethe vector. A pole crossed in one integration variable contributes its residue, with the factor $2\pi i$ and the sign fixed by the contour orientation, under the remaining integrals. Applying Eq.~(\ref{Poisson}) to each spin component and carrying out the resummation for each of the $M$ auxiliary rapidities gives
\begin{align}
&A_0(\boldsymbol\zeta)={}\sum_{\boldsymbol m\in\mathbb Z^M}e^{2\pi i\boldsymbol m\cdot\widetilde{\boldsymbol u}}\int_{\boldsymbol{\mathcal C}}d^Mu\,\mathcal B_\eta(\boldsymbol u;\boldsymbol\zeta)
e^{-\mathcal Y_{\eta,\boldsymbol m}(\boldsymbol u;\boldsymbol\zeta)},
\label{finalformstateexact}\\
&\mathcal Y_{\eta,\boldsymbol m}(\boldsymbol u;\boldsymbol\zeta)={}-\log\mathcal W_\eta(\boldsymbol u;\boldsymbol\zeta)+2\pi i\sum_{a=1}^M m_a u_a.
\label{YYtotal}
\end{align}
Each contour in $\boldsymbol{\mathcal C}=\mathcal C_1\times\cdots\times\mathcal C_M$ passes through its corresponding lattice offset. The integers $\boldsymbol m$ label the Fourier modes of the auxiliary-rapidity sum, while the winding indices of Sec.~\ref{sec:wavefunctionstructure} count shifts in the external coordinates of the particles. The Fourier convention in Eq.~(\ref{Poisson}) fixes the sign of the linear term in $\mathcal Y_{\eta,\boldsymbol m}$ and gives the weight $e^{-\mathcal Y_{\eta,\boldsymbol m}}$ in the integral. We call $\mathcal Y_{\eta,\boldsymbol m}$ the exact Yang-Yang functional of the qKZ solution. It is the logarithmic action for the auxiliary Bethe rapidities, with the physical coordinates held fixed in the integral and a consistent local branch chosen for the logarithm. Equations~(\ref{finalformstateexact}) and (\ref{YYtotal}) retain the full scalar weight in Eq.~(\ref{ansatzstatefinalform}), including its elementary pair factors. They apply for any nonzero rate $\alpha$ in the integrable time-dependent coupling for which the prescribed qKZ sum is defined. At $\alpha=0$ one uses the unscaled static matrices of Sec.~\ref{sec:Wavefunction}. To determine how the different terms in this action contribute when the coupling varies slowly, we first write its Gamma-function ratios in a form that also keeps track of the phases produced by reflection.

\subsection{Gamma-function identities and the integral kernels}
\label{sec:YYGamma}

The two types of ratios in Eq.~(\ref{YYexactweight}) depend on the dimensionless differences
\begin{align}
\Delta_{ja}=\zeta_j-u_a,\qquad v_{ab}=u_a-u_b.
\end{align}
For each particle $j$ and auxiliary rapidity $u_a$, the site factor is $\Gamma(\Delta_{ja})/\Gamma(\Delta_{ja}-i\eta)$. For a pair of auxiliary rapidities with difference $v=v_{ab}$, we use the recurrence relation $\Gamma(z+1)=z\Gamma(z)$ to write
\begin{align}
\frac{v\,\Gamma(v-i\eta)}{\Gamma(v+i\eta+1)}
=\frac{v}{v+i\eta}\frac{\Gamma(v-i\eta)}{\Gamma(v+i\eta)}.
\label{YYpairRecurrence}
\end{align}
On a consistent logarithmic branch, the elementary term $\log[v/(v+i\eta)]$ separates from the differences of logarithms of Gamma functions. This allows us to estimate their respective orders when taking the slow-variation limit. To write each Gamma difference as a single convergent integral, we start from the representation valid for $\operatorname{Re}z>0$,
\begin{align}
\log\Gamma(z)
=\int_0^\infty\frac{d\tau}{\tau}
\left[(z-1)e^{-\tau}
+\frac{e^{-z\tau}-e^{-\tau}}{1-e^{-\tau}}\right].
\label{GammaIntegral}
\end{align}
Here $\tau$ is an auxiliary integration variable. At $\tau=0$, the constant terms of the two expressions in brackets cancel, so their sum divided by $\tau$ has a finite limit. Taking the difference of Eq.~(\ref{GammaIntegral}) at two arguments $A$ and $B$ gives
\begin{align}
&G(A,B)\equiv\log\Gamma(A)-\log\Gamma(B)\nonumber\\
&=\int_0^\infty\frac{d\tau}{\tau}
\left[(A-B)e^{-\tau}
+\frac{e^{-A\tau}-e^{-B\tau}}{1-e^{-\tau}}\right],
\qquad \operatorname{Re}A,\operatorname{Re}B>0.
\label{YYGammaDifference}
\end{align}
The linear term $(A-B)e^{-\tau}$ supplies the subtraction at the lower endpoint. We therefore retain it inside the integral, including when $A-B$ is independent of the auxiliary rapidities. On a part of the contour where the Gamma arguments have positive real parts, applying Eq.~(\ref{YYGammaDifference}) to the site and pair ratios gives
\begin{align}
\log\frac{\Gamma(\Delta)}{\Gamma(\Delta-i\eta)}
&=\int_0^\infty\frac{d\tau}{\tau}
\left[i\eta e^{-\tau}+e^{-\Delta\tau}K_\eta(\tau)\right],\nonumber\\
\log\frac{v\Gamma(v-i\eta)}{\Gamma(v+i\eta+1)}
&=\log\frac{v}{v+i\eta}
+\int_0^\infty\frac{d\tau}{\tau}
\left[-2i\eta e^{-\tau}+e^{-v\tau}\widetilde K_\eta(\tau)\right],
\label{YYsitePairIntegrals}
\end{align}
where
\begin{align}
K_\eta(\tau)=\frac{1-e^{i\eta\tau}}{1-e^{-\tau}},\qquad
\widetilde K_\eta(\tau)=
\frac{e^{i\eta\tau}-e^{-i\eta\tau}}{1-e^{-\tau}}.
\label{YYkernels}
\end{align}
The shift $-i\eta$ in the denominator of the site ratio fixes the sign in $K_\eta$, and the shifts $\mp i\eta$ in the pair ratio fix the signs in $\widetilde K_\eta$. In particular, $K_\eta(\tau)=-i\eta+\mathcal O(\tau)$ and $\widetilde K_\eta(\tau)=2i\eta+\mathcal O(\tau)$ near the lower endpoint, so each kernel cancels the constant part of its subtraction in Eq.~(\ref{YYsitePairIntegrals}). Summing these convergent expressions over the particles and auxiliary rapidities gives
\begin{align}
\log\mathcal W_\eta&=\sum_{a<b}\log\frac{v_{ab}}{v_{ab}+i\eta}+\int_0^\infty\frac{d\tau}{\tau}\Bigg[i\eta M(N-M+1)e^{-\tau}\nonumber\\&\qquad +\sum_{a,j}e^{-\Delta_{ja}\tau}K_\eta(\tau)+\sum_{a<b}e^{-v_{ab}\tau}\widetilde K_\eta(\tau)\Bigg].
\label{YYexactKernelAction}
\end{align}
The $MN$ site factors contribute $i\eta MN$ to the coefficient of $e^{-\tau}$, and the $M(M-1)/2$ auxiliary pairs contribute $-i\eta M(M-1)$. The resulting expression shows how every auxiliary rapidity couples to each physical fermion through $K_\eta$ and to the other auxiliary rapidities through $\widetilde K_\eta$. Taking the negative of this logarithm and adding the Fourier term gives the exact functional in Eq.~(\ref{YYtotal}) on this part of the contour.

To extend this representation to parts of the contour where a Gamma argument leaves the right half-plane, we use Euler's reflection identity,
\begin{align}
\Gamma(z)\Gamma(1-z)=\frac{\pi}{\sin\pi z}.
\label{reflection}
\end{align}
Applying it to the numerator and denominator of a site ratio gives
\begin{align}
\frac{\Gamma(\Delta)}{\Gamma(\Delta-i\eta)}
=\frac{\sin\pi(\Delta-i\eta)}{\sin\pi\Delta}
\frac{\Gamma(1-\Delta+i\eta)}{\Gamma(1-\Delta)}.
\label{YYsiteReflection}
\end{align}
For real $\eta$ and $\operatorname{Re}\Delta\leq0$, both Gamma functions on the right have arguments with positive real part. Applying the same reflection to the Gamma ratio in Eq.~(\ref{YYpairRecurrence}) gives
\begin{align}
\frac{v\Gamma(v-i\eta)}{\Gamma(v+i\eta+1)}
=\frac{v}{v+i\eta}
\frac{\sin\pi(v+i\eta)}{\sin\pi(v-i\eta)}
\frac{\Gamma(1-v-i\eta)}{\Gamma(1-v+i\eta)}.
\label{YYpairReflection}
\end{align}
Reflection leaves each ratio unchanged as a meromorphic function, so the complete product determines which poles are encountered in continuing the weight. For example, at a positive integer $\Delta$ the zero of $1/\Gamma(1-\Delta)$ in Eq.~(\ref{YYsiteReflection}) cancels the pole of $1/\sin(\pi\Delta)$. Keeping these factors together as we continue the weight, we can use Eq.~(\ref{YYGammaDifference}) wherever the reflected Gamma arguments have positive real parts and retain the sine factors explicitly in the integrand.

Denote the site ratios reflected on this part of the contour by $\mathfrak S$ and the reflected pair ratios by $\mathfrak P$. Collecting their sine factors together with the elementary factors from all auxiliary pairs gives
\begin{align}
\mathcal R_\eta
={}&\prod_{a<b}\frac{v_{ab}}{v_{ab}+i\eta}
\prod_{(j,a)\in\mathfrak S}
\frac{\sin\pi(\Delta_{ja}-i\eta)}{\sin\pi\Delta_{ja}}\nonumber\\
&\times\prod_{(a,b)\in\mathfrak P}
\frac{\sin\pi(v_{ab}+i\eta)}{\sin\pi(v_{ab}-i\eta)}.
\label{YYreflectionPrefactor}
\end{align}
For each site or pair, we evaluate $G(A,B)$ with the original arguments if the ratio is unreflected and with the arguments of Eq.~(\ref{YYsiteReflection}) or Eq.~(\ref{YYpairReflection}) if it is reflected. With $\mathcal I_\eta=\sum G(A,B)$ denoting the sum of these contributions, the exact weight and its Yang-Yang functional take the form
\begin{align}
\mathcal W_\eta=\mathcal R_\eta e^{\mathcal I_\eta},\;\;\;
\mathcal Y_{\eta,\boldsymbol m}
=-\mathcal I_\eta-\log\mathcal R_\eta
+2\pi i\boldsymbol m\cdot\boldsymbol u.
\label{YYreflectionAction}
\end{align}
Thus the integrand $\mathcal B_\eta e^{-\mathcal Y_{\eta,\boldsymbol m}}$ can also be written as
$\mathcal R_\eta\mathcal B_\eta
\exp(\mathcal I_\eta-2\pi i\boldsymbol m\cdot\boldsymbol u)$.
Both expressions retain the full scalar weight. For each scalar factor, including those written outside the exponential, its logarithm and its derivatives in the chosen limit determine its contribution to the action and to the saddle-point equations.

We can first find the saddle-point equations for the exact scalar functional by differentiating the original weight in Eq.~(\ref{YYexactweight}). Write $\psi_0(z)=\partial_z\log\Gamma(z)$ for the digamma function and denote the logarithmic derivative of the pair factor by
\begin{align}
D_\eta(v)=\frac1v+\psi_0(v-i\eta)-\psi_0(v+i\eta+1).
\end{align}
Keeping track of the sign with which $u_a$ enters each pair difference then gives
\begin{align}
0=\partial_{u_a}\mathcal Y_{\eta,\boldsymbol m}
={}&\sum_{j=1}^N
\bigl[\psi_0(\zeta_j-u_a)-\psi_0(\zeta_j-u_a-i\eta)\bigr]
\nonumber\\
&-\sum_{b>a}D_\eta(u_a-u_b)
+\sum_{b<a}D_\eta(u_b-u_a)+2\pi i m_a.
\label{YYexactstationary}
\end{align}
Equation~(\ref{YYexactstationary}) is exact wherever the scalar weight is finite and nonzero. To approximate a component of the spin amplitude by expanding about one of these points, we need a limit in which the scalar exponent controls the integral and the Bethe vector varies slowly across the saddle.

\subsection{Slow variation and the saddle-point limit}

The required large parameter comes from comparing the driving time scale with the time a particle takes to travel the length of the system. For the decreasing time-dependent coupling, take $\alpha>0$ and introduce the slow time $s$ and the scaled coordinates and rapidities,
\begin{align}
\varepsilon=\alpha L=\eta^{-1}\longrightarrow0,
\qquad \xi_j=\varepsilon\zeta_j,
\qquad U_a=\varepsilon u_a,
\qquad s=\alpha t+\frac\beta2.
\label{YYslowScaling}
\end{align}
Since the particles propagate with unit speed, $L$ is also the time a particle takes to go once around the system. The parameter $\varepsilon=\alpha L$ thus measures the variation of the time-dependent coupling over one such trip. We take $\varepsilon\to0$ at fixed $s$, initially keeping $N$ and $M$ finite, and use $\boldsymbol U$ as the integration variables. Two features of this limit enter the derivation. First, the Gamma functions in the qKZ weight acquire large arguments proportional to $\eta$. Second, by Eq.~(\ref{relGNqKZ}), their dependence on the particle positions enters through
\begin{align}
\xi_{R,j}=\alpha x_j-s,
\qquad \xi_{L,j}=\alpha x_j+s.
\label{YYscaledSites}
\end{align}
For $0\leq x_j<L$, the inhomogeneities within each chirality span an interval of width at most $\varepsilon$. Thus they approach $-s$ for the right movers and $+s$ for the left movers as the Gamma-function arguments become large. At fixed $s>\beta/2$, the corresponding physical time satisfies $t/L=(s-\beta/2)/\varepsilon\to\infty$. Specifying what is held fixed in this long-time limit determines both the asymptotic weight and the inhomogeneities that enter the limiting Bethe equations.

We will first obtain the saddle equations at fixed admissible $s$. When we subsequently express the result in terms of the microscopic interaction, the universal relation $g(t)\simeq1/(2C(t))$ further requires weak coupling. Since the parameter of the bare two-particle $S$-matrix is $C(t)=2s$, this amounts to choosing $s\gg1$.

Consider a portion of the integration contour on which the scaled Gamma-function arguments stay away from zero and the negative real axis, after reflection wherever needed. We choose consistent logarithmic branches on this part of the contour. For example, on a closed subsector of $|\arg z|<\pi$ separated from the origin, the expansion is uniform and takes the form
\begin{align}
\log\Gamma(\eta z)={}\eta z\log\eta+\eta\ell(z)&-\frac12\log(\eta z)+\frac12\log(2\pi)+\mathcal O(\eta^{-1}),\nonumber\\
\ell(z)&=z(\log z-1).
\label{YYStirling}
\end{align}
The two arguments in each Gamma ratio differ by a quantity of order $\eta$. The terms involving $\ell$ therefore contribute to the leading rapidity dependence of the exponent, while the half-logarithmic terms yield algebraic factors. Under the same rescaling, the elementary factor accompanying a pair of rapidities becomes $v/(v+i\eta)=(U_a-U_b)/(U_a-U_b+i)$, which is of order one away from its poles and zeros.

When a Gamma ratio has first been reflected, its sine factors also enter the leading exponent. To see this, let $Z$ have a fixed nonzero imaginary part and set $\sigma_Z=\operatorname{sgn}\operatorname{Im}Z$. Then
\begin{align}
\sin(\pi\eta Z)
&=\frac{\sigma_Z i}{2}e^{-\sigma_Z i\pi\eta Z}
\left(1-e^{2\sigma_Z i\pi\eta Z}\right),\nonumber\\
\log\sin(\pi\eta Z)
&=-\sigma_Z i\pi\eta Z+\log\frac{\sigma_Z i}{2}
+\mathcal O\!\left(e^{-2\pi\eta|\operatorname{Im}Z|}\right).
\label{YYsineAsymptotic}
\end{align}
The logarithm of each sine factor in $\mathcal R_\eta$ contains a term of order $\eta$. If the arguments in a reflected ratio lie in the same half-plane, the rapidity dependence cancels between their leading terms. If the arguments lie in opposite half-planes, their difference contains a term linear in the rapidity. For example, when $0<\operatorname{Im}Z<1$, the site ratio on the continued branch gives
\begin{align}
\log\frac{\sin\pi\eta(Z-i)}{\sin\pi\eta Z}
=2\pi i\eta Z+\pi\eta+\mathcal O(1),
\label{YYsiteReflectionPhase}
\end{align}
where the bounded term includes the constant phase of the sine ratio. Here $Z=\xi_j-U_a$, so differentiating with respect to $U_a$ produces a shift of $-2\pi i\eta$. Such integer multiples of $2\pi i\eta$ enter the continued action and, where appropriate, can be included by relabeling the Fourier sector.

We include the entire leading contribution from each sine ratio, including its rapidity-independent part, in the exponent. On contour patches separated from the lines where a sine argument crosses the real axis, the remaining terms and their rapidity derivatives are subleading. A saddle that approaches one of these lines, a zero, or a pole calls for a local expansion adapted to its neighborhood. On the regular patches considered here, reflection followed by this expansion gives the same continued leading action as applying Stirling's formula directly on an admissible branch.

Collecting the terms of order $\eta\log\eta$ in $\log\mathcal W_\eta$ gives the coefficient $iM(N-M+1)$, independent of the rapidities and the site coordinates. The terms of order $\eta$ carry the remaining dependence on these variables. Including the Fourier phase in sector $\boldsymbol m$, we obtain
\begin{align}
\mathcal S_0(\boldsymbol U;\boldsymbol\xi)
={}&\sum_{a=1}^M\sum_{j=1}^N
\bigl[\ell(\xi_j-U_a)-\ell(\xi_j-U_a-i)\bigr]\nonumber\\
&+\sum_{a<b}
\bigl[\ell(U_a-U_b-i)-\ell(U_a-U_b+i)\bigr],
\label{YYaction}\\
\mathcal Y_{0,\boldsymbol m}(\boldsymbol U;\boldsymbol\xi)
={}&-\mathcal S_0(\boldsymbol U;\boldsymbol\xi)
+2\pi i\sum_{a=1}^M m_aU_a.
\label{YYleadingaction}
\end{align}
Here $\mathcal Y_{0,\boldsymbol m}$ is the rapidity-dependent coefficient of $\eta$ in the exact Yang-Yang functional. The continuation phases derived above fix its logarithmic branches.

To keep track of the relative weights of the saddles, we also retain the algebraic factors from the expansion. With $Z_{ja}=\xi_j-U_a$ and $V_{ab}=U_a-U_b$, the half-logarithmic terms on a direct Stirling branch combine with the elementary pair factors to give
\begin{align}
\mathcal P_0(\boldsymbol U;\boldsymbol\xi)
={}&\prod_{a,j}\left(\frac{Z_{ja}-i}{Z_{ja}}\right)^{1/2}
\prod_{a<b}\left[
\frac{V_{ab}}{V_{ab}+i}
\left(\frac{V_{ab}+i}{V_{ab}-i}\right)^{1/2}\right].
\label{YYleadingPrefactor}
\end{align}
The continuation of the Gamma functions fixes the branches of the square roots. With this choice, the full expansion through order one is
\begin{align}
\mathcal Y_{\eta,\boldsymbol m}(\eta\boldsymbol U;\eta\boldsymbol\xi)
={}&\eta\mathcal Y_{0,\boldsymbol m}(\boldsymbol U;\boldsymbol\xi)
-i\eta M(N-M+1)\log\eta\nonumber\\
&-\log\mathcal P_0(\boldsymbol U;\boldsymbol\xi)
+\mathcal O(\eta^{-1}).
\label{YYfunctionalExpansion}
\end{align}
The rapidity-independent term contributes a common phase to the wavefunction. Separating this phase and including the Jacobian $d^Mu=\eta^M d^MU$ in the measure gives
\begin{align}
\mathcal N_\eta
&=\exp\!\left[i\eta M(N-M+1)\log\eta\right],\nonumber\\
\mathcal P_\eta(\boldsymbol U;\boldsymbol\xi)
&=\eta^M\mathcal P_0(\boldsymbol U;\boldsymbol\xi)
\left[1+\mathcal O(\eta^{-1})\right].
\label{YYnormalizationAndMeasure}
\end{align}
At fixed $\eta,N,M$, the phase $\mathcal N_\eta$ is independent of physical time and may be absorbed into the normalization of the chosen solution. We keep it explicit to display the full scalar weight. When using reflected Gamma ratios, the remaining reflection factors enter $\mathcal P_\eta$ after their leading contributions have been included in the continued action.

Inserting the expansion into the Poisson representation gives, for the regular contributions in each Fourier sector,
\begin{align}
A_0(\boldsymbol\zeta)\sim{}&\mathcal N_\eta
\sum_{\boldsymbol m\in\mathbb Z^M}
e^{2\pi i\boldsymbol m\cdot\widetilde{\boldsymbol u}}
\int d^MU\,
\mathcal P_\eta(\boldsymbol U;\boldsymbol\xi)\nonumber\\
&\times\mathcal B_\eta(\eta\boldsymbol U;\eta\boldsymbol\xi)
e^{-\eta\mathcal Y_{0,\boldsymbol m}(\boldsymbol U;\boldsymbol\xi)}.
\label{finalformstate}
\end{align}
The exact integral specifies the contours in each Fourier sector. In Eq.~(\ref{finalformstate}) we keep the sum visible to display the relative Fourier phases. The expansion under the integral applies on regular portions of the deformed contours at fixed $N$, $M$, and $\boldsymbol m$. A summed asymptotic formula for a selected state would also require control of the contour tails, any pole contributions, and convergence uniformly in the Fourier sectors.

To determine the saddle locations from $\mathcal Y_{0,\boldsymbol m}$, we must also estimate the size and rapidity dependence of the factors multiplying the exponential. For a scalar prefactor it suffices that, in the saddle region,
\begin{align}
\log\mathcal P_\eta=o(\eta),\qquad
\partial_{U_a}\log\mathcal P_\eta=o(\eta).
\label{YYsubdominantCriterion}
\end{align}
On a nonsingular patch away from zeros of the prefactor at fixed $N,M$, these bounds follow from Eqs.~(\ref{YYleadingPrefactor}) and (\ref{YYnormalizationAndMeasure}). To estimate the Bethe vector in the same region, consider a factor of its monodromy,
\begin{align}
\widehat R_{0j}\bigl(\eta(\xi_j-U)\bigr)
=\frac{i(\xi_j-U)I_{0j}+P_{0j}}{i(\xi_j-U)+1}.
\label{YYscaledRmatrix}
\end{align}
The factors of $\eta$ cancel exactly. The finite products defining the scaled Bethe vector at fixed $N$ and $M$ are therefore rational functions of $\boldsymbol U$ and $\boldsymbol\xi$. Wherever a component is regular and nonzero, its logarithm and logarithmic derivatives are subleading on the scale of the exponent.

For an isolated nondegenerate saddle, the logarithmic derivatives of these prefactors shift the roots only at subleading order. Their values at the saddle nevertheless fix the amplitude of the corresponding spin vector. Expanding the action to second order and performing the Gaussian integral introduces its Hessian, so that a single regular saddle $\boldsymbol U^\star$ in a fixed Fourier sector contributes in proportion to
\begin{align}
\left(\frac{2\pi}{\eta}\right)^{M/2}
\frac{\mathcal P_\eta(\boldsymbol U^\star;\boldsymbol\xi)
\mathcal B_\eta(\eta\boldsymbol U^\star;\eta\boldsymbol\xi)}
{\sqrt{\det\!\left[\partial_{U_a}\partial_{U_b}
\mathcal Y_{0,\boldsymbol m}(\boldsymbol U^\star;\boldsymbol\xi)\right]}}
e^{-\eta\mathcal Y_{0,\boldsymbol m}(\boldsymbol U^\star;\boldsymbol\xi)}.
\label{YYsingleSaddle}
\end{align}
The original integration contour determines which saddles contribute, their orientations, and the branch of the square root of the Hessian determinant. For the full amplitude one restores $\mathcal N_\eta$ and the corresponding Fourier phase from Eq.~(\ref{finalformstate}). The scalar prefactor $\mathcal P_\eta$ and the inverse square root of the Hessian determinant in Eq.~(\ref{YYsingleSaddle}), together with this normalization and phase, fix the leading weight of the Bethe vector at the saddle. Taking the thermodynamic limit after this expansion requires control of the same estimates as $N$ and $M$ increase.

We can now determine the leading roots from
\begin{align}
\partial_{U_a}\mathcal Y_{0,\boldsymbol m}=0,
\qquad\text{equivalently}\qquad
\partial_{U_a}\mathcal S_0=2\pi i m_a.
\label{saddleYY}
\end{align}
Since $\ell'(z)=\log z$, differentiating a site term gives $\log[(\xi_j-U_a-i)/(\xi_j-U_a)]$. In the pair sum, $U_a$ occurs as the first argument for $a<b$ and as the second argument for $b<a$. Accounting for the sign from differentiating the second argument, each $b\ne a$ contributes $\log[(U_a-U_b-i)/(U_a-U_b+i)]$ up to the integer branch shifts represented by the Fourier sectors. Exponentiating Eq.~(\ref{saddleYY}) then gives
\begin{align}
\prod_{j=1}^N\frac{\xi_j-U_a-i}{\xi_j-U_a}
\prod_{b\ne a}\frac{U_a-U_b-i}{U_a-U_b+i}=1.
\label{YYuncenteredBethe}
\end{align}
To compare this result with the usual rational Bethe equations, introduce centered rapidities
\begin{align}
\lambda_a=U_a+\frac i2.
\label{YYcenteredRoots}
\end{align}
In terms of $\lambda_a$, the site factor in Eq.~(\ref{YYuncenteredBethe}) is the inverse of the conventional centered ratio. Moving it to the other side of the equation and grouping the particles by chirality gives
\begin{align}
\left[\prod_{j\in R}
\frac{\lambda_a-\xi_{R,j}-i/2}{\lambda_a-\xi_{R,j}+i/2}\right]
\left[\prod_{j\in L}
\frac{\lambda_a-\xi_{L,j}-i/2}{\lambda_a-\xi_{L,j}+i/2}\right]
=\prod_{b\ne a}
\frac{\lambda_a-\lambda_b-i}{\lambda_a-\lambda_b+i}.
\label{BEadiabatic}
\end{align}
The finite-$\eta$ digamma equations~(\ref{YYexactstationary}) have thus reduced, in the large-argument expansion of the exact functional, to the rational Bethe equations with inhomogeneities $\xi_j$. The accompanying spin vector is the Bethe vector evaluated at these roots, as in the saddle-point relation between qKZ solutions and Bethe vectors developed in Ref.~\cite{Tarasov:1994bb}.

To identify the spectral problem described by Eq.~(\ref{BEadiabatic}), we take the limit $\varepsilon=\alpha L\to0$ at fixed $s=\alpha t+\beta/2$, as in Eq.~(\ref{YYslowScaling}). Since $0\leq x_j<L$, the position-dependent term $\alpha x_j$ in Eq.~(\ref{YYscaledSites}) is of order $\varepsilon$. The inhomogeneities therefore approach a common value within each chirality, which can also be seen directly from the null coordinates. In the scaling limit of Eq.~(\ref{YYslowScaling}), the observation time satisfies $t\gg L$. Since $0\leq x_j<L$, the position of each particle is then small compared with $t$. Thus $x_j-t$ approaches $-t$ for every right mover and $x_j+t$ approaches $+t$ for every left mover, up to relative corrections of order $L/t$. In the scaled variables of Eq.~(\ref{YYscaledSites}), the position-dependent correction is $\alpha x_j$, which is bounded by $\varepsilon=\alpha L$ and vanishes in this limit. The leading saddle-point equations therefore contain a single inhomogeneity for each chirality. Using $C(t)=2\alpha t+\beta=2s$ and $g_{\rm B}(t)=2/C(t)$, we obtain
\begin{align}
\xi_{R,j}=-s+\mathcal O(\varepsilon),\qquad
\xi_{L,j}=+s+\mathcal O(\varepsilon),\qquad
s=\frac1{g_{\rm B}(t)}.
\label{YYinstantaneousSites}
\end{align}
This identification uses the $S$-matrix parameter directly. In the prescriptions compared here, $g_{\rm B}=4g+\mathcal O(g^3)$. Thus Eq.~(\ref{BEadiabatic}) can be written as
\begin{align}
\left(\frac{\lambda_a+s-i/2}{\lambda_a+s+i/2}\right)^{N_R}
\left(\frac{\lambda_a-s-i/2}{\lambda_a-s+i/2}\right)^{N_L}
=\prod_{b\ne a}
\frac{\lambda_a-\lambda_b-i}{\lambda_a-\lambda_b+i}.
\label{YYlimitingBethe}
\end{align}
As we show in the next subsection, Eq.~(\ref{YYlimitingBethe}) coincides with the Bethe equations corresponding to the instantaneous spectrum of the Hamiltonian at time $t$. Thus, in the slow-drive limit, the roots of a regular contributing saddle determine the isospin Bethe vector of the corresponding instantaneous eigenstate. The GBA also determines the factor multiplying this vector. It contains the Yang-Yang functional evaluated at the saddle, the prefactors from the saddle-point expansion, and the scalar factors of the exact wavefunction. The shifts $\alpha x_j$ of the inhomogeneities enter this factor through the exponent and retain the dependence on the particle positions.

For a state prepared at $t_0$ in an isolated instantaneous eigenstate $|n;t_0\rangle$, consider the slow-drive regime in which transitions to other eigenstates are suppressed. Once the coordinate dependence is included in the normalized instantaneous Bethe wavefunction, the remaining time dependence of the GBA solution gives the phase accumulated during the evolution. To leading adiabatic order, the solution therefore takes the form
\begin{align}
|\Psi_{\rm GBA}(t)\rangle
\simeq
\exp\!\left[-i\int_{t_0}^{t} E_n(t')\,dt'\right]
\exp\!\left[-\int_{t_0}^{t}
  \langle n;t'|\partial_{t'}n;t'\rangle\,dt'\right]
|n;t\rangle .
\end{align}
For a smooth choice of phase of $|n;t\rangle$, the two exponentials
give the dynamical and geometric phases, respectively. Both are
encoded in the time-dependent factors of the exact GBA solution
and emerge together in its adiabatic limit.

\subsection{Instantaneous eigenstates}
\label{sec:Instantaneous}
We now derive the Bethe equations independently by diagonalizing the Hamiltonian with its coupling held at $g(t)$. This allows us to identify the instantaneous eigenstate associated with a regular contributing
saddle. Fix a time $t$ and consider the Hamiltonian $H_{\rm GN}(t)$ with its coupling held at $g(t)$. Diagonalizing this Hamiltonian gives the instantaneous energies and eigenstates at any drive rate. The separate conservation of the numbers of left and right movers, $N_L$ and $N_R$, allows us to work in a sector with fixed $N_L,N_R$ and $N=N_L+N_R$. Within this sector the eigenvalue equation is
\begin{align}
H_{\rm GN}(t)|N_L,N_R;\nu;t\rangle_I
=E_{I,\nu}(t)|N_L,N_R;\nu;t\rangle_I,
\label{SE2}
\end{align}
where $\nu$ denotes the charge and spin quantum numbers to be determined from the Bethe ansatz. We suppress this label while constructing a single eigenstate. As in the time-dependent wavefunction, particles $1,\ldots,N_L$ move to the left with velocity $v_L=-1$, and particles $N_L+1,\ldots,N$ move to the right with velocity $v_R=1$.

For each particle $j$, let $k_j(t)$ be its signed momentum measured along its direction of propagation, so that its momentum along the $x$ axis is $p_j(t)=v_{\chi_j}k_j(t)$. The coordinate Bethe ansatz then gives the instantaneous eigenvector, up to normalization, as
\begin{align}
|N_L,N_R;t\rangle_I
={}\sum_{\boldsymbol\sigma}\int_{[0,L]^N}d^Nx\,\left\{\mathcal A\left[e^{i\sum_{j=1}^Nv_{\chi_j}k_j(t)x_j}F_I^{\{\chi_i\}}(\boldsymbol x;t)\right]\right\}_{\boldsymbol\sigma}
\prod_{i=1}^N\psi^\dagger_{\chi_i\sigma_i}(x_i)|0\rangle,
\label{npwfI1}
\end{align}
where $\mathcal A$ antisymmetrizes the fermionic wavefunction and $|0\rangle$ is the empty Fock reference for the finite-particle construction. Within each ordering of the particle coordinates, the kinetic Hamiltonian and the total momentum operator act on the plane wave to give
\begin{align}
E_I(t)=\sum_{j=1}^N k_j(t),\qquad
P_I(t)=\sum_{j=1}^N v_{\chi_j}k_j(t).
\label{instantaneousEnergyMomentum}
\end{align}
The allowed values of $k_j(t)$ are determined by applying periodic boundary conditions to the Bethe wavefunction. Just as in the case of the time-dependent wavefunction, the wavefunction consists of amplitudes corresponding to different orderings of the particles. At fixed $t$, each amplitude consists of a plane wave and a spin amplitude $A_I^{Q,\{\chi_i\}}(t)$ that is constant within the ordering $Q$. The wavefunction takes the form
\begin{align}
F_{I,\boldsymbol\sigma}^{\{\chi_i\}}(\boldsymbol x;t)
&=\sum_{Q\in S_N}\Theta_Q(\boldsymbol x)
\bigl[A_I^{Q,\{\chi_i\}}(t)\bigr]_{\boldsymbol\sigma},\nonumber\\
\Theta_Q(\boldsymbol x)
&=\prod_{r=1}^{N-1}\theta\bigl(x_{Q(r+1)}-x_{Q(r)}\bigr).
\label{npwf2i}
\end{align}
Applying the Schr\"odinger equation corresponding to the instantaneous Hamiltonian across $x_j=x_l$ relates the amplitudes for the two orderings of a right mover $j$ and a left mover $l$ through the bare two-particle $S$-matrix at coupling $g(t)$
\begin{align}
A_I^{\ldots lj\ldots}(t)=S_I^{jl}(t)A_I^{\ldots jl\ldots}(t), \;\;\;\; S_I^{jl}(t)=e^{i\phi_I(t)}\frac{iC(t)I^{jl}+P^{jl}}{iC(t)+1},
\label{smatlri}
\end{align}
where $I^{jl}$ and $P^{jl}$ act as the identity and the permutation on the spins of $j$ and $l$. The inverse $[S_I^{jl}(t)]^{-1}$ relates the amplitudes when the order of the interchange is reversed. In the exact time-dependent wavefunction, the bare two-particle $S$-matrix in Eq.~(\ref{smatlr}) depends on $d=\bar z_l-z_j$ because the right mover and the left mover scatter at time $t_c=d/2$. In the instantaneous spectral problem at time $t$, the coupling is held fixed at $g(t)$. The same coordinate-independent $S$-matrix therefore applies to every right-left pair, with parameters $C(t)=c(2t)$ and $\phi_I(t)=\phi(2t)$ obtained by evaluating Eq.~(\ref{smatlr}) at $t_c=t$. The prescription $\mathfrak r$ for regularizing the interaction at coincident positions fixes both the parameter $C$ and the scalar phase for a given microscopic coupling,
\begin{align}
C(t)=\mathcal C_{\mathfrak r}(g(t)),\qquad
e^{i\phi_I(t)}=e^{i\Phi_{\mathfrak r}(g(t))}.
\label{instantaneousContactPrescription}
\end{align}
At fixed $t$, write $g=g(t)$ and use the same two regularizations of the delta-function interaction as in the time-dependent solution. With the symmetric prescription $\theta(0)=1/2$, the wavefunction at $x_j=x_l$ is the average of its limiting values on the two sides. We obtain
\begin{align}
\mathcal C_{\rm sym}(g)
=\frac{1-3g^2/4}{2g},\qquad
e^{i\Phi_{\rm sym}(g)}=\frac{1-ig/2}{1+ig/2}.
\label{instantaneousSymmetricContact}
\end{align}
In the second prescription, we replace $\delta(x_j-x_l)$ by a narrow smooth function of unit integral and solve the two-particle Schr\"odinger equation across the interaction region before taking its width to zero. With the same normalization of $g$, we have
\begin{align}
\mathcal C_{\rm exp}(g)=\cot(2g),\qquad
e^{i\Phi_{\rm exp}(g)}=e^{-ig}.
\label{instantaneousExponentialContact}
\end{align}
Both prescriptions give the $S$-matrix in Eq.~(\ref{smatlri}). At finite coupling each prescription specifies its own pair of functions $\mathcal C_{\mathfrak r}$ and $\Phi_{\mathfrak r}$, so the spin parameter and scalar phase in the wavefunction must be taken from the same prescription. We therefore keep $C$ as the spectral parameter throughout the calculation. Close to the WZW fixed point the common expansion $C=1/(2g)+\mathcal O(g)$ allows us to relate it to the microscopic continuum coupling in the universal weak-coupling regime.

Since two particles of the same chirality move at the same velocity, the Hamiltonian does not fix the $S$-matrix which relates the amplitudes corresponding to their different orderings. We choose this $S$-matrix by requiring consistency with the opposite-chirality $S$-matrix. This gives
\begin{align}
A_I^{\ldots lj\ldots}(t)
=P^{jl}A_I^{\ldots jl\ldots}(t),
\qquad \chi_j=\chi_l.
\label{instantaneousSameChirality}
\end{align}
Consider a right mover $i$ and two left movers $j$ and $k$. The bare $S$-matrices for the two right-left pairs are $S_I^{ij}(t)$ and $S_I^{ik}(t)$, while interchanging the two left movers permutes their spins through $P^{jk}$. The two sequences of pairwise interchanges that reverse the ordering of the three particles must give the same spin amplitude. This requirement gives the Yang-Baxter equation
\begin{align}
S_I^{ij}(t)S_I^{ik}(t)P^{jk}
=P^{jk}S_I^{ik}(t)S_I^{ij}(t).
\label{YBi}
\end{align}
This relation equates the amplitudes obtained by the two possible sequences of three interchanges. The corresponding relations for each assignment of chiralities, together with the inverse matrices for reversed interchanges, allow us to obtain every $A_I^Q$ from the reference amplitude $A_I^{Q_0}$ with $Q_0=(N,\ldots,1)$.

We now determine $A_I^{Q_0}$ by imposing periodic boundary conditions. At fixed $t$, take particle $j$ once around the system in the direction of increasing $x$. Starting in the reference ordering, the particle passes the other fermions and returns to the same ordering under the periodic identification. Denote the ordered product of the corresponding two-particle $S$-matrices by $\mathcal U_{I,j}(t)$. For a right mover, this product is
\begin{align}
\mathcal U_{I,j}(t)
={}&P^{j,j+1}\cdots P^{jN}\,
S_I^{j1}(t)\cdots S_I^{jN_L}(t)\,
P^{j,N_L+1}\cdots P^{j,j-1}.
\label{transportopconst}
\end{align}
Here the rightmost factor acts first and an empty product is the identity. Moving a left mover in the same direction reverses the opposite-chirality interchange in Eq.~(\ref{smatlri}), so each passage through a right mover contributes the inverse $S$-matrix. Since the plane wave acquires the phase $e^{iv_{\chi_j}k_j(t)L}$ over the length of the system, periodicity requires
\begin{align}
&e^{iv_{\chi_j}k_j(t)L}\mathcal U_{I,j}(t)A_I^{Q_0}(t)=A_I^{Q_0}(t),\nonumber\\
&e^{ik_j(t)L}A_I^{Q_0}(t)=Z_j(t)A_I^{Q_0}(t),\qquad
Z_j(t)=\mathcal U_{I,j}(t)^{-v_{\chi_j}}.
\label{eigenconst}
\end{align}
The second line expresses the boundary condition in terms of $k_j$, measured along the direction of propagation for either chirality. The sign of the plane wave and the orientation of the interchange fix $Z_j=\mathcal U_{I,j}^{-1}$ for a right mover and $Z_j=\mathcal U_{I,j}$ for a left mover. In the time-dependent construction the two-particle $S$-matrices depend on the spacetime coordinates of the particles, and transporting a particle around the system gives rise to the qKZ difference equations. Here the spin amplitudes are independent of the positions of the particles, and since we are diagonalizing the Hamiltonian at a particular time $t$, we obtain an eigenvalue problem for $A_I^{Q_0}$. At a given time the Yang-Baxter relations imply
\begin{align}
[\mathcal U_{I,i}(t),\mathcal U_{I,j}(t)]=0,
\qquad [Z_i(t),Z_j(t)]=0.
\label{commutation}
\end{align}
The coupling entering these relations has the same value in every factor. When the coupling changes between $t$ and $t'$, the commutator $[Z_i(t),Z_j(t')]$ can be nonzero. For the instantaneous problem, a real coupling and a regularization that preserves unitarity give a family of commuting unitary operators at each $t$. We can therefore obtain all the momenta from their common eigenvectors. To diagonalize the spin part of the transport operators, we introduce the Bethe-ansatz coupling
\begin{align}
g_{\rm B}(t)\equiv\frac{2}{C(t)},\qquad
g_{\rm B}(t)=4g(t)+\mathcal O(g^3)
\quad\hbox{in the universal weak-coupling regime}.
\label{BetheCoupling}
\end{align}
Its definition in terms of $C(t)$ is exact. The expansion in $g(t)$ gives its universal leading relation to the microscopic coupling, with higher-order terms set by the regularization of the interaction. In particular, the allowed time dependence $C(t)=2\alpha t+\beta$ gives
\begin{align}
g_{\rm B}(t)=\frac{1}{\alpha t+\beta/2}.
\label{BetheProtocol}
\end{align}
Equation~(\ref{BetheCoupling}) relates these results to the microscopic coupling $g$ near the WZW fixed point, in the normalization of the fermionic Hamiltonian in Eq.~(\ref{bridgeCurrentHamiltonian}).

In the instantaneous problem, all particles of the same chirality have the same inhomogeneity. We denote these two values by $\xi_L(t)$ and $\xi_R(t)$ and assign them to the sites according to
\begin{align}
\xi_j(t)=
\begin{cases}
\xi_L(t)=+1/g_{\rm B}(t),& j\leq N_L,\\
\xi_R(t)=-1/g_{\rm B}(t),& j>N_L.
\end{cases}
\label{wwbarlimit}
\end{align}
Particles of the same chirality have equal inhomogeneities, while the difference $\xi_L-\xi_R=C(t)=2/g_{\rm B}(t)$ is precisely the parameter in the opposite-chirality $S$-matrix. We can therefore write both types of two-particle matrix in terms of the rational $R$-matrix
\begin{align}
R_{ab}(v)=\frac{ivI_{ab}+P_{ab}}{iv+1}.
\label{instantaneousRmatrix}
\end{align}
Since $R(0)=P$ and $R(v)^{-1}=R(-v)$, the factor for moving $j$ past $l$ in the direction of increasing $x$ is $R_{jl}(\xi_l-\xi_j)$. Its argument is $+C$ for a right mover passing a left mover and $-C$ for the reverse passage. The spin part of the product for taking particle $j$ around the system becomes
\begin{align}
\mathcal U^{\rm spin}_{I,j}
={}&R_{j,j+1}(\xi_{j+1}-\xi_j)\cdots
R_{jN}(\xi_N-\xi_j)\nonumber\\
&\times R_{j1}(\xi_1-\xi_j)\cdots
 R_{j,j-1}(\xi_{j-1}-\xi_j).
\label{instantaneousSpinTransport}
\end{align}
Each passage through a fermion of the opposite chirality also contributes a scalar phase. Writing $N_{\bar\chi_j}$ for the number of such fermions and taking account of the orientation of each interchange, we obtain the complete transport operator
\begin{align}
\mathcal U_{I,j}(t)
=e^{iv_{\chi_j}N_{\bar\chi_j}\phi_I(t)}
\mathcal U^{\rm spin}_{I,j}(t).
\label{instantaneousTransportPhase}
\end{align}
To obtain its eigenvectors, we introduce an auxiliary spin space labeled by $0$ and form the monodromy matrix in the order fixed by Eq.~(\ref{instantaneousSpinTransport}). Taking the trace over the auxiliary spin space gives the transfer matrix,
\begin{align}
\mathcal T_{I,0}(U;t)
&=R_{01}(\xi_1-U)\cdots
R_{0N}(\xi_N-U)\nonumber\\
&=\begin{pmatrix}
A_{I,0}(U;t)&B_{I,0}(U;t)\\
C_{I,0}(U;t)& D_{I,0}(U;t)
\end{pmatrix},\nonumber\\
 t_I(U;t)&=\operatorname{tr}_0\mathcal T_{I,0}(U;t).
\label{instantaneousMonodromy}
\end{align}
At fixed $t$, these transfer matrices commute for different values of $U$. At $U=\xi_j$, the factor acting on particle $j$ reduces to $R(0)=P$. Taking the auxiliary trace then gives precisely the ordered product for that particle,
\begin{align}
 t_I(\xi_j;t)=\mathcal U^{\rm spin}_{I,j}(t).
\label{instantaneousTransferTransport}
\end{align}
Thus a common eigenvector of the transfer matrices supplies the reference spin amplitude for every periodic boundary condition in Eq.~(\ref{eigenconst}).

We construct this amplitude from the fully polarized vector $|\Omega\rangle=|\uparrow\cdots\uparrow\rangle$. Each application of $B_{I,0}$ lowers its total $S^z$ by one unit. With $B'_0(\lambda;t)=B_{I,0}(\lambda-i/2;t)$ in the centered rapidity convention, a state with $M$ reversed spins has the form
\begin{align}
A_I^{Q_0}(t)=\prod_{a=1}^M B'_0(\lambda_a;t)|\Omega\rangle,
\qquad S^z=\frac N2-M.
\label{InstantaneousES}
\end{align}
We usually leave the time dependence of the rapidities $\lambda_a$ implicit. Acting with the transfer matrix gives a term proportional to this vector together with terms in which its spectral argument replaces one of the roots. For Eq.~(\ref{InstantaneousES}) to be an eigenvector, the coefficients of these additional terms must vanish. This gives the instantaneous spin Bethe equations
\begin{align}
\left(
\frac{\lambda_a+1/g_{\rm B}(t)-i/2}
{\lambda_a+1/g_{\rm B}(t)+i/2}
\right)^{N_R}
\left(
\frac{\lambda_a-1/g_{\rm B}(t)-i/2}
{\lambda_a-1/g_{\rm B}(t)+i/2}
\right)^{N_L}
=\prod_{\substack{b=1\\b\ne a}}^M
\frac{\lambda_a-\lambda_b-i}{\lambda_a-\lambda_b+i}.
\label{BE2}
\end{align}
Let us compare the Bethe equations~(\ref{BE2}), obtained here by diagonalizing $H(t)$ at fixed $t$, with those obtained from the Yang-Yang action in the limit of slow variation [Eq.~(\ref{YYlimitingBethe})]. The saddle-point equations of the Yang-Yang action take the Bethe form in Eq.~(\ref{BEadiabatic}), with one inhomogeneity for each particle. In the slow-variation limit of Eq.~(\ref{YYslowScaling}), the scaled inhomogeneities $w_j/\eta$ of the right movers approach $\xi_R(t)$, while $\bar w_l/\eta$ for the left movers approach $\xi_L(t)$, as given in Eq.~(\ref{wwbarlimit}). In this limit the saddle-point equations reduce to Eq.~(\ref{YYlimitingBethe}), which coincides with the instantaneous Bethe equations~(\ref{BE2}) derived independently by diagonalizing the Hamiltonian at fixed $t$. Thus, we have shown that in the limit of slow variation, the Bethe vector at each regular contributing saddle approaches the instantaneous Bethe vector with the same rapidities. Together with the suppression of transitions between instantaneous eigenstates under the conditions derived above, this identifies the adiabatic regime of the time-dependent model. For a system prepared in an isolated instantaneous eigenstate at $t_0$, the exact GBA wavefunction in this regime is proportional, to leading order, to the corresponding instantaneous eigenstate at time $t$. For normalized states, the factor of proportionality is the dynamical phase $\exp[-i\int_{t_0}^{t} E_I(t')\,dt']$ multiplied by the geometric phase acquired along the evolution.

 The Bethe equations admit real roots and complex configurations, including strings and quartets~\cite{DestriLowenstein}. All roots should be distinct for the wavefunction to be nonvanishing, and the Bethe eigenstates correspond to highest-weight states. The descendants of each $SU(2)$ multiplet follow by applying the global spin-lowering operator.

For an admissible solution the eigenvalue of the spin transport operator is
\begin{align}
\tau_j^I(t)=\prod_{a=1}^M
\frac{\lambda_a-\xi_j(t)+i/2}
{\lambda_a-\xi_j(t)-i/2}.
\label{instantaneousSpinEigenvalue}
\end{align}
Substituting this eigenvalue into Eqs.~(\ref{eigenconst}) and (\ref{instantaneousTransportPhase}) restores the scalar phase and gives the eigenvalue of $Z_j(t)$,
\begin{align}
z_j^I(t)
=e^{-iN_{\bar\chi_j}\phi_I(t)}
\bigl[\tau_j^I(t)\bigr]^{-v_{\chi_j}}.
\label{instantaneousCompleteEigenvalue}
\end{align}
The scalar phase cancels from the spin Bethe equations but enters the momentum of each fermion through $z_j^I(t)$. For a unitary $S$-matrix, $|z_j^I|=1$, and the periodic boundary condition quantizes the momenta as
\begin{align}
e^{ik_j(t)L}=z_j^I(t),\qquad
k_j(t)=\frac{2\pi n_j+\arg z_j^I(t)}{L},
\qquad n_j\in\mathbb Z.
\label{eigenen}
\end{align}
On any interval free of singularities we can follow $\arg z_j^I(t)$ continuously, with a change of branch absorbed into the integer $n_j$. Summing the momenta along the propagation directions as in Eq.~(\ref{instantaneousEnergyMomentum}) then gives the total instantaneous energy,
\begin{align}
E_I(t)=\frac{2\pi}{L}\sum_{j=1}^N n_j
+\frac1L\sum_{j=1}^N\arg z_j^I(t).
\label{eninst}
\end{align}
The integers $n_j$ label the charge branches, and the phases contain the full dependence on the two-particle scattering and the spin rapidities. To obtain a physical excitation energy we subtract the energy of the filled instantaneous ground state at the same coupling and ultraviolet cutoff. For the filled ground state considered below the spin roots are real, and their distribution is most readily obtained by taking a logarithm of Eq.~(\ref{BE2}). We use
\begin{align}
\Theta(x,b)=\arctan(x/b),\qquad b>0,
\end{align}
with the real arctangent in $(-\pi/2,\pi/2)$. For complex roots, the logarithms must instead be continued on specified branches, or one may use the equations for the centers of the complex configurations. On the real branch the identity $(x-ib)/(x+ib)=-e^{2i\Theta(x,b)}$ gives
\begin{align}
N_R\Theta\!\left(\lambda_a+\frac1{g_{\rm B}(t)},\frac12\right)
+N_L\Theta\!\left(\lambda_a-\frac1{g_{\rm B}(t)},\frac12\right)
=\pi m_a^I+\sum_{b=1}^M\Theta(\lambda_a-\lambda_b,1).
\label{Logbae2}
\end{align}
The $b=a$ term vanishes. The real-arctangent convention fixes the spin branch numbers to $m_a^I\in\mathbb Z+(N-M+1)/2$, modulo integer shifts. Their integer or half-integer values therefore depend on the parity of $N-M+1$. The Fourier-sector integers $\boldsymbol m$ in the exact Yang-Yang representation label the logarithmic branches before the saddle-point equations are exponentiated. Returning to real logarithms in Eq.~(\ref{Logbae2}) introduces the fixed phase shifts included in $m_a^I$.

The Bethe equations restrict the spin branch numbers, while the linear dispersion leaves the integers $n_j$ unbounded below. To define the Fermi sea we impose a cutoff on the unshifted charge momenta and fill the allowed states subject to fermionic exclusion for the full set of charge, spin, and chirality labels
\begin{align}
\frac{\pi|n_j|}{L}<\pi\Lambda,\qquad
\Lambda=\frac NL.
\label{instantaneousCutoff}
\end{align}
Since $N$ counts the bare fermions filling the Fermi sea, their density defines the ultraviolet scale $\Lambda$, independently of the density of physical excitations above the sea~\cite{AndreiLowenstein79}. In the convention of Eq.~(\ref{eigenen}), the cutoff bounds the magnitude of the unshifted charge momentum $2\pi n_j/L$ by $2\pi\Lambda$, with the scattering phases retained explicitly. 

In the thermodynamic limit the Bethe roots form a dense set, and Eq.~(\ref{Logbae2}) determines their density, which we use in the next section to calculate the energy of a spinon excitation and the instantaneous mass gap. For the decreasing  time-dependent coupling, keeping this gap finite as the cutoff is removed will fix a characteristic time $t_0$. This identifies a time about which to compare the instantaneous spectrum with the evolution of a prepared state over the interval satisfying Eq.~(\ref{YYslowScaling}).

\section[The quantum mass gap and time-dependent dimensional transmutation]{The quantum mass gap and time-dependent dimensional transmutation}
\label{sec:massgap}

In Sec.~\ref{sec:Analysis}, we showed how the instantaneous Bethe equations emerge from the exact wavefunction in the slow-drive limit and identified the conditions under which a prepared state follows an
instantaneous eigenstate. We now use these equations to determine how the dimensionless current-current interaction generates a time-dependent mass gap in the isospin sector. Holding the coupling fixed at its value
at time $t$, we first determine the distribution of Bethe rapidities that minimizes the energy. This gives the ground state relative to which we calculate the energy of an elementary isospin excitation.
We obtain its dispersion at finite particle density and then take the continuum scaling limit. The minimum of this dispersion gives the mass gap as a function of the Bethe coupling $g_{\rm B}(t)$ and hence
determines how the gap changes during the drive.

\subsection{Ground-state roots and spin excitations}

To determine the instantaneous spectrum, we fix the time $t$ and consider the Hamiltonian at the corresponding value of its coupling. On the massive branch, $g_{\rm B}(t)>0$, the parameter entering the Bethe equations is
\begin{align}
s(t)=\frac{1}{g_{\rm B}(t)}=\frac{C(t)}{2}
    =\alpha t+\frac{\beta}{2}.
\label{massScaledCoupling}
\end{align}
An eigenstate is specified by the charge integers $n_j$ and the isospin quantum numbers $m_a^I$. We fill the lowest charge levels allowed by the cutoff in Eq.~(\ref{instantaneousCutoff}) and keep the occupied $n_j$ fixed. We can then construct an excitation in the isospin sector by changing the $m_a^I$ and calculate its energy and momentum relative to the ground state at the same $t$. Writing $s=s(t)$, the logarithmic Bethe equations for $M$ real isospin rapidities $\lambda_a$ take the form
\begin{align}
N_R\Theta\!\left(\lambda_a+s,\tfrac12\right)
+N_L\Theta\!\left(\lambda_a-s,\tfrac12\right)
=\pi m_a^I+\sum_{b=1}^{M}\Theta(\lambda_a-\lambda_b,1),
\label{BetheLogDynamics}
\end{align}
where $\Theta(x,n)=\arctan(x/n)$. The shifts $+s$ and $-s$ in the two phases follow from the inhomogeneities $\xi_R=-s$ and $\xi_L=s$ of the right and left movers. We take an even number $N$ of fermions and work in the sector
\begin{align}
N_R=N_L=\frac{N}{2}.
\end{align}
The ground state is a singlet with $M=N/2$ real roots \cite{AndreiLowenstein79,DestriLowenstein}. To identify the quantum numbers occupied by these roots, define the counting function
\begin{align}
\nu(\lambda;t)=\frac{1}{\pi}\left[
N_R\Theta\!\left(\lambda+s,\tfrac12\right)
+N_L\Theta\!\left(\lambda-s,\tfrac12\right)
-\sum_{b=1}^{M}\Theta(\lambda-\lambda_b,1)\right].
\end{align}
By construction $\nu(\lambda_a;t)=m_a^I$ at every occupied root, while $\nu(\pm\infty;t)=\pm(N-M)/2$. On the logarithmic branch of Eq.~(\ref{Logbae2}), $m_a^I\in\mathbb Z+(N-M+1)/2$.\footnote{In the saddle-point construction the Fourier sectors label the branches before exponentiation. Taking the real logarithm of the Bethe equations fixes the offset of $m_a^I$ used here.} The allowed values for finite real roots thus run in steps of one from $-(N-M-1)/2$ to $(N-M-1)/2$. In the singlet there are $N/2$ such values and $N/2$ roots to occupy them. This consecutive filling of the isospin quantum numbers defines the ground-state Bethe sea.

In the thermodynamic limit $N,L\to\infty$ at fixed density $\Lambda=N/L$, the change in $\nu$ across a rapidity interval counts all the isospin quantum numbers in that interval. Denoting the densities of occupied and vacant quantum numbers by $\rho$ and
$\rho^{\rm h}$, respectively, we therefore have
\begin{align}
\partial_\lambda\nu(\lambda;t)
=\rho(\lambda;t)+\rho^{\rm h}(\lambda;t).
\label{massCountingDensity}
\end{align}
Here $\int d\lambda\,\rho=M$, so $\rho/L$ is the density of occupied roots per unit length of the system. In the filled sea every quantum number is occupied, and the derivative of $\nu$ gives the ground-state density directly. Introducing the derivative of the scattering phase,
\begin{align}
a_n(x)=\frac{1}{\pi}\partial_x\Theta(x,n)
=\frac{n}{\pi(n^2+x^2)},\qquad n>0,\qquad
\int_{-\infty}^{\infty} dx\,a_n(x)=1,
\label{massBetheKernel}
\end{align}
and replacing the sum over the occupied roots in $\nu$ by an integral, we obtain
\begin{align}
\rho_{\rm gs}(\lambda;t)
+\int_{-\infty}^{\infty}d\mu\,a_1(\lambda-\mu)\rho_{\rm gs}(\mu;t)
=\frac{N}{2}\sum_{\sigma=\pm1}a_{1/2}(\lambda+\sigma s).
\label{GroundRootEq}
\end{align}
The integral is a convolution, so we can solve for the density by Fourier transformation. With the convention
\begin{align}
\widetilde f(\omega)=\int_{-\infty}^{\infty}d\lambda\,
e^{i\omega\lambda}f(\lambda),\qquad
f(\lambda)=\int_{-\infty}^{\infty}\frac{d\omega}{2\pi}\,
e^{-i\omega\lambda}\widetilde f(\omega),
\label{massFourierConvention}
\end{align}
we have $\widetilde a_n(\omega)=e^{-n|\omega|}$. The two shifted terms on the right of Eq.~(\ref{GroundRootEq}) give a factor
$\cos(\omega s)$, and we find
\begin{align}
\widetilde\rho_{\rm gs}(\omega;t)
 &=\frac{Ne^{-|\omega|/2}\cos(\omega s)}{1+e^{-|\omega|}}
   =\frac{N\cos(\omega s)}{2\cosh(\omega/2)},
\label{massGroundFourier}\\
\rho_{\rm gs}(\lambda;t)
 &=\frac{N}{4}\sum_{\sigma=\pm1}
        \frac{1}{\cosh[\pi(\lambda+\sigma s)]}.
\label{GroundRootDensity}
\end{align}
The normalization $\widetilde\rho_{\rm gs}(0;t)=N/2$ recovers the number of roots in the singlet.

Starting from this state, let us construct the highest-weight state of a triplet. Since a regular highest-weight Bethe state has total isospin $S=N/2-M$, this requires reducing $M$ from $N/2$ to $N/2-1$. Removing one root also increases the number $N-M$ of available quantum numbers by one. There are now $N/2+1$ allowed values and only $N/2-1$ occupied roots, leaving two holes in the sequence \cite{trieste,AndreiLowenstein79}. Choosing the two omitted quantum numbers $m_r^{\rm h}$ specifies the excitation. The rapidity of each hole is the value $\lambda_r^{\rm h}$ at which the counting function of this excited state satisfies $\nu(\lambda_r^{\rm h};t)=m_r^{\rm h}$, giving the hole density
\begin{align}
\rho^{\rm h}(\lambda;t)=\sum_{r=1}^2\delta(\lambda-\lambda_r^{\rm h}).
\end{align}
The holes affect the other roots through the sum of scattering phases in Eq.~(\ref{BetheLogDynamics}). To find how those roots shift, we apply Eq.~(\ref{massCountingDensity}) to the excited state. With $(a_1*f)(\lambda)=\int d\mu\,a_1(\lambda-\mu)f(\mu)$, its density obeys
\begin{align}
\rho_{\rm exc}(\lambda;t)+(a_1*\rho_{\rm exc})(\lambda;t)
+\rho^{\rm h}(\lambda;t)
=\frac{N}{2}\sum_{\sigma=\pm1}a_{1/2}(\lambda+\sigma s).
\label{ExcitedRootEq}
\end{align}
The source on the right is the same as for the singlet since both states have the same numbers of right and left movers. Subtracting Eq.~(\ref{GroundRootEq}) thus gives the change in the occupied-root density, $\delta\rho=\rho_{\rm exc}-\rho_{\rm gs}$,
\begin{align}
\delta\rho+a_1*\delta\rho&=-\rho^{\rm h},\qquad
\widetilde{\delta\rho}(\omega;t)
=-\sum_{r=1}^2\frac{e^{i\omega\lambda_r^{\rm h}}}{1+e^{-|\omega|}} .
\label{massHoleBackflow}
\end{align}
Each hole contributes $-1/2$ to the change in the number of occupied roots, as follows by setting $\omega=0$. It therefore carries isospin projection $+1/2$ in the highest-weight triplet.
The charge filling is unchanged, and the two holes describe the neutral isospin-$1/2$ excitations called spinons.

We now calculate the energy and momentum carried by this pair. The charge integers $n_j$ are the same in the singlet and triplet, but the values of $k_j$ in the two states differ because each occupied root contributes to the phase in $k_j=(2\pi n_j+\arg z_j^I)/L$. The scalar phase in Eq.~(\ref{instantaneousCompleteEigenvalue}) depends only on the
coupling and the number of fermions of the opposite chirality, so it cancels when we subtract the singlet values from the triplet values. In our convention $k_j$ is measured along the direction of propagation of fermion $j$, so its momentum along the $x$ axis is $v_{\chi_j}k_j$. We can therefore obtain the change in both observables by inserting the isospin phases into the total energy $E=\sum_j k_j$ and momentum $P=\sum_j v_{\chi_j}k_j$ of Eq.~(\ref{instantaneousEnergyMomentum}).

Consider one occupied root of rapidity $\lambda$ and write $\Theta_\pm=\Theta(\lambda\pm s,1/2)$. On the branches used for the filled charge levels, this root contributes $2\Theta_+-\pi$ to the phase $\arg z_j^I$ of a right mover and $-2\Theta_- -\pi$ to that of a left mover.\footnote{In Eq.~(\ref{instantaneousSpinEigenvalue}) the argument of the root
factor $(\lambda-\xi_j+i/2)/(\lambda-\xi_j-i/2)$ is $v_{\chi_j}\pi-2\Theta(\lambda-\xi_j,1/2)$ on these branches. Eq.~(\ref{instantaneousCompleteEigenvalue}) raises the factor to the power $-v_{\chi_j}$. Changing a logarithmic branch shifts the corresponding charge integer in Eq.~(\ref{eigenen}) when the same physical state is described.}
There are $N/2$ fermions of each chirality. Adding their phase shifts in $E$ and weighting them by $v_{\chi_j}$ in $P$ gives the following contribution of the occupied root to the two observables,
\begin{align}
\epsilon_0(\lambda;t)&=\Lambda\bigl[\Theta_+-\Theta_- -\pi\bigr],
&
p_0(\lambda;t)&=\Lambda\bigl[\Theta_++\Theta_-\bigr].
\label{massBareRootEnergy}
\end{align}
In particular, the constant parts of the phases add in the energy and cancel in the momentum. Summing over occupied roots and subtracting the two states now gives
\begin{align}
\Delta E=\int d\lambda\,\epsilon_0(\lambda;t)\,\delta\rho(\lambda;t),
\qquad
\Delta P=\int d\lambda\,p_0(\lambda;t)\,\delta\rho(\lambda;t).
\end{align}
Both integrals involve the change in the entire distribution because each of the two holes displaces the other occupied roots.

Equation~(\ref{massHoleBackflow}) gives $\delta\rho$ as the sum of the responses to holes at $\lambda_1^{\rm h}$ and $\lambda_2^{\rm h}$.
Consequently, each hole contributes separately to $\Delta E$ and $\Delta P$. To evaluate its contribution, let $\epsilon_{\rm h}$ and
$p_{\rm h}$ solve
\begin{align}
\epsilon_{\rm h}+a_1*\epsilon_{\rm h}&=-\epsilon_0,
&
p_{\rm h}+a_1*p_{\rm h}&=-p_0 .
\label{massDressedEnergyEquation}
\end{align}
For example, substituting the first equation into the energy integral and using $a_1(\lambda-\mu)=a_1(\mu-\lambda)$ gives
$\Delta E=-\int d\lambda\,\epsilon_{\rm h}
(\delta\rho+a_1*\delta\rho)=\int d\lambda\,\epsilon_{\rm h}\rho^{\rm h}$.
Applying the same argument to the momentum yields
\begin{align}
\Delta E&=\epsilon_{\rm h}(\lambda_1^{\rm h};t)
         +\epsilon_{\rm h}(\lambda_2^{\rm h};t),\qquad
\Delta P=p_{\rm h}(\lambda_1^{\rm h};t)+p_{\rm h}(\lambda_2^{\rm h};t).
\end{align}
Thus $\epsilon_{\rm h}$ and $p_{\rm h}$ are the energy and momentum of a single spinon, dressed by the response of the sea. To solve Eq.~(\ref{massDressedEnergyEquation}), we differentiate with respect to $\lambda$. The derivative of $-\epsilon_0$ gives the difference of the two kernels centered at $\pm s$, while the derivative of $-p_0$ gives their negative sum. Fourier transforming the left-hand sides gives the same factor $1+e^{-|\omega|}$ as in the ground-state equation. Each $a_{1/2}(\lambda\pm s)$ in the source thus becomes $1/(2\cosh[\pi(\lambda\pm s)])$ upon solving for the derivative. It follows that
\begin{align}
\partial_\lambda\epsilon_{\rm h}(\lambda;t)
 &=\frac{\pi\Lambda}{2}
   \left[\frac{1}{\cosh[\pi(\lambda-s)]}
        -\frac{1}{\cosh[\pi(\lambda+s)]}\right],\nonumber\\
\partial_\lambda p_{\rm h}(\lambda;t)
 &=-\frac{\pi\Lambda}{2}
   \left[\frac{1}{\cosh[\pi(\lambda-s)]}
        +\frac{1}{\cosh[\pi(\lambda+s)]}\right]
 =-\frac{2\pi}{L}\rho_{\rm gs}(\lambda;t).
\label{massDressedEnergyDerivative}
\end{align}
The energy constant is fixed by the undifferentiated equation. As $|\lambda|\to\infty$, both scattering phases approach the same value and $\epsilon_0\to-\pi\Lambda$. Since the integral of $a_1$ is one, Eq.~(\ref{massDressedEnergyEquation}) then gives $\epsilon_{\rm h}(\pm\infty;t)=\pi\Lambda/2$. For the momentum, the oddness of $p_0$ and the evenness of $a_1$ give $p_{\rm h}(0;t)=0$. Integrating Eq.~(\ref{massDressedEnergyDerivative}) with these values yields the dispersion at finite cutoff,
\begin{align}
\epsilon_{\rm h}(\lambda;t)
 &=\Lambda\arctan\!\left[\frac{\cosh(\pi\lambda)}{\sinh(\pi s)}\right],
\label{massRegulatedDispersion}\\
p_{\rm h}(\lambda;t)
 &=-\Lambda\arctan\!\left[\frac{\sinh(\pi\lambda)}{\cosh(\pi s)}\right].
\label{massRegulatedMomentum}
\end{align}
The sign of the momentum also follows directly from the Bethe quantum numbers. Summing Eq.~(\ref{BetheLogDynamics}) over occupied roots cancels the double sum of the odd phase $\Theta(\lambda_a-\lambda_b,1)$ and gives $\sum_a p_0(\lambda_a;t)=(2\pi/L)\sum_a m_a^I$. The complete set of available quantum numbers sums to zero in either
state. The singlet fills this set, while the triplet leaves $m_1^{\rm h}$ and $m_2^{\rm h}$ vacant. Their momentum difference is therefore $\Delta P=-(2\pi/L)(m_1^{\rm h}+m_2^{\rm h})$,
in agreement with the decrease of $p_{\rm h}$ with increasing hole rapidity in Eq.~(\ref{massRegulatedMomentum}).

The dressed energy attains its minimum at $\lambda=0$, where the spinon momentum vanishes. Its rest energy at finite cutoff is therefore $\Lambda\arctan[1/\sinh(\pi s)]
=2\Lambda\arctan(e^{-\pi s})$. This identifies the physical quantity to hold fixed in the continuum limit. We let the density cutoff $\Lambda$ tend to infinity and the Bethe coupling
$g_{\rm B}=1/s$ tend to zero with $2\Lambda e^{-\pi s}$ fixed \cite{AndreiLowenstein79,DestriLowenstein,WiegmannGN}. The relation to the microscopic current-current interaction in this regime is $g_{\rm B}=4g+\mathcal O(g^3)$, as in Eq.~(\ref{BetheCoupling}). At fixed $\lambda$, the arguments of the two arctangents in Eqs.~(\ref{massRegulatedDispersion}) and~(\ref{massRegulatedMomentum}) are then asymptotic to $2e^{-\pi s}\cosh(\pi\lambda)$ and $2e^{-\pi s}\sinh(\pi\lambda)$, respectively. The spinon dispersion consequently takes the relativistic form
\begin{align}
&E_{\rm spinon}(\lambda;t)=m(t)\cosh(\pi\lambda),\qquad
p_{\rm spinon}(\lambda;t)=-m(t)\sinh(\pi\lambda),\nonumber\\
& \hspace{30mm} m(t)=2\Lambda\exp\!\left[-\frac{\pi}{g_{\rm B}(t)}\right].
\label{TimedependentGap}
\end{align}
With $\vartheta=-\pi\lambda$ as the physical rapidity, $E_{\rm spinon}=m(t)\cosh\vartheta$ and $p_{\rm spinon}=m(t)\sinh\vartheta$, giving $E_{\rm spinon}^2-p_{\rm spinon}^2=m(t)^2$. The mass gap $m(t)$ is the rest energy of one spinon in the isospin sector. In the sector with fixed even fermion number the spinons occur in pairs, and the two-spinon
continuum begins at $2m(t)$ in the thermodynamic limit. The decoupled $U(1)$ charge sector remains gapless.

The fermions in the Hamiltonian are massless. Their dimensionless current-current interaction generates the finite isospin mass gap in Eq.~(\ref{TimedependentGap}) through dimensional transmutation. In the static model, prescribing this gap determines how the coupling must scale as the cutoff is removed. In the time-dependent model, $1/g_{\rm B}(t)=\alpha t+\beta/2$ fixes when the coupling reaches that value. We use this condition in the following subsection to identify the characteristic time for a prescribed physical mass gap.

\subsection{Continuum scaling and the dynamical interpretation}
In the static model, keeping the mass gap finite as the ultraviolet cutoff is removed determines how the Bethe coupling must scale with $\Lambda$. In the time-dependent model, integrability specifies the coupling as a function of time, so the same requirement determines when the system reaches a prescribed instantaneous mass gap. To see this, consider the decreasing time-dependent coupling with fixed drive rate $\alpha=\alpha_0>0$ and offset $\beta$. Substituting the Bethe coupling into Eq.~(\ref{TimedependentGap}) gives
\begin{align}
\frac{1}{g_{\rm B}(t)}=\alpha_0 t+\frac{\beta}{2},
\qquad
m(t)=2\Lambda e^{-\pi(\alpha_0 t+\beta/2)}.
\label{massProtocolClock}
\end{align}
Choose a physical mass gap $m_0>0$ and define the characteristic time $t_0$ by $m(t_0)=m_0$. Writing $t=t_0+\Delta t$, we obtain
\begin{align}
&m_0=2\Lambda e^{-\pi(\alpha_0 t_0+\beta/2)},\nonumber\\ &m(t_0+\Delta t)=m_0e^{-\pi\alpha_0\Delta t}.
\label{GapEvolution}
\end{align}
The first relation fixes $t_0$ for a given cutoff. The second gives the instantaneous gap at elapsed time $\Delta t$ and is independent of the choice of time origin. Indeed, in the shifted time coordinate $t'=t+\beta/(2\alpha_0)$, the time-dependent coupling reads $1/g_{\rm B}=\alpha_0 t'$ and the reference gap is $m_0=2\Lambda e^{-\pi\alpha_0 t'_0}$, with $t'-t'_0=t-t_0$. As $\Lambda$ increases, the coupling required to obtain the same $m_0$ becomes weaker. The decreasing time-dependent coupling reaches this value at progressively later times. The continuum limit thus takes the form
\begin{align}
\Lambda\to\infty,\qquad t_0\to\infty,\qquad
m_0,\ \alpha_0,\ \beta\ \text{fixed}.
\label{massContinuumLimit}
\end{align}
Solving the first line of Eq.~(\ref{GapEvolution}) gives the cutoff dependence of the characteristic time,
\begin{align}
t_0(\Lambda)
=\frac{1}{\pi\alpha_0}\ln\!\left(\frac{2\Lambda}{m_0}\right)
-\frac{\beta}{2\alpha_0}.
\label{CharacteristicTime}
\end{align}
Thus $t_0$ grows logarithmically with $\Lambda$, while $\beta$ sets the time origin.\footnote{At finite cutoff the spinon rest energy is $2\Lambda\arctan(e^{-\pi s(t)})$, with $s(t)=1/g_{\rm B}(t)$. Setting this energy equal to $m_0$ gives $t_0^{\rm reg}=-(\pi\alpha_0)^{-1} \ln\tan[m_0/(2\Lambda)]-\beta/(2\alpha_0)$. Its difference from Eq.~(\ref{CharacteristicTime}) is of order
$m_0^2/(\alpha_0\Lambda^2)$ and vanishes in the continuum limit.} At this time the Bethe coupling satisfies
\begin{align}
g_{\rm B}(t_0)=\frac{\pi}{\ln(2\Lambda/m_0)}
\longrightarrow 0.
\label{massReferenceCoupling}
\end{align}

Equation~(\ref{massReferenceCoupling}) fixes the coupling at the characteristic time $t_0(\Lambda)$ by requiring that the mass gap remain equal to $m_0$ in the continuum limit. To determine how the gap changes from this value, we evaluate it at a fixed elapsed time $\Delta t$ from $t_0(\Lambda)$. Over this interval the inverse Bethe coupling increases by $\alpha_0\Delta t$, so the finite-cutoff gap at $t=t_0(\Lambda)+\Delta t$ is $2\Lambda\arctan\!\left[(m_0/2\Lambda)e^{-\pi\alpha_0\Delta t}\right]$. Taking $\Lambda\to\infty$ at fixed $\Delta t$, we obtain the instantaneous mass gap $m(t_0+\Delta t)=m_0e^{-\pi\alpha_0\Delta t}$.

 When the drive suppresses transitions out of a prepared instantaneous eigenstate, the state follows the corresponding eigenstate of this continuum spectrum. In this sense, adiabatic evolution near $t_0$ is the dynamical realization of the scaling regime of the static model. To determine the time interval over which this description applies, we compare the rate of change of the gap with the gap itself.\begin{align}
\frac{|\dot m(t)|}{m(t)^2}=\frac{\pi\alpha_0}{m(t)}.
\label{massAdiabaticRatio}
\end{align}
This ratio compares the drive rate with the instantaneous mass gap. Along the decreasing time-dependent coupling, $|\dot g_{\rm B}|=\alpha_0g_{\rm B}^2$ decreases while the gap closes exponentially, so the ratio in Eq.~(\ref{massAdiabaticRatio}) grows. To determine how long a prepared state follows an instantaneous eigenstate, one therefore also needs the matrix elements of $\partial_t H$ and the relevant level separations. The saddle-point comparison was made first at fixed $N,M$ and $\alpha L\ll1$, whereas the dressed dispersion above is a bulk result obtained from the Bethe equations in the thermodynamic limit. At large finite $L$, both scales can be resolved when $m(t)L\gg1$ and $\alpha_0 L\ll1$. A joint limit of the full wavefunction would also require the saddle and adiabatic estimates to remain uniform as $N$ and $L$ increase. This distinction can also be seen by comparing two drive rates. At the same cutoff and with the same offset $\beta$, the equality
\begin{align}
\alpha_\ast t_\ast=\alpha_0t_0
\label{massEqualScaleTimes}
\end{align}
places the two instantaneous Hamiltonians at the same Bethe coupling and mass gap. With different offsets this requires $\alpha_\ast t_\ast+\beta_\ast/2=\alpha_0t_0+\beta_0/2$. Even at this common value of the gap, the two rates give different adiabaticity conditions, and the state produced by each time-dependent coupling depends on its preparation and subsequent evolution.

\subsection{Time-dependent coupling strength and the renormalization-group flow}
\label{sec:RG}

The continuum mass gap derived above allows us to compare two independent determinations of the coupling. In the static model, requiring the physical mass gap to remain fixed as the ultraviolet cutoff is removed determines how the coupling varies with the cutoff. In the time-dependent model, the Yang-Baxter equation determines how the coupling varies with physical time. By expressing both results in terms of the Bethe coupling $g_{\rm B}=2/C$ defined in Eq.~(\ref{BetheCoupling}), we can identify the regime in which they agree and explain what this agreement implies for the mass gap.

Consider first the static model, with the fermion density $\Lambda=N/L$ serving as the ultraviolet cutoff. In the continuum scaling regime, the mass gap is
\begin{align}
m=2\Lambda e^{-\pi/g_{\rm B}}.
\label{StaticGap}
\end{align}
To keep a prescribed physical gap $m>0$ fixed as $\Lambda$ increases, the coupling must vary according to
\begin{align}
g_{\rm B}(\Lambda)=\frac{\pi}{\ln(2\Lambda/m)}.
\label{StaticRunning}
\end{align}
The logarithmic decrease of $g_{\rm B}$ as the cutoff is removed expresses the asymptotic freedom of the static model. Differentiating Eq.~(\ref{StaticRunning}) at fixed $m$ gives the corresponding beta function in the normalization of the Bethe coupling,
\begin{align}
\left.\frac{d g_{\rm B}}{d\ln\Lambda}\right|_m
=-\frac{g_{\rm B}^2}{\pi}.
\label{StaticBeta}
\end{align}

For the time-dependent model, we obtain the evolution of the same coupling directly from the bare two-particle $S$-matrices. Their Yang-Baxter equation requires $C(t)=2\alpha t+\beta$, so that
$1/g_{\rm B}(t)=\alpha t+\beta/2$ and hence
\begin{align}
\frac{dg_{\rm B}}{dt}=-\alpha g_{\rm B}^2.
\label{DynamicBeta}
\end{align}
The relation to the static flow is most transparent in the inverse coupling. Equation~(\ref{StaticRunning}) makes $1/g_{\rm B}$ linear in $\ln\Lambda$, while the Yang-Baxter equation makes it linear in $t$. Choosing reference values $t_{\rm ref}$ and $\Lambda_{\rm ref}$ at which the couplings agree, we can write the correspondence as
\begin{align}
\ell_{\rm RG}=\pi\alpha(t-t_{\rm ref}),\qquad
\frac{dg_{\rm B}}{d\ell_{\rm RG}}
=-\frac{g_{\rm B}^2}{\pi},\qquad
\ell_{\rm RG}\longleftrightarrow
\ln\!\left(\frac{\Lambda}{\Lambda_{\rm ref}}\right).
\label{TimeRGMap}
\end{align}
With this choice of time scale and origin, the time-dependent Bethe coupling has the same inverse-affine form as the running coupling obtained from the static mass gap.

To compare this result with the classical Lax construction, we return to the microscopic interaction strength $g$. Near the WZW fixed point, the relation $g_{\rm B}=4g+\mathcal O(g^3)$ gives
\begin{align}
\left.\frac{dg}{d\ln\Lambda}\right|_m
=-\frac{4}{\pi}g^2+\cdots,\qquad
\frac{dg}{dt}=-4\alpha g^2+\cdots.
\label{massMicroscopicLeadingFlows}
\end{align}
The second equation is precisely the leading time dependence found from classical Lax flatness once the current normalization and time scale are fixed. Thus the classical Lax condition and the quantum Yang-Baxter equation independently select a coupling whose time dependence reproduces the static one-loop flow in the universal weak-coupling regime.

Beyond leading order, it matters how the coupling and the cutoff are defined. The Bethe coupling $g_{\rm B}$ parametrizes the bare two-particle $S$-matrix, and its relation to $g$ at finite coupling depends on the regularization of the delta-function interaction. For the two prescriptions in Eqs.~(\ref{instantaneousSymmetricContact}) and~(\ref{instantaneousExponentialContact}), expressing Eqs.~(\ref{StaticBeta}) and~(\ref{DynamicBeta}) in terms of $g$ produces the first correction to Eq.~(\ref{massMicroscopicLeadingFlows}) at order $g^4$. This correction comes from the nonlinear relation between the two bare couplings. In a field theory with one marginal renormalized coupling, an analytic redefinition $g_{\rm R}'=g_{\rm R}+a g_{\rm R}^2+\cdots$ preserves the first two beta-function coefficients when the linear normalization is fixed~\cite{LudwigWiese2003}. Extracting the two-loop coefficient from the Bethe ansatz would therefore require matching both $g_{\rm B}$ and the density cutoff $\Lambda$ to the coupling and regulator of the renormalized field theory at the corresponding order. The density-cutoff derivative in Eq.~(\ref{StaticBeta}) establishes the universal leading comparison.

The physical significance of Eq.~(\ref{TimeRGMap}) becomes clear when we follow the mass gap along the two trajectories. In the static model, the coupling varies with the cutoff precisely so that the gap remains fixed. During the time evolution, the cutoff is fixed, so the change in the coupling produces a change in the instantaneous mass gap. Substituting the two trajectories into Eq.~(\ref{StaticGap}) gives
\begin{align}
\left.\frac{d\ln m}{d\ln\Lambda}\right|_{\rm static\ RG}=0,
\qquad
\left.\frac{d\ln m(t)}{dt}\right|_\Lambda=-\pi\alpha.
\label{massRGAndPhysicalTime}
\end{align}
The characteristic time $t_0(\Lambda)$ connects these two statements. For a prescribed gap $m_0$, Eq.~(\ref{StaticRunning}) fixes the coupling required at cutoff $\Lambda$. The time-dependent coupling reaches precisely this value at $t_0(\Lambda)$, as shown in Eq.~(\ref{massReferenceCoupling}). Measuring time from this point gives $m(t_0+\Delta t)=m_0e^{-\pi\alpha_0\Delta t}$, which remains finite as the cutoff is removed at fixed $\Delta t$. In this way the time-dependent interaction generates a finite mass scale through the same dimensional transmutation that determines the gap in the static model.

This mass gap is a property of the instantaneous Bethe spectrum, obtained from the energy of a dressed hole in the filled Bethe sea. Its relevance to the evolving state follows from the wavefunction analysis. In the limit of slow variation, the saddle-point equations of the Yang-Yang action reproduce the instantaneous Bethe equations. Under the adiabatic conditions derived above, a state prepared in an isolated instantaneous eigenstate then follows the corresponding eigenstate as the gap changes. The next section examines the fast-drive limit of the same time-dependent interaction, in which the isospin sector approaches the $SU(2)_1$ WZW model with a weak current-current perturbation.

\section{Discussion of the fast-drive limit}
\label{sec:fast}

The exact GBA wavefunction describes the evolution for any nonzero drive rate. To understand what happens to the interaction as the rate increases, we start from the fermionic Hamiltonian. Separating the charge and spin currents as in Sec.~\ref{sec:fastWZW} gives
\begin{align}
H_{\rm GN}(t)
&= H_c+H_{\rm WZW}
   +u_\alpha(t)\int_0^L dx\,\mathbf J_R\cdot\mathbf J_L,
   \nonumber\\
H_{\rm WZW}
&= \frac{2\pi}{3}\int_0^L dx\,
   \bigl(:\mathbf J_R^2:_0+:\mathbf J_L^2:_0\bigr).
\label{fastWZWHamiltonian}
\end{align}
The charge sector remains decoupled throughout the evolution. The drive therefore affects the spin dynamics entirely through the coefficient $u_\alpha(t)=8g_\alpha(t)$ of the interaction between the two chiral currents.

We now examine how the spin Hamiltonian approaches the $SU(2)_1$ WZW fixed point in the fast-drive limit $\alpha L\to+\infty$, with the system length $L$ and the offset $\beta$ held fixed. On the decreasing positive-coupling branch, the inverse Bethe coupling is $1/g_{\rm B}(t)=C_\alpha(t)/2=\alpha t+\beta/2$. At $t=0$, the value $C_\alpha(0)=\beta$ is independent of the drive rate, so we consider a fixed interval $\delta\leq t\leq T$ with $\delta>0$. Since $C_\alpha(t)$ increases with time for $\alpha>0$, the Bethe coupling $g_{\rm B}(t)=2/C_\alpha(t)$ is largest at $t=\delta$. Its value there tends to zero in the fast-drive limit, so the weak-coupling relation between $g_{\rm B}(t)$ and $g_\alpha(t)$ in Eq.~(\ref{BetheCoupling}) applies throughout the interval and gives\begin{align}
g_\alpha(t)= \frac{1}{4\alpha t+2\beta}+&\mathcal O\!\left(C_\alpha(t)^{-3}\right),
\nonumber\\
\sup_{t\in[\delta,T]}|u_\alpha(t)|&\longrightarrow 0 .
\label{fastcoupling}
\end{align}
Thus, at large but finite $\alpha$, the spin Hamiltonian is a weak current-current perturbation of the $SU(2)_1$ WZW model throughout the interval. The regularization affects the higher-order terms in the relation between $C_\alpha$ and $g_\alpha$, while the leading interaction is universal. The equal-time current algebra has level one at every drive rate. As the interaction vanishes, the Sugawara Hamiltonian in Eq.~(\ref{fastWZWHamiltonian}) generates independent propagation of the right- and left-moving currents,
\begin{align}
(\partial_t+\partial_x)\mathbf J_R &= 0, &
(\partial_t-\partial_x)\mathbf J_L &= 0 .
\end{align}

This approach to the WZW fixed point can also be seen in the instantaneous mass gap. At fixed ultraviolet cutoff $\Lambda$, the dressed energy in Eq.~(\ref{massRegulatedDispersion}) gives the rest energy of a single bulk spinon as
\begin{align}
\epsilon_{\rm h}(0;t)
&=2\Lambda\arctan\!\left[e^{-\pi/g_{\rm B}(t)}\right],
\qquad
\frac{1}{g_{\rm B}(t)}=\alpha t+\frac{\beta}{2}.
\end{align}
The bulk mass gap therefore tends to zero in the fast-drive limit, while a system of finite length retains the conformal level spacings of order $1/L$. To relate this result to the characteristic time $t_0$ of Sec.~\ref{sec:massgap}, consider the reference drive rate $\alpha_0$ at the same cutoff and offset $\beta$. The continuum scaling form of the gap and its value $m_0=m_{\alpha_0}(t_0)$ give
\begin{align}
m_\alpha(t)
&=2\Lambda\exp\!\left[-\pi\left(\alpha t+\frac{\beta}{2}\right)\right],
\nonumber\\
\frac{m_\alpha(t)}{m_0}
&=\exp\!\left[-\pi\left(\alpha t-\alpha_0t_0\right)\right].
\label{fastReferenceMass}
\end{align}
The gap is exponentially small compared with $m_0$ once $\alpha t-\alpha_0t_0\gg1$. Increasing the drive rate can bring the Hamiltonian into this regime even at times $0<t<L$, while at fixed $\alpha_0$ the same weak-coupling regime is reached at late times. In both cases the instantaneous Hamiltonian approaches the WZW fixed point. The many-body state also depends on its preparation and on the evolution through which this regime is reached.

The Hamiltonian thus identifies the WZW model as the conformal theory about which the fast-drive expansion is taken. To compare its Yang-Baxter structure with that of the GBA, consider first a chiral conformal block $F_{\rm CFT}$ of the $SU(2)_1$ WZW model. Let $z_i$ denote the complex Euclidean coordinate of the $i$th primary field and $s_i^A$ the $SU(2)$ generator acting on its representation. By the current Ward identity, inserting a chiral current at $z$ gives
$\left(\sum_i\frac{s_i^A}{z-z_i}\right)F_{\rm CFT}$. Applying this identity to the two currents in the Sugawara stress tensor expresses the simple-pole residues of the stress-tensor insertion in terms of pairs of generators. The conformal Ward identity relates these residues to derivatives of the block, giving the differential Knizhnik-Zamolodchikov equations
\begin{align}
3\,\partial_{z_i}F_{\rm CFT}&=H_i^{\rm G}F_{\rm CFT},\nonumber\\
H_i^{\rm G}=\sum_{j\ne i}\frac{\Omega_{ij}}{z_i-z_j},&\qquad\Omega_{ij}=2\sum_A s_i^A s_j^A .
\label{fastKZ}
\end{align}
Here the coefficient $3$ is the level-one Sugawara factor $k+2$. The operators $H_i^{\rm G}$ are the rational Gaudin Hamiltonians. Their commutativity follows from the classical Yang-Baxter equation for the rational $r$-matrix. Writing $z_{ij}=z_i-z_j$ and
$r^{\rm G}_{ij}=\Omega_{ij}/z_{ij}$, we have
\begin{align}
[r^{\rm G}_{12},r^{\rm G}_{13}]
+[r^{\rm G}_{12},r^{\rm G}_{23}]
+[r^{\rm G}_{13},r^{\rm G}_{23}]=0 .
\end{align}
This identity follows from $z_{13}=z_{12}+z_{23}$ and the $SU(2)$ invariance of $\Omega$, which gives $[\Omega_{12},\Omega_{13}+\Omega_{23}]=0$ and its permutations. Here the $r$-matrices act on the representations of the primary fields in a quantum WZW conformal block \cite{KnizhnikZamolodchikov,FeiginFrenkelReshetikhin1994,BabujianFlume1994}. The Maillet kernels of Sec.~\ref{sec:classicalMaillet} act in the auxiliary spaces of the classical Lax matrix and determine its equal-time Poisson brackets.

The same rational spin tensor appears at weak but finite interaction in the GBA. For two spin-$1/2$ fermions, $\Omega_{ij}=P^{ij}-I^{ij}/2$. To see how this tensor enters the qKZ transport, take the quasi-classical expansion of each quantum $R$-matrix in Eq.~(\ref{uniformQKZTransport}). At fixed nonzero argument $v$, Eq.~(\ref{dimensionlessQKZData}) gives
\begin{align}
\widehat R^{ij}(v)
&=I^{ij}+\eta\,r^{\rm GBA}_{ij}(v)+\mathcal O(\eta^2),
\qquad \eta=\frac{1}{\alpha L},\\
r^{\rm GBA}_{ij}(v)
&=\frac{P^{ij}-I^{ij}}{iv}
 =\frac{\Omega_{ij}-I^{ij}/2}{iv}.
\end{align}
The factor $1/i$ follows from the spectral convention in the GBA, and the term proportional to the identity drops out of the commutators. The coefficient of the leading correction therefore satisfies the same rational classical Yang-Baxter equation. In the transport operator we keep the factors in their original order, expanding each at the argument it has in Eq.~(\ref{uniformQKZTransport}), including the unit shift where it occurs. At fixed particle number, retaining the first correction in every factor gives an error of order $\eta^2$ in the product when all arguments stay bounded away from zero. Since the shift $\zeta_i\to\zeta_i-1$ remains finite as $\eta\to0$, the GBA spin amplitudes satisfy finite-difference equations, which we call the discrete KZ equations. Determining whether these amplitudes approach the WZW conformal blocks of Eq.~(\ref{fastKZ}) requires an asymptotic analysis of the full qKZ wavefunction that retains this shift. A detailed analysis of the fast drive limit will be undertaken elsewhere.

\section{Conclusion and outlook}
\label{sec:conclusion}

In this work we considered an integrable sigma model and promoted its coupling strengths to be dependent on time. We asked whether the conditions on the time-dependent coupling imposed by integrability on the classical model and those imposed by integrability on the quantized model are the same. Following Hoare, Levine, and Tseytlin \cite{Benhoare}, we first constructed a generalized Lax connection for the isotropic current-current deformation of the $SU(2)$ WZW model directly in terms of its classical currents. Requiring this connection to be flat fixes the time dependence of the interaction and reproduces the result for the $\lambda$-model after a coupling redefinition and a reparametrization of time. We then quantized the current Hamiltonian at level one, where non-Abelian bosonization identifies it with the interacting isospin sector of the chiral-invariant $SU(2)$ Gross-Neveu model. Using its exact generalized Bethe ansatz solution \cite{PasnooriGrossNeveu}, we found that Yang-Baxter consistency of the bare two-particle $S$-matrices independently selects the same leading inverse-affine dependence on physical time as classical Lax flatness in the universal weak-coupling regime. We then show that these integrability-preserving time-dependent strengths follow the same trajectories as the RG equations corresponding to the static model upon identification of time with the logarithm of the cutoff. We then construct the exact wavefunction of the Gross-Neveu model and analyze it and extract physical consequences of the time-dependent coupling. 

Our classical construction starts from the WZW current Hamiltonian, with the level $k$ and the current Poisson brackets held fixed as the coupling $q(t)$ between the two chiral currents varies. We construct the generalized Lax connection directly from these currents. Its flatness determines the spatial and temporal derivatives of the spectral variable $y(t,x;w)$ at fixed spectral label $w$. Requiring these derivatives to be compatible gives $\dot q=Kq^2(1-q)$, where $K$ sets the rate and orientation of the drive. Integrating this equation fixes the coupling trajectory, along which we obtain $y(t,x;w)$ implicitly. The coupling redefinition $q=2\ell/(1+\ell^2)$ and the reparametrization of time $d\tau/dt=(1-\ell^2)/(1+\ell^2)$ relate this trajectory to that of the local-coupling $\lambda$-model \cite{Benhoare,Sfetsos2014,HollowoodMiramontesSchmidtt2014}. Expressing the Lax matrix in terms of the WZW currents also allows us to derive its equal-time Poisson algebra. The spatial dependence of $y$ contributes an additional derivative term to the generalized Maillet bracket, and the resulting $r/s$ kernels satisfy the extended classical Yang-Baxter relation. This spatial dependence also enters the construction of the monodromy. For a periodic system of length $L$, continuing $y$ along the same branch gives $y(t,x+L;w)=y(t,x;w+KL)$. A construction of global conserved charges from the monodromy must therefore account for the shift of the spectral label under transport through the length of the system.

At level one, non-Abelian bosonization identifies the current-current deformed WZW Hamiltonian with the interacting isospin sector of the chiral-invariant $SU(2)$ Gross-Neveu model. The level-one current algebra fixes the coefficient of the normal-ordered Sugawara Hamiltonian, while the coupling between the left and right currents becomes the time-dependent fermion interaction. In the generalized Bethe ansatz, this interaction enters the bare two-particle $S$-matrix through its value at the time a right mover and a left mover meet \cite{PasnooriGrossNeveu}. To construct the many-body wavefunction, different sequences of pairwise interchanges leading to the same particle ordering must give the same amplitude. This consistency requirement, expressed by the quantum Yang-Baxter equation, fixes the parameter in the bare $S$-matrix to be affine in the collision time, $C(t)=c(2t)=2\alpha t+\beta$. Its relation to the microscopic coupling then determines the allowed time-dependent strength for the chosen regularization. 

To compare the condition on the time-dependent coupling strength imposed by the quantum Yang-Baxter equation with the one obtained from classical Lax flatness, we express both in terms of the microscopic coupling $g(t)$. In the Gross-Neveu model, the relation between $g(t)$ and the parameter $C(t)$ of the bare two-particle $S$-matrix depends on how the delta-function interaction is regularized. All regularizations agree in the universal weak-coupling regime near the WZW fixed point, where $C\sim 1/(2g)$ in the normalization $u=8g$. In this regime, the affine dependence $C(t)=2\alpha t+\beta$ required by the quantum Yang-Baxter equation gives $\dot g\simeq-4\alpha g^2$. On the classical side, we identify the coefficient of $\mathbf J_R\cdot\mathbf J_L$ in the two Hamiltonians through $u=8g=4\pi q/k$ and fix the normalization of time by setting $K=-2\pi\alpha/k$. The classical Lax condition then gives the same equation for $g(t)$ to leading order. The inverse-affine time-dependent coupling thus follows
independently from classical flatness and from the consistency of the exact level-one quantum wavefunction. Under the time rescaling in Eq.~(\ref{TimeRGMap}), the dependence of the time-dependent coupling on physical time has the same form as the dependence of the static coupling on the logarithm of the ultraviolet scale. 

In the Gross-Neveu model, for the integrability-preserving time-dependent coupling strength imposed by the quantum Yang-Baxter equation, periodic boundary conditions on the wavefunction lead to the quantum Knizhnik-Zamolodchikov (qKZ) equations. Their off-shell Bethe-ansatz solution determines the explicit form of the exact many-body wavefunction, which admits an exact Yang-Yang representation at any nonzero drive rate. Its relation to the instantaneous spectrum emerges in the slow-variation regime near the characteristic time $t_0\gg L$, where the leading saddle-point equations and the associated isospin vectors at regular contributing saddles reproduce the Bethe equations and eigenvectors of the Hamiltonian at fixed time. When transitions out of a prepared, isolated eigenstate are suppressed, the wavefunction follows the corresponding instantaneous eigenstate to adiabatic accuracy and carries both its dynamical and geometric phases. The instantaneous spectrum exhibits dimensional transmutation: the dimensionless current-current coupling generates a mass gap $m(t)$ for spinon excitations, while the charge sector remains gapless. In the static model, fixing the physical mass gap determines how the coupling runs as the ultraviolet cutoff is removed. In the time-dependent model, the coupling varies in time at fixed cutoff, so the same mass relation determines how the instantaneous gap evolves. For a decreasing coupling with drive rate $\alpha_0>0$, requiring a finite reference mass gap $m(t_0)=m_0$ in the continuum limit fixes the characteristic time $t_0$ and gives $m(t_0+\Delta t)=m_0e^{-\pi\alpha_0\Delta t}$ at fixed elapsed time $\Delta t$.

The fast-drive limit probes the approach to the WZW fixed point directly through the Hamiltonian. For $\alpha L\to\infty$ at fixed $L$ and $\beta$, the interaction becomes uniformly small on finite positive-time intervals bounded away from $t=0$. At large but finite $\alpha$ the spin sector is a weakly perturbed $SU(2)_1$ WZW model, and its two chiral currents propagate independently in the limit. At fixed cutoff the bulk mass gap tends to zero, with conformal level spacings of order $1/L$ in a finite system. Comparison with the reference mass gap at $t_0$ shows how a faster drive reaches the weakly interacting regime earlier. Late evolution at fixed rate approaches the same limiting Hamiltonian, with the state in each case determined by its preparation and evolution. The Ward identities of the limiting WZW model give the rational Gaudin Hamiltonians in the differential KZ equations for conformal blocks. The same invariant spin tensor occurs in the quasi-classical expansion of the individual $R$-matrices in the GBA transport operators at separated spectral arguments. Here the coordinate shift remains finite, giving the discrete KZ equations for the many-body amplitudes. These two appearances of the rational tensor motivate studying the asymptotics of the full qKZ state to relate its solutions to the conformal blocks.

Open boundary conditions raise a further question for the classical sigma model. On an interval, bulk Lax flatness determines the time evolution of parallel transport in terms of the temporal Lax connection at the two ends. In the static $\lambda$-model, Driezen, Sevrin, and Thompson imposed boundary conditions that relate this connection at opposite values of the spectral parameter at each end. As a result, the boundary monodromy matrix constructed by transporting the auxiliary vector along the interval and back evolves by conjugation, so the traces of its powers are conserved. Geometrically, these boundary conditions constrain the open-string endpoints to D-branes that wrap twisted conjugacy classes of the group \cite{DriezenSevrinThompson2018}. In the time-dependent model studied here, the spectral variable itself depends on position and time. The classical question is therefore how it transforms under reflection and which boundary conditions allow the resulting boundary monodromy matrix to evolve by conjugation. On the quantum side, the open-boundary GBA gives the reflection equations for the time-dependent $SU(2)$ Gross-Neveu and $U(1)$ Thirring models \cite{PasnooriBGBA}.  Following a particle through successive reflections at the two ends then gives the boundary qKZ equations \cite{RishetikhinbqKZ} for the many-body amplitudes \cite{PasnooriBGBA}. This would allow a direct comparison of the boundary time-dependent couplings selected by the classical Lax construction and the quantum GBA. The case of open boundaries also has an important application in condensed matter physics: The open-boundary GBA would also give access to topological edge states when the current-current interaction of the $SU(2)_1$ WZW model is anisotropic. In this case, the interacting spin sector of the $U(1)$ Thirring model admits a sine-Gordon description in which the longitudinal and transverse current interactions are related to the boson stiffness and the cosine potential \cite{LeClairRomanSierra2003,PasnooriAnisotropic2026,PAA1,PAA2,SUSYpaper}. For charge-conjugation-symmetric Dirichlet boundary conditions, the sign of the cosine potential relative to the field values fixed at the two ends selects between a trivial phase and a symmetry-protected topological phase with zero-energy states localized at the ends in the thermodynamic limit \cite{circuitSPT,pasnoori2025duality,pasnoori2025interplay}. The boundary can also carry a quantum degree of freedom, such as a magnetic impurity whose coupling to the bulk current has its own renormalization-group flow \cite{parmeshkondo1,parmeshkondo2,SUSYpaper}. For a given integrable bulk time-dependent interaction strength, the boundary GBA can therefore be used to determine which time-dependent impurity couplings preserve integrability and to study the evolution of the impurity spin from the resulting many-body wavefunction.

At higher level $k$, the question is whether quantum Yang-Baxter consistency selects the same leading time dependence of the current-current coupling as classical Lax flatness in the universal weak-coupling regime. For a fixed integer $k>1$, the starting point is the static isospin Hamiltonian in Eq.~(\ref{quantumCurrentHamiltonian}), with $u$ the coefficient of $\mathbf J_R\cdot\mathbf J_L$. A fermionic realization uses $k$ flavors of isospin-$\tfrac12$ Dirac fermions, as in the multiflavor construction used by Polyakov and Wiegmann to study the $SU(2)$ principal chiral model \cite{PolyakovWiegmann1983}. For each chirality, summing the $k$ level-one isospin currents gives the level-$k$ current. The interaction therefore couples the total right-moving isospin current to the total left-moving one. Although the $U(1)$ charge and $SU(k)_2$ flavor sectors decouple from this interaction, the fermionic Hilbert space fixes which charge and flavor representations can accompany each $SU(2)_k$ isospin representation. The level also restricts the fusion of the isospin primaries, whose labels range from $j=0$ to $j=k/2$ in steps of one half. In the proposed physical $S$-matrix of the $SU(2)$ $\lambda$-model, this restriction appears in the scattering of kinks that connect neighboring labels in this range \cite{HollowoodMiramontesSchmidtt2016}. For the GBA, the starting point would be the bare two-particle $S$-matrices derived from the fermionic Hamiltonian with coupling $u(t)$. Requiring these matrices to satisfy the Yang-Baxter equations would determine which time dependences are compatible with the many-body wavefunction and allow a direct comparison with classical Lax flatness at the same level. Imposing periodic boundary conditions would then give matrix difference equations for the reference amplitude. If their solution admits a Yang-Yang representation, one could ask whether its saddle-point equations reduce, in a controlled limit of slow variation, to the Bethe equations derived independently from the Hamiltonian at fixed time. The dressed excitation energies obtained from those equations would determine the instantaneous isospin mass gap and its dependence on $k$. Matching the coupling and time coordinate to those of the $\lambda$-model at the same level would then allow one to investigate, in the semiclassical large-$k$ limit, how the coupling governing this quantum mass scale enters the time-dependent target-space metric and antisymmetric tensor \cite{Sfetsos2014,HollowoodMiramontesSchmidtt2014,Benhoare}.

A natural example for extending the GBA to sigma models based on noncompact Lie groups is the $SL(2,\mathbb R)$ WZW model on the universal cover of the group, which describes the $\mathrm{AdS}_3$ sector of a bosonic string with NS-NS flux. The undeformed WZW spectrum contains affine representations built on the discrete and principal continuous series, together with their images under integer spectral flow. After including the internal conformal field theory, one obtains the physical closed-string states by imposing the left- and right-moving Virasoro constraints and quotienting by physical null states \cite{MaldacenaOoguri2001}. For a time-dependent current-current deformation, the first step would be to identify the elementary degrees of freedom whose many-body states realize these WZW sectors. One could then derive the bare two-particle $S$-matrices from the deformed Hamiltonian and determine which time dependences of the coupling satisfy the Yang-Baxter equations. This would give a quantum condition to compare with classical Lax flatness. Constructing the wavefunction would further require specifying the measure and integration contours for continuous labels and showing that transporting one particle around the system maps an admissible amplitude to another amplitude in the same space. A string interpretation of the time-dependent theory would also require a target-space metric, antisymmetric tensor, and dilaton that satisfy the conditions for worldsheet Weyl invariance. The null backgrounds discussed by Hoare, Levine, and Tseytlin illustrate how an explicitly time-dependent coupling can arise in this way. In their construction, the metric and antisymmetric tensor of the transverse sigma model depend on a null target-space coordinate $u$, and the dilaton contains a term linear in the conjugate null coordinate $v$. Weyl invariance of the higher-dimensional background relates the $u$-dependence of the transverse couplings to their RG flow. Choosing light-cone gauge $u=\tau$ then turns this dependence into explicit dependence on worldsheet time \cite{Benhoare}. For the noncompact deformation, the question is whether the coupling selected by the GBA can be realized in a Weyl-invariant string background.

Four-dimensional Chern-Simons theory offers a complementary approach to time-dependent integrability. In the construction of Costello, Witten, and Yamazaki, Wilson lines extend along the topological plane at fixed points of the spectral plane. Perturbative quantization of the gauge theory assigns a quantum $R$-matrix to a crossing of two Wilson lines, and the equivalence of the two arrangements of three pairwise crossings gives the Yang-Baxter equation \cite{Costello1,Costello2}. In the same framework, coupling the gauge field to surface defects gives two-dimensional field theories with classical Lax connections \cite{costello3}. Komatsu, Sakamoto, Wallberg, and Yamazaki extended the classical construction by allowing the one-form in the action and the positions of the defects to depend on spacetime. Their examples include the Gross-Neveu model and the massless non-Abelian Thirring model. In the latter, the coupling between the left and right fermionic currents is the inverse of the separation of two defects in the spectral plane. This separation varies linearly with physical time, giving an inverse-affine coupling whose time dependence follows the one-loop RG flow of the static model up to a rescaling of time \cite{KomatsuSakamotoWallbergYamazaki2026}. For $SU(2)$, this is the same isotropic current-current interaction that governs the isospin sector of the chiral-invariant Gross-Neveu model studied here. The next step is to quantize the four-dimensional gauge theory in the presence of these spacetime-dependent defects. The GBA gives a concrete test of such a quantization. One can ask whether Wilson-line crossings reproduce the bare two-particle $S$-matrices, including their scalar phases and their dependence on the particle coordinates. With periodic boundary conditions, one can then ask whether transporting a Wilson line around the system reproduces the finite shifts in the qKZ equations and which boundary conditions in the gauge theory select the corresponding off-shell Bethe ansatz solution. Conversely, the gauge theory could suggest other representations, interactions, and boundary conditions for which the GBA can be used to construct exact wavefunctions. 

\section{Acknowledgments}

We thank N. Andrei, P. Azaria, C. Rylands and J. D. Sau for discussions related to the fermionic model, and R. Thorngren and P. Kraus for discussions on conformal field theory, nonlinear sigma models and their applications to string theory. We acknowledge the use of ChatGPT in our literature review and in improving the language and presentation of this work.

 \bibliographystyle{JHEP}
 
\bibliography{refpaper_v7}

\end{document}